\documentclass[11pt, letterpaper, onecolumn, peerreview]{IEEEtran}
\usepackage{exscale}
\usepackage{color}
\usepackage{colordvi}
\usepackage{fancybox,epsfig}
\usepackage{graphics}
\usepackage{amsmath}
\usepackage{amssymb}
\usepackage{epsfig}
\usepackage{epic}
\usepackage{afterpage}

\newtheorem{theorem}{Theorem}[section]
\newtheorem{lemma}{Lemma}[section]

\newtheorem{definition}{Definition}[section]
\newtheorem{corollary}{Corollary}[section]

\DeclareMathOperator*{\argmax}{argmax}

\begin{document}
\title{\large \bf Why block length and delay behave differently if feedback is present} 

\author{Anant Sahai\footnote{Wireless Foundations, Department of
Electrical Engineering and Computer Science at the University of
California at Berkeley.}  \\ {\small
sahai@eecs.berkeley.edu}}


\maketitle

\begin{abstract} 
  For output-symmetric DMCs at even moderately high rates, fixed-block-length
  communication systems show no improvements in their error exponents
  with feedback. In this paper, we study systems with fixed end-to-end
  delay and show that feedback generally provides dramatic gains in
  the error exponents.

  A new upper bound (the uncertainty-focusing bound) is given on
  the probability of symbol error in a fixed-delay communication
  system with feedback. This bound turns out to have a similar form to
  Viterbi's bound used for the block error probability of
  convolutional codes as a function of the fixed constraint length.
  The uncertainty-focusing bound is shown to be asymptotically
  achievable with noiseless feedback for erasure channels as well as
  any output-symmetric DMC that has strictly positive zero-error
  capacity. Furthermore, it can be achieved in a delay-universal
  (anytime) fashion even if the feedback itself is delayed by a small 
  amount. Finally, it is shown that for end-to-end delay, it is
  generally possible at high rates to beat the sphere-packing bound
  for general DMCs --- thereby providing a counterexample to a
  conjecture of Pinsker.
\end{abstract}


\IEEEpeerreviewmaketitle

\section{Introduction}
The channel coding theorems studied in information theory are not just
interesting as mathematical results, they also provide insights into
the underlying tradeoffs in reliable communication systems. While in
practice there are many different parameters of interest such as
power, complexity, and robustness, perhaps the most fundamental two
are end-to-end system delay and the probability of error. Error
probability is fundamental because a low probability of bit error lies
at the heart of the digital revolution justified by the source/channel
separation theorem. Delay is important because it is the most basic
cost that must be paid in exchange for reliability --- it allows the
laws of large numbers to be harnessed to smooth out the variability
introduced by random communication channels.

In our entire discussion, the assumption is that information naturally
arises as a stream generated in real time at the source (e.g.~voice,
video, or sensor measurements) and it is useful to the destination in
finely grained increments (e.g.~a few milliseconds of voice, a single
video frame, etc.). The acceptable end-to-end delay is determined by
the application and can often be much larger than the natural
granularity of the information being communicated (e.g.~voice may
tolerate a delay of hundreds of milliseconds despite being useful in
increments of a few milliseconds). This is different from cases in
which information arises in large bursts with each burst needing to be
received by the destination before the next burst even becomes
available at the source.

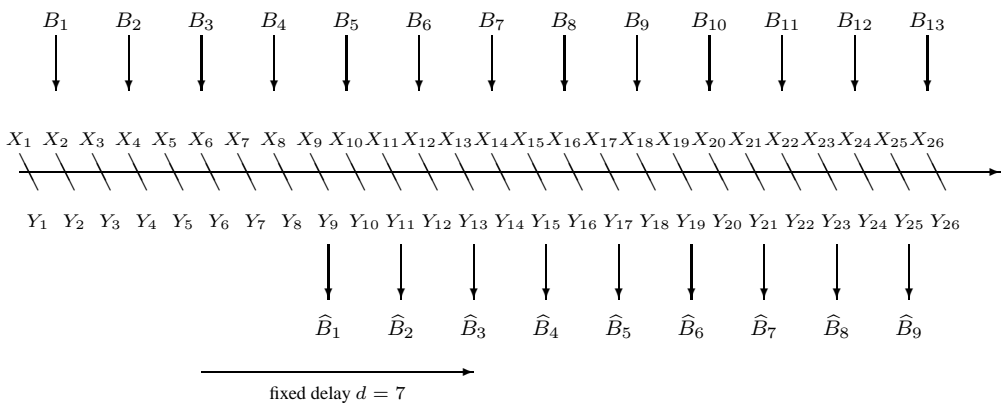
\begin{figure}[htbp]
\begin{center}
\setlength{\unitlength}{3000sp}%
\begingroup\makeatletter\ifx\SetFigFont\undefined%
\gdef\SetFigFont#1#2#3#4#5{%
  \reset@font\fontsize{#1}{#2pt}%
  \fontfamily{#3}\fontseries{#4}\fontshape{#5}%
  \selectfont}%
\fi\endgroup%
\begin{picture}(8124,3289)(439,-2594)
\thinlines
{\color[rgb]{0,0,0}\put(451,-661){\vector( 1, 0){8100}}
}%
{\color[rgb]{0,0,0}\put(451,-511){\line( 1,-2){150}}
}%
{\color[rgb]{0,0,0}\put(751,-511){\line( 1,-2){150}}
}%
{\color[rgb]{0,0,0}\put(1051,-511){\line( 1,-2){150}}
}%
{\color[rgb]{0,0,0}\put(1351,-511){\line( 1,-2){150}}
}%
{\color[rgb]{0,0,0}\put(1651,-511){\line( 1,-2){150}}
}%
{\color[rgb]{0,0,0}\put(1951,-511){\line( 1,-2){150}}
}%
{\color[rgb]{0,0,0}\put(2251,-511){\line( 1,-2){150}}
}%
{\color[rgb]{0,0,0}\put(2551,-511){\line( 1,-2){150}}
}%
{\color[rgb]{0,0,0}\put(2851,-511){\line( 1,-2){150}}
}%
{\color[rgb]{0,0,0}\put(3151,-511){\line( 1,-2){150}}
}%
{\color[rgb]{0,0,0}\put(3451,-511){\line( 1,-2){150}}
}%
{\color[rgb]{0,0,0}\put(3751,-511){\line( 1,-2){150}}
}%
{\color[rgb]{0,0,0}\put(4051,-511){\line( 1,-2){150}}
}%
{\color[rgb]{0,0,0}\put(4351,-511){\line( 1,-2){150}}
}%
{\color[rgb]{0,0,0}\put(4651,-511){\line( 1,-2){150}}
}%
{\color[rgb]{0,0,0}\put(4951,-511){\line( 1,-2){150}}
}%
{\color[rgb]{0,0,0}\put(5251,-511){\line( 1,-2){150}}
}%
{\color[rgb]{0,0,0}\put(5551,-511){\line( 1,-2){150}}
}%
{\color[rgb]{0,0,0}\put(5851,-511){\line( 1,-2){150}}
}%
{\color[rgb]{0,0,0}\put(6151,-511){\line( 1,-2){150}}
}%
{\color[rgb]{0,0,0}\put(6451,-511){\line( 1,-2){150}}
}%
{\color[rgb]{0,0,0}\put(6751,-511){\line( 1,-2){150}}
}%
{\color[rgb]{0,0,0}\put(7051,-511){\line( 1,-2){150}}
}%
{\color[rgb]{0,0,0}\put(7351,-511){\line( 1,-2){150}}
}%
{\color[rgb]{0,0,0}\put(7651,-511){\line( 1,-2){150}}
}%
{\color[rgb]{0,0,0}\put(7951,-511){\line( 1,-2){150}}
}%
{\color[rgb]{0,0,0}\put(1951,-2311){\vector( 1, 0){2250}}
}%
{\color[rgb]{0,0,0}\put(3001,-1261){\vector( 0,-1){450}}
}%
{\color[rgb]{0,0,0}\put(3601,-1261){\vector( 0,-1){450}}
}%
{\color[rgb]{0,0,0}\put(4201,-1261){\vector( 0,-1){450}}
}%
{\color[rgb]{0,0,0}\put(4801,-1261){\vector( 0,-1){450}}
}%
{\color[rgb]{0,0,0}\put(5401,-1261){\vector( 0,-1){450}}
}%
{\color[rgb]{0,0,0}\put(6001,-1261){\vector( 0,-1){450}}
}%
{\color[rgb]{0,0,0}\put(6601,-1261){\vector( 0,-1){450}}
}%
{\color[rgb]{0,0,0}\put(7201,-1261){\vector( 0,-1){450}}
}%
{\color[rgb]{0,0,0}\put(7801,-1261){\vector( 0,-1){450}}
}%
{\color[rgb]{0,0,0}\put(6751,464){\vector( 0,-1){450}}
}%
{\color[rgb]{0,0,0}\put(7351,464){\vector( 0,-1){450}}
}%
{\color[rgb]{0,0,0}\put(7951,464){\vector( 0,-1){450}}
}%
{\color[rgb]{0,0,0}\put(751,464){\vector( 0,-1){450}}
}%
{\color[rgb]{0,0,0}\put(1351,464){\vector( 0,-1){450}}
}%
{\color[rgb]{0,0,0}\put(1951,464){\vector( 0,-1){450}}
}%
{\color[rgb]{0,0,0}\put(2551,464){\vector( 0,-1){450}}
}%
{\color[rgb]{0,0,0}\put(3151,464){\vector( 0,-1){450}}
}%
{\color[rgb]{0,0,0}\put(3751,464){\vector( 0,-1){450}}
}%
{\color[rgb]{0,0,0}\put(4351,464){\vector( 0,-1){450}}
}%
{\color[rgb]{0,0,0}\put(4951,464){\vector( 0,-1){450}}
}%
{\color[rgb]{0,0,0}\put(5551,464){\vector( 0,-1){450}}
}%
{\color[rgb]{0,0,0}\put(6151,464){\vector( 0,-1){450}}
}%
\put(451,-436){\makebox(0,0)[b]{\smash{\SetFigFont{7}{6}{\rmdefault}{\mddefault}{\updefault}{\color[rgb]{0,0,0}$X_{1}$}%
}}}
\put(751,-436){\makebox(0,0)[b]{\smash{\SetFigFont{7}{6}{\rmdefault}{\mddefault}{\updefault}{\color[rgb]{0,0,0}$X_{2}$}%
}}}
\put(1051,-436){\makebox(0,0)[b]{\smash{\SetFigFont{7}{6}{\rmdefault}{\mddefault}{\updefault}{\color[rgb]{0,0,0}$X_{3}$}%
}}}
\put(1351,-436){\makebox(0,0)[b]{\smash{\SetFigFont{7}{6}{\rmdefault}{\mddefault}{\updefault}{\color[rgb]{0,0,0}$X_{4}$}%
}}}
\put(1651,-436){\makebox(0,0)[b]{\smash{\SetFigFont{7}{6}{\rmdefault}{\mddefault}{\updefault}{\color[rgb]{0,0,0}$X_{5}$}%
}}}
\put(1951,-436){\makebox(0,0)[b]{\smash{\SetFigFont{7}{6}{\rmdefault}{\mddefault}{\updefault}{\color[rgb]{0,0,0}$X_{6}$}%
}}}
\put(2251,-436){\makebox(0,0)[b]{\smash{\SetFigFont{7}{6}{\rmdefault}{\mddefault}{\updefault}{\color[rgb]{0,0,0}$X_{7}$}%
}}}
\put(2551,-436){\makebox(0,0)[b]{\smash{\SetFigFont{7}{6}{\rmdefault}{\mddefault}{\updefault}{\color[rgb]{0,0,0}$X_{8}$}%
}}}
\put(2851,-436){\makebox(0,0)[b]{\smash{\SetFigFont{7}{6}{\rmdefault}{\mddefault}{\updefault}{\color[rgb]{0,0,0}$X_{9}$}%
}}}
\put(3151,-436){\makebox(0,0)[b]{\smash{\SetFigFont{7}{6}{\rmdefault}{\mddefault}{\updefault}{\color[rgb]{0,0,0}$X_{10}$}%
}}}
\put(3451,-436){\makebox(0,0)[b]{\smash{\SetFigFont{7}{6}{\rmdefault}{\mddefault}{\updefault}{\color[rgb]{0,0,0}$X_{11}$}%
}}}
\put(3751,-436){\makebox(0,0)[b]{\smash{\SetFigFont{7}{6}{\rmdefault}{\mddefault}{\updefault}{\color[rgb]{0,0,0}$X_{12}$}%
}}}
\put(4051,-436){\makebox(0,0)[b]{\smash{\SetFigFont{7}{6}{\rmdefault}{\mddefault}{\updefault}{\color[rgb]{0,0,0}$X_{13}$}%
}}}
\put(4351,-436){\makebox(0,0)[b]{\smash{\SetFigFont{7}{6}{\rmdefault}{\mddefault}{\updefault}{\color[rgb]{0,0,0}$X_{14}$}%
}}}
\put(4651,-436){\makebox(0,0)[b]{\smash{\SetFigFont{7}{6}{\rmdefault}{\mddefault}{\updefault}{\color[rgb]{0,0,0}$X_{15}$}%
}}}
\put(4951,-436){\makebox(0,0)[b]{\smash{\SetFigFont{7}{6}{\rmdefault}{\mddefault}{\updefault}{\color[rgb]{0,0,0}$X_{16}$}%
}}}
\put(5251,-436){\makebox(0,0)[b]{\smash{\SetFigFont{7}{6}{\rmdefault}{\mddefault}{\updefault}{\color[rgb]{0,0,0}$X_{17}$}%
}}}
\put(5551,-436){\makebox(0,0)[b]{\smash{\SetFigFont{7}{6}{\rmdefault}{\mddefault}{\updefault}{\color[rgb]{0,0,0}$X_{18}$}%
}}}
\put(5851,-436){\makebox(0,0)[b]{\smash{\SetFigFont{7}{6}{\rmdefault}{\mddefault}{\updefault}{\color[rgb]{0,0,0}$X_{19}$}%
}}}
\put(6151,-436){\makebox(0,0)[b]{\smash{\SetFigFont{7}{6}{\rmdefault}{\mddefault}{\updefault}{\color[rgb]{0,0,0}$X_{20}$}%
}}}
\put(6451,-436){\makebox(0,0)[b]{\smash{\SetFigFont{7}{6}{\rmdefault}{\mddefault}{\updefault}{\color[rgb]{0,0,0}$X_{21}$}%
}}}
\put(6751,-436){\makebox(0,0)[b]{\smash{\SetFigFont{7}{6}{\rmdefault}{\mddefault}{\updefault}{\color[rgb]{0,0,0}$X_{22}$}%
}}}
\put(7051,-436){\makebox(0,0)[b]{\smash{\SetFigFont{7}{6}{\rmdefault}{\mddefault}{\updefault}{\color[rgb]{0,0,0}$X_{23}$}%
}}}
\put(7351,-436){\makebox(0,0)[b]{\smash{\SetFigFont{7}{6}{\rmdefault}{\mddefault}{\updefault}{\color[rgb]{0,0,0}$X_{24}$}%
}}}
\put(7651,-436){\makebox(0,0)[b]{\smash{\SetFigFont{7}{6}{\rmdefault}{\mddefault}{\updefault}{\color[rgb]{0,0,0}$X_{25}$}%
}}}
\put(7951,-436){\makebox(0,0)[b]{\smash{\SetFigFont{7}{6}{\rmdefault}{\mddefault}{\updefault}{\color[rgb]{0,0,0}$X_{26}$}%
}}}
\put(601,-1111){\makebox(0,0)[b]{\smash{\SetFigFont{7}{6}{\rmdefault}{\mddefault}{\updefault}{\color[rgb]{0,0,0}$Y_{1}$}%
}}}
\put(901,-1111){\makebox(0,0)[b]{\smash{\SetFigFont{7}{6}{\rmdefault}{\mddefault}{\updefault}{\color[rgb]{0,0,0}$Y_{2}$}%
}}}
\put(1201,-1111){\makebox(0,0)[b]{\smash{\SetFigFont{7}{6}{\rmdefault}{\mddefault}{\updefault}{\color[rgb]{0,0,0}$Y_{3}$}%
}}}
\put(1501,-1111){\makebox(0,0)[b]{\smash{\SetFigFont{7}{6}{\rmdefault}{\mddefault}{\updefault}{\color[rgb]{0,0,0}$Y_{4}$}%
}}}
\put(1801,-1111){\makebox(0,0)[b]{\smash{\SetFigFont{7}{6}{\rmdefault}{\mddefault}{\updefault}{\color[rgb]{0,0,0}$Y_{5}$}%
}}}
\put(2101,-1111){\makebox(0,0)[b]{\smash{\SetFigFont{7}{6}{\rmdefault}{\mddefault}{\updefault}{\color[rgb]{0,0,0}$Y_{6}$}%
}}}
\put(2401,-1111){\makebox(0,0)[b]{\smash{\SetFigFont{7}{6}{\rmdefault}{\mddefault}{\updefault}{\color[rgb]{0,0,0}$Y_{7}$}%
}}}
\put(2701,-1111){\makebox(0,0)[b]{\smash{\SetFigFont{7}{6}{\rmdefault}{\mddefault}{\updefault}{\color[rgb]{0,0,0}$Y_{8}$}%
}}}
\put(3001,-1111){\makebox(0,0)[b]{\smash{\SetFigFont{7}{6}{\rmdefault}{\mddefault}{\updefault}{\color[rgb]{0,0,0}$Y_{9}$}%
}}}
\put(3301,-1111){\makebox(0,0)[b]{\smash{\SetFigFont{7}{6}{\rmdefault}{\mddefault}{\updefault}{\color[rgb]{0,0,0}$Y_{10}$}%
}}}
\put(3601,-1111){\makebox(0,0)[b]{\smash{\SetFigFont{7}{6}{\rmdefault}{\mddefault}{\updefault}{\color[rgb]{0,0,0}$Y_{11}$}%
}}}
\put(3901,-1111){\makebox(0,0)[b]{\smash{\SetFigFont{7}{6}{\rmdefault}{\mddefault}{\updefault}{\color[rgb]{0,0,0}$Y_{12}$}%
}}}
\put(4201,-1111){\makebox(0,0)[b]{\smash{\SetFigFont{7}{6}{\rmdefault}{\mddefault}{\updefault}{\color[rgb]{0,0,0}$Y_{13}$}%
}}}
\put(4501,-1111){\makebox(0,0)[b]{\smash{\SetFigFont{7}{6}{\rmdefault}{\mddefault}{\updefault}{\color[rgb]{0,0,0}$Y_{14}$}%
}}}
\put(4801,-1111){\makebox(0,0)[b]{\smash{\SetFigFont{7}{6}{\rmdefault}{\mddefault}{\updefault}{\color[rgb]{0,0,0}$Y_{15}$}%
}}}
\put(5101,-1111){\makebox(0,0)[b]{\smash{\SetFigFont{7}{6}{\rmdefault}{\mddefault}{\updefault}{\color[rgb]{0,0,0}$Y_{16}$}%
}}}
\put(5401,-1111){\makebox(0,0)[b]{\smash{\SetFigFont{7}{6}{\rmdefault}{\mddefault}{\updefault}{\color[rgb]{0,0,0}$Y_{17}$}%
}}}
\put(5701,-1111){\makebox(0,0)[b]{\smash{\SetFigFont{7}{6}{\rmdefault}{\mddefault}{\updefault}{\color[rgb]{0,0,0}$Y_{18}$}%
}}}
\put(6001,-1111){\makebox(0,0)[b]{\smash{\SetFigFont{7}{6}{\rmdefault}{\mddefault}{\updefault}{\color[rgb]{0,0,0}$Y_{19}$}%
}}}
\put(6301,-1111){\makebox(0,0)[b]{\smash{\SetFigFont{7}{6}{\rmdefault}{\mddefault}{\updefault}{\color[rgb]{0,0,0}$Y_{20}$}%
}}}
\put(6601,-1111){\makebox(0,0)[b]{\smash{\SetFigFont{7}{6}{\rmdefault}{\mddefault}{\updefault}{\color[rgb]{0,0,0}$Y_{21}$}%
}}}
\put(6901,-1111){\makebox(0,0)[b]{\smash{\SetFigFont{7}{6}{\rmdefault}{\mddefault}{\updefault}{\color[rgb]{0,0,0}$Y_{22}$}%
}}}
\put(7201,-1111){\makebox(0,0)[b]{\smash{\SetFigFont{7}{6}{\rmdefault}{\mddefault}{\updefault}{\color[rgb]{0,0,0}$Y_{23}$}%
}}}
\put(7501,-1111){\makebox(0,0)[b]{\smash{\SetFigFont{7}{6}{\rmdefault}{\mddefault}{\updefault}{\color[rgb]{0,0,0}$Y_{24}$}%
}}}
\put(7801,-1111){\makebox(0,0)[b]{\smash{\SetFigFont{7}{6}{\rmdefault}{\mddefault}{\updefault}{\color[rgb]{0,0,0}$Y_{25}$}%
}}}
\put(8101,-1111){\makebox(0,0)[b]{\smash{\SetFigFont{7}{6}{\rmdefault}{\mddefault}{\updefault}{\color[rgb]{0,0,0}$Y_{26}$}%
}}}
\put(3076,-2536){\makebox(0,0)[b]{\smash{\SetFigFont{7}{6}{\rmdefault}{\mddefault}{\updefault}{\color[rgb]{0,0,0}fixed delay $d=7$}%
}}}
\put(6001,-2011){\makebox(0,0)[b]{\smash{\SetFigFont{8}{6}{\rmdefault}{\mddefault}{\updefault}{\color[rgb]{0,0,0}$\widehat{B}_6$}%
}}}
\put(7201,-2011){\makebox(0,0)[b]{\smash{\SetFigFont{8}{6}{\rmdefault}{\mddefault}{\updefault}{\color[rgb]{0,0,0}$\widehat{B}_8$}%
}}}
\put(3001,-2011){\makebox(0,0)[b]{\smash{\SetFigFont{8}{6}{\rmdefault}{\mddefault}{\updefault}{\color[rgb]{0,0,0}$\widehat{B}_1$}%
}}}
\put(3601,-2011){\makebox(0,0)[b]{\smash{\SetFigFont{8}{6}{\rmdefault}{\mddefault}{\updefault}{\color[rgb]{0,0,0}$\widehat{B}_2$}%
}}}
\put(4201,-2011){\makebox(0,0)[b]{\smash{\SetFigFont{8}{6}{\rmdefault}{\mddefault}{\updefault}{\color[rgb]{0,0,0}$\widehat{B}_3$}%
}}}
\put(4801,-2011){\makebox(0,0)[b]{\smash{\SetFigFont{8}{6}{\rmdefault}{\mddefault}{\updefault}{\color[rgb]{0,0,0}$\widehat{B}_4$}%
}}}
\put(5401,-2011){\makebox(0,0)[b]{\smash{\SetFigFont{8}{6}{\rmdefault}{\mddefault}{\updefault}{\color[rgb]{0,0,0}$\widehat{B}_5$}%
}}}
\put(6601,-2011){\makebox(0,0)[b]{\smash{\SetFigFont{8}{6}{\rmdefault}{\mddefault}{\updefault}{\color[rgb]{0,0,0}$\widehat{B}_7$}%
}}}
\put(7801,-2011){\makebox(0,0)[b]{\smash{\SetFigFont{8}{6}{\rmdefault}{\mddefault}{\updefault}{\color[rgb]{0,0,0}$\widehat{B}_9$}%
}}}
\put(751,539){\makebox(0,0)[b]{\smash{\SetFigFont{8}{6}{\rmdefault}{\mddefault}{\updefault}{\color[rgb]{0,0,0}$B_1$}%
}}}
\put(1351,539){\makebox(0,0)[b]{\smash{\SetFigFont{8}{6}{\rmdefault}{\mddefault}{\updefault}{\color[rgb]{0,0,0}$B_2$}%
}}}
\put(1951,539){\makebox(0,0)[b]{\smash{\SetFigFont{8}{6}{\rmdefault}{\mddefault}{\updefault}{\color[rgb]{0,0,0}$B_3$}%
}}}
\put(2551,539){\makebox(0,0)[b]{\smash{\SetFigFont{8}{6}{\rmdefault}{\mddefault}{\updefault}{\color[rgb]{0,0,0}$B_4$}%
}}}
\put(3151,539){\makebox(0,0)[b]{\smash{\SetFigFont{8}{6}{\rmdefault}{\mddefault}{\updefault}{\color[rgb]{0,0,0}$B_5$}%
}}}
\put(3751,539){\makebox(0,0)[b]{\smash{\SetFigFont{8}{6}{\rmdefault}{\mddefault}{\updefault}{\color[rgb]{0,0,0}$B_6$}%
}}}
\put(4351,539){\makebox(0,0)[b]{\smash{\SetFigFont{8}{6}{\rmdefault}{\mddefault}{\updefault}{\color[rgb]{0,0,0}$B_7$}%
}}}
\put(4951,539){\makebox(0,0)[b]{\smash{\SetFigFont{8}{6}{\rmdefault}{\mddefault}{\updefault}{\color[rgb]{0,0,0}$B_8$}%
}}}
\put(5551,539){\makebox(0,0)[b]{\smash{\SetFigFont{8}{6}{\rmdefault}{\mddefault}{\updefault}{\color[rgb]{0,0,0}$B_9$}%
}}}
\put(6151,539){\makebox(0,0)[b]{\smash{\SetFigFont{8}{6}{\rmdefault}{\mddefault}{\updefault}{\color[rgb]{0,0,0}$B_{10}$}%
}}}
\put(6751,539){\makebox(0,0)[b]{\smash{\SetFigFont{8}{6}{\rmdefault}{\mddefault}{\updefault}{\color[rgb]{0,0,0}$B_{11}$}%
}}}
\put(7351,539){\makebox(0,0)[b]{\smash{\SetFigFont{8}{6}{\rmdefault}{\mddefault}{\updefault}{\color[rgb]{0,0,0}$B_{12}$}%
}}}
\put(7951,539){\makebox(0,0)[b]{\smash{\SetFigFont{8}{6}{\rmdefault}{\mddefault}{\updefault}{\color[rgb]{0,0,0}$B_{13}$}%
}}}
\end{picture}
\end{center}
\caption{The timeline in a rate-$\frac{1}{2}$ code with decoding delay
  $7$. Both the encoder and decoder must be causal in that the channel
  inputs $X_i$ and decoded bits $\widehat{B}_i$ are functions only of
  quantities to the left of them on the timeline. If noiseless
  feedback is available, the $X_i$ can also have an explicit
  functional dependence on the channel outputs $Y_1^{i-1}$ that 
  lie to the left on the timeline.}
\label{fig:timeline}
\end{figure}

Rather than worrying about what the appropriate granularity of
information should be, the formal problem is specified at the
individual bit level. (See Figure~\ref{fig:timeline}.) If a bit is not
delivered correctly by its deadline, it is considered to be
erroneous. The upper and lower bounds of this paper turn out to not
depend on the choice of information granularity, only on the fact that
the granularity is much finer than the tolerable end-to-end delay.

In the next section of this introduction, the example of the binary
erasure channel at $R = \frac{1}{2}$ bits per channel use is used to
constructively show how fixed-delay codes can dramatically outperform
fixed-block-length codes at the same rates when feedback is
present. Existing information-theoretic views of feedback and
reliability are then reviewed in
Section~\ref{sec:background}. Section~\ref{sec:mainresults} states the
main results of the paper, with the constructions and proofs following
in subsequent sections. Numerical examples and plots are also given in
Section~\ref{sec:mainresults} to illustrate these results.

Section~\ref{sec:nofeedback} generalizes Pinsker's result from
\cite{PinskerNoFeedback} for non-block-code performance with fixed
delay and also explains why, contrary to Pinsker's assertion, this
argument {\em does not} generalize to the case when feedback is
present. The new upper bound (the ``uncertainty-focusing bound'') on
fixed-delay performance is proved in Section~\ref{sec:feedbackbound}
by reviving Forney's inverse concatenation construction to serve this
new purpose. Asymptotic achievability of this new bound with noiseless
feedback is shown in Section~\ref{sec:erasurefeedback} for erasure
channels. These results are extended in
Section~\ref{sec:fortifiedfeedback} to general DMCs. It turns out that
for channels with strictly positive feedback-zero-error capacity, a
low-rate error-free path can be constructed with very little overhead
thereby attaining the performance of the uncertainty-focusing
bound. For generic channels at high message rates, the overhead of
this approach is non-negligible but the error probability still
asymptotically beats that predicted by the sphere-packing bound for
the same end-to-end delay.

\subsection{A simple example using the BEC} \label{sec:becexample}

The natural question of end-to-end delay in situations with finely
grained information was considered by Pinsker in
\cite{PinskerNoFeedback}. He explicitly treats the BSC case, while
asserting that the results hold for any DMC. The main result
(Theorem~5 in \cite{PinskerNoFeedback}) is that the sphere-packing
bound $E_{sp}(R)$ is an upper bound to the fixed-delay error exponent
for any nonblock code. Theorem~8 in \cite{PinskerNoFeedback} asserts
that the same bound continues to hold even with feedback. As reviewed
in Section~\ref{sec:fixedlengthreview}, these theorems parallel what
is already known to hold for fixed-block-length codes.

The binary erasure channel (BEC) with erasure probability $\beta <
\frac{1}{2}$ used at rate $R' = \frac{1}{2}$ bits per channel use
gives a counterexample to Pinsker's generalized conjecture. The BEC is
so simple that everything can be understood with a minimum of
overhead. A counterexample that covers the BSC itself is given later
in Section~\ref{sec:comments} (plotted in
Figure~\ref{fig:beatspherebsc}) and others are given in
\cite{TuncThesis, SimsekJainVaraiya} using much more involved codes
built around control-theoretic ideas.

The sphere-packing bound in the BEC case corresponds to the
probability that the channel erases more than $\frac{1}{2}$ of the
inputs during the block:
\begin{equation} \label{eqn:becspherebound}
 E_{sp}(\frac{1}{2}) = D(\frac{1}{2}||\beta) = 
                     - \frac{\ln(4\beta(1-\beta))}{2}.
\end{equation}
For $\beta = 0.4$, this yields an error exponent of about $0.02$. Even
with feedback, there is no way for a fixed-block-length code to beat
this exponent. If the channel lets fewer than $\frac{n}{2}$ bits
through, it is impossible to reliably communicate an $\frac{n}{2}$-bit
message! Bit-error vs block-error considerations alone do not change
the overall picture since they buy at most a factor of $\frac{2}{n}$
in the average probability of error --- nothing on an exponential
scale.

With noiseless feedback, the natural nonblock code just retransmits a
bit over the BEC until it is correctly received. To be precise, as
bits arrive steadily at the rate $R' = \frac{1}{2}$ bits per channel
use, they enter a FIFO queue of bits awaiting transmission. At time
$0$, both the encoder and decoder know that there are no bits
waiting. From that time onward, the bit arrivals are modeled here as
deterministic and come every other channel use. Since both the encoder
and decoder know when a bit arrives as well as when a bit is
successfully received, there is no ambiguity in how to interpret a
channel output.

\begin{figure}[htbp]
\begin{center}
\setlength{\unitlength}{3000sp}%
\begingroup\makeatletter\ifx\SetFigFont\undefined%
\gdef\SetFigFont#1#2#3#4#5{%
  \reset@font\fontsize{#1}{#2pt}%
  \fontfamily{#3}\fontseries{#4}\fontshape{#5}%
  \selectfont}%
\fi\endgroup%
\begin{picture}(6458,816)(518,-494)
{\color[rgb]{0,0,0}\thinlines
\put(5626,-61){\circle{750}}
}%
{\color[rgb]{0,0,0}\put(4051,-61){\circle{750}}
}%
{\color[rgb]{0,0,0}\put(2476,-61){\circle{750}}
}%
{\color[rgb]{0,0,0}\put(901,-61){\circle{750}}
}%
{\color[rgb]{0,0,0}\put(6001, 89){\vector( 1, 0){900}}
}%
{\color[rgb]{0,0,0}\put(6826,-211){\vector(-1, 0){825}}
}%
{\color[rgb]{0,0,0}\put(4426, 89){\vector( 1, 0){900}}
}%
{\color[rgb]{0,0,0}\put(5251,-211){\vector(-1, 0){825}}
}%
{\color[rgb]{0,0,0}\put(2851, 89){\vector( 1, 0){900}}
}%
{\color[rgb]{0,0,0}\put(3676,-211){\vector(-1, 0){825}}
}%
{\color[rgb]{0,0,0}\put(1276, 89){\vector( 1, 0){900}}
}%
{\color[rgb]{0,0,0}\put(2101,-211){\vector(-1, 0){825}}
}%
\put(6376,164){\makebox(0,0)[b]{\smash{\SetFigFont{7}{6}{\rmdefault}{\mddefault}{\updefault}{\color[rgb]{0,0,0}$\beta^2$}%
}}}
\put(6376,-436){\makebox(0,0)[b]{\smash{\SetFigFont{7}{6}{\rmdefault}{\mddefault}{\updefault}{\color[rgb]{0,0,0}$(1-\beta)^2$}%
}}}
\put(4801,164){\makebox(0,0)[b]{\smash{\SetFigFont{7}{6}{\rmdefault}{\mddefault}{\updefault}{\color[rgb]{0,0,0}$\beta^2$}%
}}}
\put(4801,-436){\makebox(0,0)[b]{\smash{\SetFigFont{7}{6}{\rmdefault}{\mddefault}{\updefault}{\color[rgb]{0,0,0}$(1-\beta)^2$}%
}}}
\put(3226,164){\makebox(0,0)[b]{\smash{\SetFigFont{7}{6}{\rmdefault}{\mddefault}{\updefault}{\color[rgb]{0,0,0}$\beta^2$}%
}}}
\put(3226,-436){\makebox(0,0)[b]{\smash{\SetFigFont{7}{6}{\rmdefault}{\mddefault}{\updefault}{\color[rgb]{0,0,0}$(1-\beta)^2$}%
}}}
\put(1651,164){\makebox(0,0)[b]{\smash{\SetFigFont{7}{6}{\rmdefault}{\mddefault}{\updefault}{\color[rgb]{0,0,0}$\beta^2$}%
}}}
\put(1651,-436){\makebox(0,0)[b]{\smash{\SetFigFont{7}{6}{\rmdefault}{\mddefault}{\updefault}{\color[rgb]{0,0,0}$(1-\beta)^2$}%
}}}
\put(2476,-136){\makebox(0,0)[b]{\smash{\SetFigFont{7}{6}{\rmdefault}{\mddefault}{\updefault}{\color[rgb]{0,0,0}1}%
}}}
\put(901,-136){\makebox(0,0)[b]{\smash{\SetFigFont{7}{6}{\rmdefault}{\mddefault}{\updefault}{\color[rgb]{0,0,0}0}%
}}}
\put(4051,-136){\makebox(0,0)[b]{\smash{\SetFigFont{7}{6}{\rmdefault}{\mddefault}{\updefault}{\color[rgb]{0,0,0}2}%
}}}
\put(5626,-136){\makebox(0,0)[b]{\smash{\SetFigFont{7}{6}{\rmdefault}{\mddefault}{\updefault}{\color[rgb]{0,0,0}3}%
}}}
\put(6976,-136){\makebox(0,0)[b]{\smash{\SetFigFont{7}{6}{\rmdefault}{\mddefault}{\updefault}{\color[rgb]{0,0,0}$\cdots$}%
}}}
\end{picture}
\end{center}
\caption{The birth-death Markov chain governing the rate-$\frac{1}{2}$
communication system over an erasure channel with feedback. The
scheme merely retransmits bits until successful reception.}
\label{fig:birthdeath}
\end{figure}
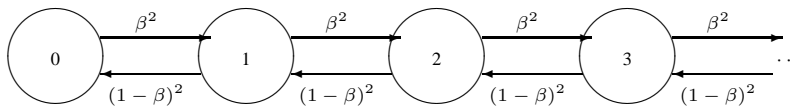

If the queue length is examined every two channel uses, exactly one
new bit has arrived while the channel may have successfully served 0,
1, or 2 bits in this period. Thus, the length of the queue can either
increase by one, stay the same, or decrease by one. The queue length
can be modeled (see Figure~\ref{fig:birthdeath}) as a birth-death
Markov chain with a $\beta^2$ probability of birth and a $(1-\beta)^2$
probability of death. The steady state distribution of the queue
length is therefore $\pi_i = \kappa (\frac{\beta}{1-\beta})^{2i}$
where $\kappa$ is the normalization constant $(1 -
(\frac{\beta}{1-\beta})^2)$.

To understand the probability of error with end-to-end delay, just
notice that the only way a bit can miss its deadline is if it is still
waiting in the queue. If it was a bit from $d$ time steps ago, the
queue must currently hold at least $\frac{d}{2}$ bits. The steady
state distribution reveals that the asymptotic probability of this is:
$$\kappa \sum_{i=\frac{d}{2}}^\infty (\frac{\beta}{1-\beta})^{2i} = 
(\frac{\beta}{1-\beta})^d 
 \kappa \sum_{i=0}^\infty (\frac{\beta}{1-\beta})^{2i} = 
(\frac{\beta}{1-\beta})^d.$$
Converting that into an error exponent with delay $d$ gives
\begin{equation} \label{eqn:halfbecfeedback}
 E^{bec}_{a}(\frac{1}{2}) = \ln(1-\beta) - \ln(\beta).
\end{equation}
Plugging in $\beta=0.4$ reveals an exponent of more than $0.40$. This
is about twenty times higher than the sphere-packing bound! Simple
computations can verify that the ratio of (\ref{eqn:halfbecfeedback}) to
(\ref{eqn:becspherebound}) goes to infinity as $\beta \rightarrow
\frac{1}{2}$.

To help get an intuitive idea for why this happens, it is worthwhile
to consider an idealized feedback-free code for erasure channels (the
reader may find it helpful to think of packet erasure channels with
large alphabets). Suppose that the encoder causally generated
``parities'' of all the message symbols so far with the property that
symbols could be decoded whenever the receiver had as many unerased
parities as there were undecoded symbols.\footnote{This is in the
  style of rateless block coding \cite{LTCodes}, except that the
  message bits are revealed to the encoder in time rather than being
  known all at the beginning.} The queue size can be reinterpreted in
this setting as the number of additional parities required before the
decoder could solve for the currently uncertain message symbols. The
queue's renewal times correspond to the times at which the decoder can
solve for the current set of undecoded message symbols.

\begin{figure}[ht]
\begin{center}
\includegraphics[width=4in,height=3in]{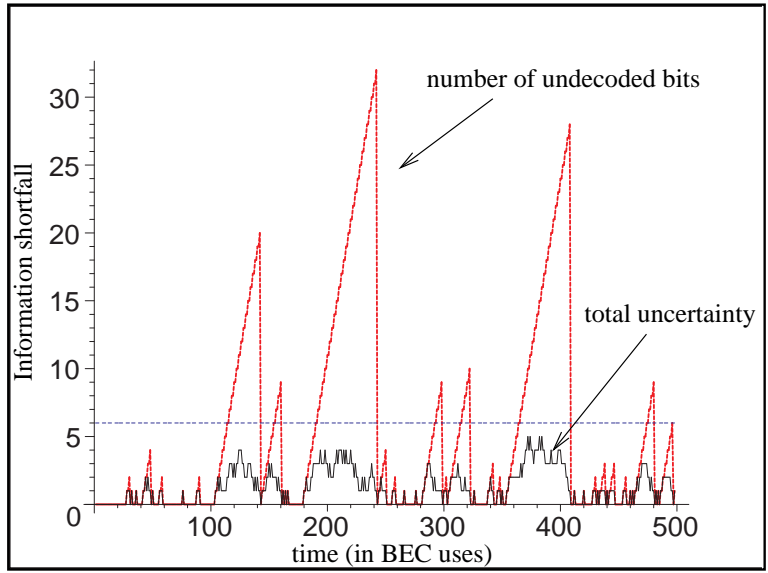}
\end{center}
\caption{A simulated run of an $0.4$ erasure channel using an
  idealized linear causal code without feedback. The red upper
  sawtooth represents the number of current message symbols that are
  still ambiguous at the decoder while the lower curve represents the
  number of additional parities that would enable it to resolve the
  current ambiguity. The lower curve is not coincidentally also the
  queue size for the natural FIFO-based code with feedback. The dotted
  line at 6 represents a potential delay deadline of 12 time units.}
\label{fig:backlogsimulation}
\end{figure}

\begin{figure}[hb]
\begin{center}
\includegraphics[width=4in,height=3in]{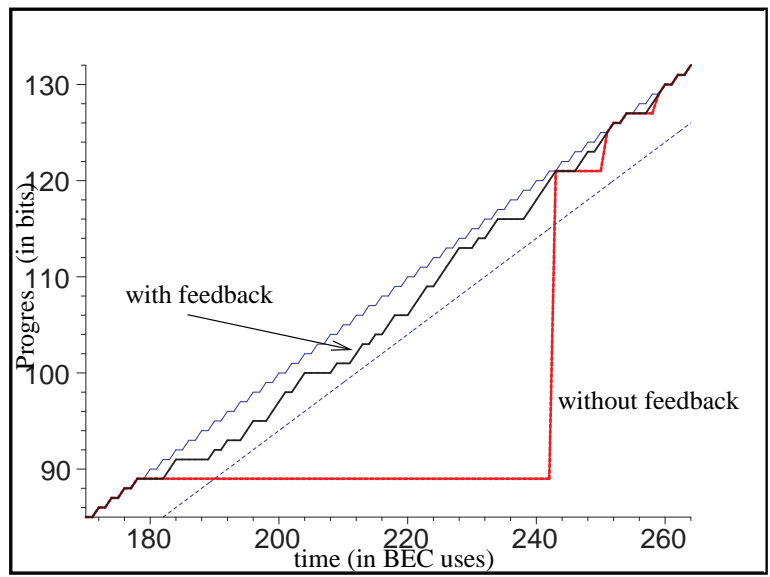}
\end{center}
\caption{A zoomed-in look at the simulation of
  Figure~\ref{fig:backlogsimulation} showing the total number of
  decoded symbols as a function of time. The thin upper curve is the
  total number of symbols that have been received at the
  rate-$\frac{1}{2}$ encoder. The next lower line is the total number
  of symbols decoded by the code with feedback. The lowest curve
  corresponds to the code without feedback. The thin dotted line
  represents the deadline of $12$ time steps. Whenever the decoder
  curves are below this curve, they are missing the deadline.}
\label{fig:timedomainzoom}
\end{figure}

\afterpage{\clearpage}

Figure~\ref{fig:backlogsimulation} illustrates the backlog of
undecoded bits in a simulated run of a rate-$\frac{1}{2}$ code over a
channel with erasure probability
$0.4$. Figure~\ref{fig:timedomainzoom} zooms in on a particular
segment of time corresponding to an ``error event'' and shows the
differences between how the feedback-free code and feedback code make
progress. During an error event in which the channel is erasing too
many symbols, progress at the decoder seems to stop entirely in the
feedback-free code, only catching up in a sudden burst when the error
event ends. By contrast, the code with feedback makes visible, but
slower, progress at the decoder even during these error events. As a
result, it is able to meet the target delay deadline whereas the code
without feedback misses it. This example also shows how the delays in
the feedback-free code are related to the inter-renewal times of the
queue, while the delays in the code with feedback are related to the
length of the queue itself.

Stepping back, this example illustrates that Pinsker's bound with
delay does not generally apply when feedback is available. Instead,
fixed-delay nonblock codes can dramatically outperform
fixed-block-length codes with feedback. Moreover, it is possible to
glimpse why this occurs. Reliable communication always takes place at
message rates $R$ that are less than the capacity $C$. In a fixed-delay
setting with feedback, the encoder has the flexibility to do flow
control based on what the channel has been doing in the past. It can
vary the short-term operational rate $R$ --- in effect stealing
channel uses from later bits to make sure that earlier bits meet their
looming deadlines, while still hoping that the later bits will be able
to meet their later deadlines. This flexibility is missing in the
fixed-block-length setting because all the bits in the block are
forced to share a common deadline.

This can also be seen by contrasting the total conditional entropy
$H(B_{iR'}^{(i+d)R'}|Y_i^{i+d})$ of the message bits
$B_{iR'}^{(i+d)R'}$ given the channel outputs
$Y_i^{i+d}$ to the sum $\sum_{k=iR'}^{(i+d)R'}
H(B_{k}|Y_i^{i+d})$ of the marginal conditional entropies of the bits
given the channel outputs. If the channel misbehaves slightly and
makes it hard to distinguish only a single pair of bit strings, the
marginal entropies $H(B_{k}|Y_i^{i+d})$ can become large even as the
total conditional entropy is small. Such situations are common without
feedback. From the decoder's perspective, the feedback encoder's
strategy should be to focus the uncertainty
$H(B_{iR'}^{(i+d)R'}|Y_i^{i+d})$ onto later bits
$B_{(i+d)R' - \Delta}^{(i+d)R'}$ to pay for reducing it on
earlier bits. The sum of the marginal conditional entropies can then
be made the same as the total conditional entropy.

The total delay experienced by a bit can also be broken into two
components: queuing delay and transmission delay. For the erasure
channel, the transmission delay is just a geometric random variable
governed by an exponent of $-\ln(\beta)$.  This transmission exponent
does not change with the message rate. The queuing delay is the dominant
term, and its exponent does change with the message rate.

Finally, it is interesting to examine the computational burden of
implementing this simple code. At the encoder, all that is needed is a
FIFO queue that costs a constant (assuming memory is free) per unit
time to operate. The decoder has similar complexity since it too just
tracks how many bits it has received so far in comparison with the
number of bits known to have arrived at the encoder. The computational
burden does not change with either the target delay or the quality of
the channel!


\section{Background} \label{sec:background}

\subsection{Fixed-length codes} \label{sec:fixedlengthreview}
Traditionally, reliable communication was first explored in the
context of block codes \cite{ShannonOriginalPaper}. If physical
information sources are considered to produce bits steadily at $R'$
bits per second, then the use of a block code of length 
$n$ channel uses (with channel uses assumed to occur once per second)
contributes to end-to-end delay in two ways.
\begin{itemize}
 \item Enough bits must first be buffered up to even compute the
       codeword. This takes no more than $n$ seconds and can
       take less if the block code is systematic in nature. 

 \item The decoder must wait for $n$ seconds to get the $n$
       channel outputs needed to decode the block. This second delay
       would be present even if the source bits were realized entirely
       in advance of the use of the channel.
\end{itemize}
In this context, the fundamental lower bound on error probability
comes from the sphere-packing bound. To understand this bound, it is
helpful to think about the message block as representing a certain
{\em volume} of entropic uncertainty that the decoder has about the
message. The objective of using the channel is to reduce this
uncertainty. Let $P$ be the transition matrix ($p_{y|x}$ is the
probability of seeing output $y$ given input $x$) for the DMC. Each
channel use can reduce the uncertainty on average by no more than the
capacity
\begin{equation} \label{eqn:capacitydefinition}
C(P) = \max_{\vec{q}} I(\vec{q},P)
\end{equation}
where $I(\vec{q},P)$ is the mutual information between input and
output of channel $P$ when $\vec{q}$ is the input distribution and is
defined by
\begin{equation}\label{eqn:mutualinformationdef}
I(\vec{q},P) = 
\sum_x q_x \sum_y p_{y|x} \ln \frac{p_{y|x}}{\sum_{k}q_k p_{y|k}}.
\end{equation}

With or without feedback, successful communication is not possible if
during the block, the memoryless channel acts like one whose capacity
is less than the target message rate. Following \cite{csiszarkorner,
  Haroutunian}, for fixed-block-length codes this idea immediately
gives the following upper bound (referred to as the {\em Haroutunian
  bound} throughout this paper) on the block-coding error exponent
($\limsup_{n \rightarrow \infty} \frac{-\ln P_e}{n}$):
\begin{eqnarray} 
E^+(R) & = &  \inf_{G: C(G) < R} 
            \sup_{\vec{r}} D\left(G||P |
              \vec{r}\right) \label{eqn:primitiveupperbound} \\
& = &  \min_{G: C(G) \leq R} 
        \max_x \sum_y g_{y|x} \ln
        \frac{g_{y|x}}{p_{y|x}} \label{eqn:Haroutunianbound}
\end{eqnarray}
where $D(G || P | \vec{r})$ is the divergence term that governs the
exponentially small probability of the true channel $P$ behaving like
channel $G$ when facing the input distribution $\vec{r}$. The
divergence is defined as
\begin{equation}\label{eqn:divergencedef}
D(G||P | \vec{r})  = 
\sum_x r_x \sum_y g_{y|x} \ln \frac{g_{y|x}}{p_{y|x}}.
\end{equation}

\begin{figure}[htbp]
\begin{center}
\includegraphics[width=4in,height=3in]{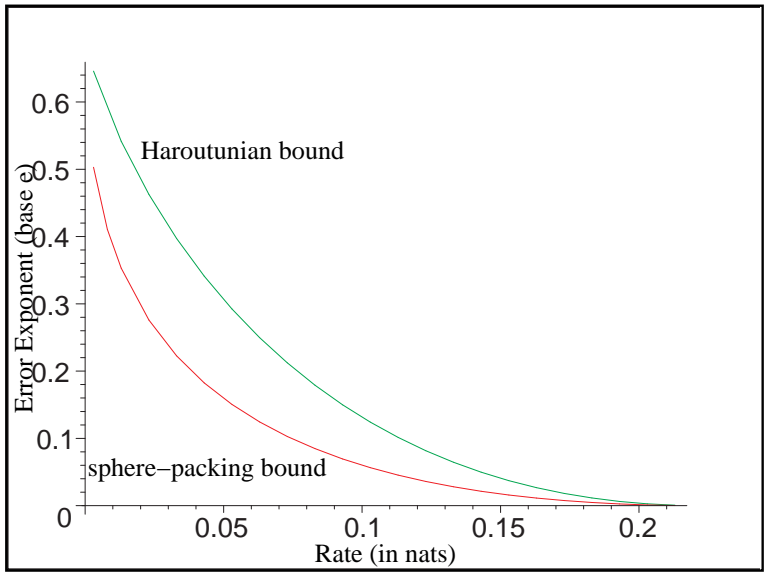}
\end{center}
\caption{The sphere-packing and Haroutunian bounds for the Z-channel
  with nulling probability $0.5$. The upper curve is the Haroutunian
  upper bound for the error exponent of block codes with feedback and
  the lower curve is the classical sphere-packing bound. Both approach
  zero very rapidly around the capacity of $0.223$ nats per channel
  use. Due to the asymmetry of the Z-channel, the capacity-achieving
  distribution is not the same as the sphere-packing-bound-achieving
  distribution.}
\label{fig:zchannelbounds}
\end{figure}

Without feedback, the encoder does not have the flexibility to change
the input distribution in response to the channel's behavior. The
optimization can take this into account to get the bound traditionally
known as the sphere-packing bound
\begin{equation} \label{eqn:spherepackingbound}
    E_{sp}(R) = \max_{\vec{r}} \min_{G: I(\vec{r},G) \leq R } D\left(G||P|\vec{r}\right).
\end{equation}
It is clear that $E_{sp}(R) \leq E^+(R)$ and
Figure~\ref{fig:zchannelbounds} illustrates that the inequality can be 
strict. 

It is often useful to use an alternate form for $E_{sp}(R)$ given
by \cite{gallager} 
\begin{equation} \label{eqn:rhospherepack}
  E_{sp}(R) = \max_{\rho \geq 0} \big[ E_0(\rho) - \rho R \big]
\end{equation}
with the Gallager function $E_0(\rho)$ defined as:
\begin{eqnarray} 
  E_0(\rho) &=& \max_{\vec{q}} E_0(\rho,\vec{q}), \nonumber \\
  E_0(\rho, \vec{q}) &=&  -\ln \sum_y \bigg[ \sum_x q_x
  p_{y|x}^\frac{1}{1+\rho} \bigg]^{(1+\rho)}. \label{eqn:enought}
\end{eqnarray}
Since the random-coding error exponent is given by
\begin{equation} \label{eqn:rhorandom}
  E_{r}(R) = \max_{0 \leq \rho \leq 1} \big[ E_0(\rho) - \rho R \big],
\end{equation}
it is clear that the sphere-packing bound is achievable, even without
feedback, at message rates close to $C$ since for those rates, $\rho <
1$ optimizes both expressions \cite{gallager}.  

It is less well appreciated that the points on the sphere-packing
bound where $\rho > 1$ are also achievable by random coding if the
sense of ``correct decoding'' is relaxed. Rather than forcing the
decoder to emit a single estimated codeword, list decoding allows the
decoder to emit a small list of guessed codewords. The decoding is
considered correct if the true codeword is on the list. For
list decoding with list size $\ell$ in the context of random codes,
Problem 5.20 in \cite{gallager} reveals that
\begin{equation} \label{eqn:listrhorandom}
  E_{r,\ell}(R) = \max_{0 \leq \rho \leq \ell} \big[ E_0(\rho) - \rho R \big]
\end{equation}
is achievable. At high message rates (where the maximizing $\rho$ is
small), there is no benefit from relaxing to list decoding, but it
makes a difference at low rates.

\begin{figure}[htbp]
\begin{center}
\includegraphics[width=4in,height=3in]{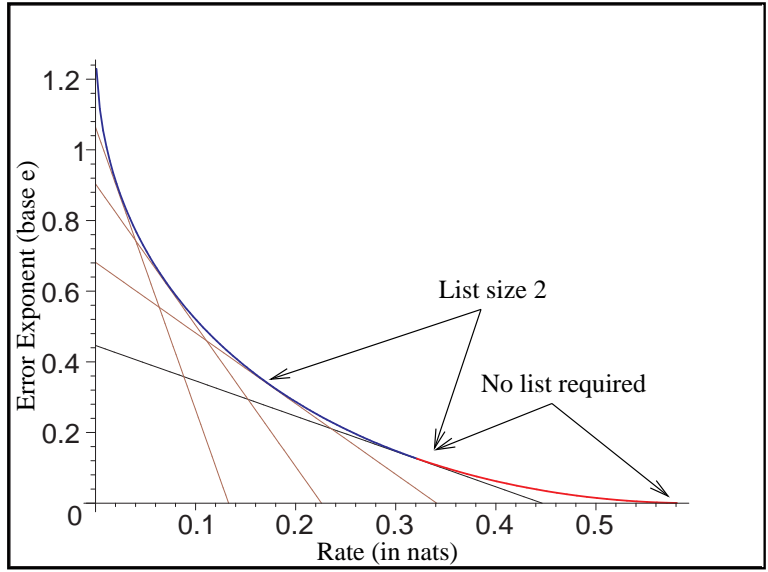}
\end{center}
\caption{The sphere-packing bound divided up into two sections: a blue
  segment where list decoding is needed for random codebooks to
  achieve it, and a red segment where lists are not needed. The
  tangents represent list sizes of $8$, $4$, $2$, and $1$. }
\label{fig:listnolist}
\end{figure}

Figure~\ref{fig:listnolist} illustrates the range of exponents for
which list decoding is required for a BSC. The blue part of the
sphere-packing curve shows where list decoding is important and the
red part shows where lists are not required. Four tangents are
illustrated corresponding to list sizes of $8$, $4$, $2$, and $1$. The
y-intercepts of these tangents represent the maximum error
exponents possible using those list sizes and random codes.

For output-symmetric channels (see Definition~\ref{def:channels}), it
is clear that $E^+(R) = E_{sp}(R)$ since the input distribution
$\vec{r}$ can always be chosen to be uniform
\cite{DobrushinReliability}. Thus, for fixed-block-length codes and
output-symmetric DMCs, not only does causal feedback not improve
capacity, it does not improve reliability either, at least at high
rates.\footnote{Notice how the situation for unconstrained DMCs is
  dramatically different from the behavior of the AWGN channel with
  noiseless feedback for which Schalkwijk and Kailath showed
  double-exponential reliability with block length \cite{Kailath66,
    Schalkwijk66}. However, those results rely crucially on the {\em
    variable} nature of an input constraint that only has to hold on
  average. An unconstrained DMC is more like an AWGN channel with just
  a hard amplitude constraint on the channel inputs
  \cite{WynerSKLimit}.}

The extreme limit of reliability in the fixed-block-length setting is
given by the study of zero-error capacity, in which the probability of
decoding error is required to be exactly zero. As pointed out in
\cite{ShannonZeroError}, this can be different with and without
feedback. For zero-error capacity, the details of the channel matrix
$P$ are not important as it clearly only depends on which entries are
zero. The true zero-error capacity without feedback $C_0$ is very hard
to evaluate, but the zero-error capacity with feedback $C_{0,f}$ can
be easily evaluated when it is greater than zero \cite{ZeroErrorSurvey}.

Although there is an explicit expression for $C_{0,f}$ in
\cite{ShannonZeroError}, the interpretation is more straightforward in
the context of
(\ref{eqn:rhospherepack}). 
\begin{equation} \label{eqn:zeroerrorexpression}
C_{0,f} = \lim_{\rho \rightarrow \infty} \frac{E_0(\rho)}{\rho}
\end{equation}
was established in \cite{ShannonGallagerBerlekampI} by evaluating the
limit and showing that it is identical to the expression for $C_{0,f}$
from \cite{ShannonZeroError}. If $C_{0,f}$ is nonzero, both the
sphere-packing bound (\ref{eqn:rhospherepack}) and Haroutunian bound
(\ref{eqn:primitiveupperbound}) are infinite at message rates below
$C_{0,f}$ and finite above it.


\subsection{Variable-length codes} \label{sec:varlength}

Since feedback neither improves the capacity nor significantly
improves the fixed-block-length reliability function, it seemed that
this particular reliability somehow represented the wrong technical
question to ask. After all, it was unable to answer why feedback
seemed to be so useful in practice. The traditional response to this
was to fall back to the issue of complexity.

Because classical decoding of fixed-block-length codes has a
complexity that is not linear in the block length, the block length
was viewed as a proxy for implementation complexity rather than only
for end-to-end delay. Just as in variable-length source-coding, the
idea in variable-block-length channel-coding is to extend use of the
channel when the channel is behaving atypically. This way, the
presumed complexity of increased block lengths is only experienced
rarely and on average, the system can be simpler to operate.

Without feedback, a variable-length mode of operation is impossible
since the encoder has no way to know if the channel is behaving
typically or atypically. With noiseless feedback, the length of the
codeword can be made to vary based on what the channel has done so far
--- as long as this variation depends only on the received channel
symbols. This is the counterpart to the unique decodability
requirement in source coding in that both are needed to prevent an
irrecoverable loss of synchronization between the encoder and decoder.

One proposed error exponent for variable-length channel codes divides
the negative log of the probability of block error $\epsilon$ by the
expected block length $E[N_\epsilon]$ of an average rate-$\bar{R}$
variable-length code \cite{burnashev}.

$$E_{vl}(\bar{R}) = \limsup_{\epsilon \rightarrow 0} -\frac{\ln(\epsilon)}{E[N_\epsilon]}.$$

Burnashev gave an upper bound to this exponent by using martingale
arguments treating the ending of a block as a stopping time and
studying the rate of decrease in the conditional entropy of the
message at the receiver \cite{burnashev}. This gives
\begin{equation} \label{eqn:burnashevbound}
E_{v}(\bar{R}) = C_1 \bigg(1-\frac{\bar{R}}{C}\bigg) 
\end{equation}
where $C$ is the Shannon capacity of the channel and
\begin{equation} \label{eqn:c1def}
C_1 = \max_{x,x'} D(P(\cdot|x)||P(\cdot|x'))
    = \max_{x,x'} \sum_{y} p_{y|x} \ln \frac{p_{y|x}}{p_{y|x'}}
\end{equation}
represents the maximum divergence possible between channel output
distributions given choice of two input letters.

While Burnashev gives an explicit variable-length scheme in
\cite{burnashev} that asymptotically attains the exponent of
(\ref{eqn:burnashevbound}), the scheme of Yamamoto and Itoh in
\cite{yamamoto} is simpler and makes clear the idea of separating
reliability from efficiency. Suppose there is a single message of $nR$
nats to send:
\begin{enumerate}
 \item Transmit the message using any reliable block code at a rate
       $\widetilde{R} < C$ close to capacity but larger than the target average
       rate $\bar{R}$. This will consume $n \frac{R}{\widetilde{R}}$ channel
       uses.

 \item Use the noiseless feedback to decide at the encoder whether the
       message was received correctly or incorrectly.

 \item If the message was received correctly, send a ``confirm''
       signal by sending input $x$ from (\ref{eqn:c1def})
       repeated $n(1-\frac{R}{\widetilde{R}})$ times. Otherwise, use the
       channel to send a ``deny'' signal by repeating input $x'$
       the same number of times. 

       This part can be interpreted as a sort of punctuation: a
       ``deny'' is a backspace telling the decoder to erase what it
       has seen so far while a ``confirm'' is a comma telling the
       decoder that this block is finished. 

 \item The decoder performs a simple binary hypothesis test on the
       received confirm/deny channel outputs to decide whether to
       accept the current message block. If it rejects the block, then
       the encoder will retransmit it until it is accepted. Since
       errors only occur when the message is falsely accepted, the
       decoder minimizes the probability of false alarm while holding
       the probability of missed detection to some acceptably low
       level.
\end{enumerate}
Since retransmissions can be made as rare as desired as long as
$\widetilde{R} < C$, the overall average rate $\bar{R}$ of the scheme
approaches $R$. Since the number of slots for the ``confirm/deny''
message can be made to approach $n(1-\frac{R}{C})$, the reliability
approaches 
(\ref{eqn:burnashevbound}) by Stein's Lemma \cite{coverthomas}. {\em
  Our approach to generic channels in Section~\ref{sec:comments} can
  be considered as using variable-block-length codes to achieve good
  fixed-delay performance by combining an alternative approach to
  punctuation with a softer sense of retransmission.}  

The Burnashev exponent is dramatically higher than the
fixed-block-length exponents (see Figure~\ref{fig:straightbsc}) and
thus seems to demonstrate the advantage of feedback. However, it is
unclear what the significance of average delay or block length really
is in a system. The block length under the Yamamoto and Itoh scheme is
distributed like a scaled geometric random variable. Consequently, the
block length will exceed a target deadline (like an underlying
channel's coherence time or an application-specific latency
requirement) far more often than the scheme makes an undetected
error. There are also no known nontrivial separation theorems
involving either average block length or average delay.

\subsection{Nonblock codes} \label{sec:intrononblock}

Another classical approach to the problem of reliable communication is
to consider codes without any block structure. Convolutional and tree
codes represent the prototypical examples. It was realized early on
that in an infinite-constraint-length convolutional code under ML
decoding, all bits will eventually be decoded correctly
\cite{gallager}. Given that this asymptotic probability of error is
zero, there are two possible ways to try to understand the underlying
tradeoffs: look at complexity or look at the delay.

The traditional approach was to focus on complexity by examining the
case of finite constraint lengths. The per-symbol encoding complexity
of a convolutional code is linear in the constraint length, and if
sequential decoding algorithms are used and the message rate is below
the cutoff rate $E_0(1)$, so is the average decoding complexity 
\cite{ForneySeq}. With a fixed constraint length $\nu$, the
probability of error cannot go to zero and so it is natural to
consider the tradeoff between the error probability and constraint
length $\nu$. Viterbi used a genie-aided argument to map the
sphere-packing bound for block codes into an upper bound for
fixed-constraint-length convolutional codes. (A variant of this
argument is used in Section~\ref{sec:feedbackbound} to bound
performance with delay.)  This gives the following parametric upper
bound for the exponent governing how fast the bit error probability
can improve with the constraint length:\cite{viterbi}
\begin{equation} \label{eqn:viterbibound}
 E_{c}(R) = E_0(\rho)\,\,;\,\, R = \frac{E_0(\rho)}{\rho}
\end{equation}
where $\rho \geq 0$. The ``inverse concatenation construction''
(illustrated in Figure~\ref{fig:erasurebounds}) is the graphical
representation of the above curve --- it is the envelope of the
$(R,E)$ intercepts traced out by the tangents to the sphere-packing
bound. Thus, this upper bound can be tightened in the low-rate regime
by using the ``straight-line bound'' from
\cite{ShannonGallagerBerlekampII}. The bound (\ref{eqn:viterbibound})
is also achievable in the high-rate regime ($R > E_0(1)$)
\cite{ForneySeq}.

The $E_c(R)$ from (\ref{eqn:viterbibound}) for fixed
constraint lengths is substantially higher than $E_{sp}(R)$ from
(\ref{eqn:rhospherepack}) for fixed block-lengths. This was used to
argue for the superiority of convolutional codes over block codes from
an implementation point of view. However, it is important to remember
that this favorable comparison does not hold when end-to-end delay,
rather than complexity, is considered. 

If the end-to-end delay is forced to be bounded, then the bit-error
probability with delay is governed by $E_{r}(R)$ for random
convolutional codes, even when the constraint lengths are unbounded
\cite{ForneyML}. This performance with delay is also achievable using
an appropriately biased sequential decoder \cite{JelinekSequential}. A
nice feature of sequential decoders is that they are not tuned to any
target delay --- they can be prompted for estimates at any time and
they will give the best estimate that they have. Thus an
infinite-constraint-length convolutional code with appropriate
sequential decoding achieves the exponent $E_r(R)$ delay universally
over all (sufficiently long) delays. This property turns out to be
important for this paper since such codes are used in place of 
two-point block codes to encode punctuation information in
Section~\ref{sec:comments}. 

The role of feedback in nonblock codes has also been investigated
considerably by considering a variety of different schemes
\cite{Horstein, Fang, HashimotoARQ, Veugen, SchalkwijkPost,
  KudryashovPPI, KudryashovIT}, each with an idiosyncratic way of
defining a relevant error exponent. The simplest approach is to
consider a variable-constraint-length model in which complexity is
counted by the expected number of multiply-accumulate operations that
are required to encode a new channel symbol. This is done in
Appendix~\ref{app:feedbackconvolution}. The result is that for all
rates below the computational cutoff rate, a finite amount of expected
computation per input bit is enough to get an arbitrarily low
probability of error --- that the computational error exponent is
infinite.

At first glance, this infinite exponent seems to show the 
superiority of variable-constraint-length codes over
variable-block-length codes with feedback. After all, the Burnashev
bound (\ref{eqn:burnashevbound}) is only infinite for channels whose
probability matrices $P$ contains a zero. However, this is not a fair
comparison since it is comparing expected per-channel-use
computational complexity here with expected block length in the
variable-block-length case.

The variable-block-length schemes of Ooi and Wornell \cite{ooiwornell,
  Ooi} achieve linear complexity in the block length for the
message-communication part. Once complexity is linear in the expected
length, it is constant on an average per-symbol basis. Thus block
codes can also achieve any desired probability of error by adjusting
the length of the confirm/deny phase in the same way that a large
enough terminator $d$ can be chosen for the variable-constraint-length
convolutional codes of Appendix~\ref{app:feedbackconvolution}. So both
have infinite computational error exponents with feedback.

An infinite exponent just means that the asymptotic tradeoff of
probability of error with expected per-symbol computation is
uninteresting when noiseless feedback is allowed. As a result, it is
very natural to consider the tradeoff with end-to-end delay
instead. The open questions that are addressed in this paper are
whether the end-to-end delay performance can generally be improved
using feedback, and if so, what are the limits to such improvements.

\section{Main results and examples} \label{sec:mainresults}

First, some basic definitions are needed. Vector notation $\vec{x}$ is
used to denote sequences $x_1^n$ where the indices are obvious from
the context.
\begin{definition} \label{def:channels}
  A {\em discrete time discrete memoryless channel} (DMC) is a
  probabilistic system with an input and an output. At every time step
  $t$, it takes an input $x_t \in {\cal X}$ and produces an output
  $y_t \in {\cal Y}$ with probability ${\cal P}(Y_t = y|X_t = x) =
  p_{y|x}$. Both ${\cal X},{\cal Y}$ are finite sets and the transition
  probability matrix $P$ containing the $p_{y|x}$ entries is a
  stochastic matrix. The current channel output is independent of all
  past random variables in the system conditioned on the current
  channel input.

  Following \cite[page 94]{gallager}, a DMC is called {\em
    output-symmetric} if the set of outputs ${\cal Y}$ can be
  partitioned into disjoint subsets\footnote{Notice how Gallager's
    definition of output-symmetric channels slightly generalizes the
    symmetric channel definitions of Dobrushin
    \cite{DobrushinReliability} and Csisz\'{a}r and K\"{o}rner \cite[page
    114]{csiszarkorner}. Such output-symmetric channels can
    be understood as convex combinations of symmetric channels, each
    with its own distinct output alphabet. Knowledge of the partition
    the output lands in just tells the decoder which of the symmetric
    channels it happens to be encountering, but does not reveal
    anything about the channel input itself.} in such a way that for
  each subset, the matrix of transition probabilities has the property
  that each row is a permutation of each other row and each column is
  a permutation of each other column.
\end{definition}
\vspace{0.1in}

\begin{definition} \label{def:nofeedbackencoder} A {\em rate-$R$
    encoder ${\cal E}$ without feedback} is a sequence of maps
  $\{{\cal E}_t\}$. Each ${\cal E}_t: \{0,1\}^{\lfloor R't \rfloor}
  \rightarrow {\cal X}$ where the range is the finite set of channel
  inputs $\cal X$. The $t$-th map takes as input the available message
  bits $B_1^{\lfloor R't \rfloor}$ where $R' = \frac{R}{\ln 2}$ is the
  encoder's rate in bits rather than nats per channel use.

For a {\em rate-$R$ encoder with noiseless feedback},  the maps
${\cal E}_t: {\cal Y}^{t-1} \times \{0,1\}^{\lfloor R't \rfloor}
\rightarrow {\cal X} $ also get access to all the past channel outputs  
$Y_1^{t-1}$. 

A {\em delay-$d$ rate-$R$ decoder} is a sequence of maps $\{ {\cal
  D}_i \}$. Each ${\cal D}_i: {\cal Y}^{\lceil \frac{i}{R'} \rceil + d}
\rightarrow \{0,1\}$ where the output of each map is the estimate
$\widehat{B}_i$ for the $i$-th bit. The $i$-th map takes as input the
available channel outputs $Y_1^{\lceil \frac{i}{R'} \rceil + d}$. This
means that it can see $d$ time units (channel uses) beyond when the
bit to be estimated first had the potential to influence the channel
inputs.

{\em Randomized encoders and decoders} also have access to random
variables $W_t$ denoting common randomness available in the
system. 

\end{definition}
\vspace{0.1in}

\begin{definition} \label{def:achievable} The fixed-delay error
  exponent $\alpha$ is asymptotically {\em achievable} at message rate
  $R$ across a noisy channel if for every delay $d_j$ in some strictly
  increasing sequence indexed by $j$ there exist rate-$R$ encoders
  ${\cal E}^{j}$ and delay-$d_j$ rate-$R$ decoders ${\cal D}^{j}$ that
  satisfy the following properties when used with input bits $B_i$
  drawn from iid fair coin tosses.
\begin{enumerate} 
 \item For the $j$-th code, there exists an $\epsilon_j < 1$ so that
   ${\cal P}(B_i
       \neq \widehat{B}_i(d_j)) \leq \epsilon_j$ for every bit
       position $i \geq 1$. The $\widehat{B}_i(d_j)$ represents the
       delay-$d_j$ estimate of $B_i$ produced by the $({\cal E}^j,
       {\cal D}^j)$ pair connected through the channel in question. 

 \item $\lim_{j \rightarrow \infty} \frac{-\ln \epsilon_j}{d_j} \geq
       \alpha$
\end{enumerate}

The exponent $\alpha$ is asymptotically {\em achievable universally
over delay} or in an {\em anytime fashion} if a single encoder ${\cal
E}$ can be used simultaneously for all sufficiently long delays $d$.
\end{definition}
\vspace{0.1in}

\subsection{Main results}
With these definitions, the five main results of this paper can be stated:

\begin{theorem} \label{thm:nofeedbackbound} For a DMC, no fixed-delay
  exponent greater than the Haroutunian bound ($\alpha > E^+(R)$ from
  (\ref{eqn:Haroutunianbound})) is asymptotically achievable without
  feedback.
\end{theorem}
\vspace{0.1in}
\begin{theorem} {\em Uncertainty-focusing bound:}  \label{thm:generalfeedbackbound}
  For a DMC, no delay exponent $\alpha > E_{a}(R)$ is asymptotically
  achievable even if the encoders are allowed access to noiseless
  feedback.
\begin{equation} \label{eqn:generalfeedbackbound}
E_a(R)  = \inf_{0 \leq \lambda < 1} \frac{E^+(\lambda R)}{1-\lambda}
\end{equation}
  where $E^+$ is the Haroutunian bound from
  (\ref{eqn:Haroutunianbound}). Whenever $E^+(R) = E_{sp}(R)$ (e.g. the DMC
  is output-symmetric), $E_a(R) = E_{a,s}(R)$ where the latter is expressed
  parametrically as 
\begin{eqnarray} \label{eqn:symmetricfeedbackbound}
 E_{a,s}(R) & = & E_0(\eta),\\
 R & = & \frac{E_0(\eta)}{\eta} \nonumber
\end{eqnarray}
where $E_0(\eta)$ is the Gallager function from (\ref{eqn:enought}),
and $\eta$ ranges from $0$ to $\infty$.

  The curve (\ref{eqn:symmetricfeedbackbound}) has negative slope of
  at least $2C/ \frac{\partial^2 E_0(0)}{\partial \eta^2}$ in the
  vicinity of the $(C,0)$ point where the derivatives of $E_0$ are
  taken fixing the capacity-achieving distribution. 
\end{theorem}
\vspace{0.1in}
\begin{theorem} \label{thm:erasurecase} For the binary erasure channel
  with erasure probability $\beta > 0$, there exists a code using
  noiseless feedback with a delay error exponent that asymptotically
  approaches the uncertainty-focusing bound $E_a(R)$ for all message
  rates $R < C$. Viewed as a reliability-dependent capacity, the
  tradeoff is given by
\begin{equation} \label{eqn:becanytimecapacity}
C'(\alpha) = \frac{\alpha}{\alpha +
\log_2 \left(\frac{1-\beta}{1-2^\alpha \beta}\right)}
\end{equation}
where $\alpha$ is the desired reliability (in base 2) with fixed delay
and $C'(\alpha)$ is the supremal rate (in bits per channel use) at
which reliable communication can be sustained with fixed-delay
reliability $\alpha$. 

Furthermore, for every $r \geq \frac{2 - \log_2 \log_2 \beta^{-1}}{\log_2
  \beta^{-1}}$ (in particular: any $r \geq 0$ as long as $\beta \leq
\frac{1}{16}$), at all rates $R' < \frac{1}{1+2r}$ bits per channel
use, the error exponent (in base 2) with respect to delay is $\geq
\log_2  \beta^{-1} - 2\beta^r$. 
\end{theorem}
\vspace{0.1in}
 
\begin{theorem} \label{thm:withzeroerror} For any DMC with strictly
  positive zero-error capacity $C_{0,f} > 0$, it is possible to
  asymptotically approach all delay exponents within the region
  $\alpha < E_{a,s}(R)$ defined by (\ref{eqn:symmetricfeedbackbound})
  using noiseless feedback and randomized encoders, even if the
  feedback is delayed by a constant $\phi$ channel uses.

  This rate/reliability region can also be asymptotically achieved for
  any DMC by an encoder/decoder pair that has access to noiseless
  feedback if it also has access to an error-free forward communication 
  channel with any strictly positive rate.

  Furthermore, the delay exponents can be achieved in a
  delay-universal or ``anytime'' sense.
\end{theorem}
\vspace{0.1in}

As is shown in Section~\ref{sec:fortifiedfeedback}, the scheme that
approaches the uncertainty-focusing bound is built around a
variable-length channel code with the zero-error aspects used to
convey unambiguous ``punctuation'' information that allows the decoder
to stay synchronized with the encoder. Without any zero-error
capacity, this punctuation information can be encoded in a separate
parallel stream of channel uses to give the following result.

\begin{theorem} \label{thm:genericachieve} For any DMC, it is possible
  with noiseless feedback and randomized encoders to asymptotically
  achieve all delay exponents $\alpha < E'(R)$ where the tradeoff
  curve is given parametrically by varying $\rho \in (0,\infty)$:
\begin{eqnarray} \label{eqn:timesharingregion}
E'(\rho) & = & \left(\frac{1}{E_0(\rho)} + \frac{1}{E_0(1)}\right)^{-1}, \\
R(\rho)  & = & \frac{E'(\rho)}{\rho}. \nonumber
\end{eqnarray}

The curve (\ref{eqn:timesharingregion}) has strictly negative slope
$-E_0(1) / (C - \frac{E_0(1)}{2C} \left(\frac{\partial^2
    E_0(0)}{\partial \rho^2} \right))$  in the vicinity of the $(C,0)$
point.

Furthermore, these delay exponents are also achievable in a
delay-universal or ``anytime'' sense.
\end{theorem} 
\vspace{0.1in} 

The fact that this achievable region (\ref{eqn:timesharingregion})
generically has strictly negative slope in the vicinity of $(C,0)$
while the Haroutunian bound $E^+$ and sphere-packing bound $E_{sp}$
both generically approach $(C,0)$ only quadratically with zero slope
establishes that noiseless feedback generally improves the tradeoff
between end-to-end delay and the probability of error.

The above results relate to the strict interior of the region defined
by $E_{a,s}(R)$ or $E'(R)$ for achievability and the strict exterior
region corresponding to $E_a(R)$ for the converse. Unlike the case of
fixed-block-length codes where the sphere-packing bound is known to be
achievable at high rates, the results above do not cover points on the
$E_{a,s}(R)$ curve itself at any rates.

The results of Theorems \ref{thm:erasurecase}, \ref{thm:withzeroerror}
and \ref{thm:genericachieve} are also stated using asymptotic language
--- they apply in the limit of large end-to-end delays. In the case of
Theorems \ref{thm:withzeroerror} and \ref{thm:genericachieve}, the
parameters defining the randomized codes are also allowed to get
asymptotically large in order to approach the delay-error-exponent
frontier. However, the proofs use techniques that make it possible to
evaluate the performance of schemes with finite parameters.

\subsection{Numerical examples} \label{sec:regularexamples}

\begin{figure}[htbp]
\begin{center}
\includegraphics[width=4in,height=3in]{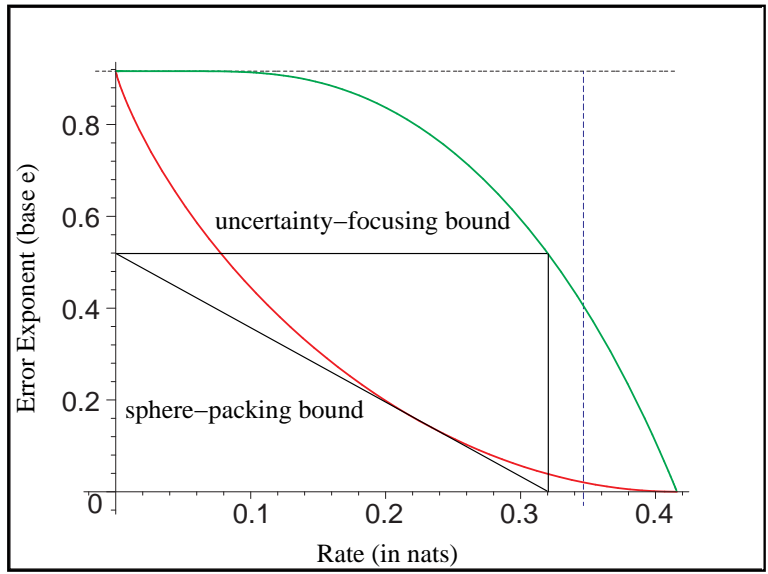}
\end{center}
\caption{The binary erasure channel with $\beta=0.4$. The vertical
  dashed line represents the rate of $\frac{1}{2}$ bits per channel
  use while the horizontal dashed line is the ultimate limit of
  $-\ln(0.4)$ for the reliability function. Notice how the
  uncertainty-focusing bound gets very close to that ultimate bound
  even at moderately small rates. The triangle illustrates the
  ``inverse concatenation construction'' connecting the two bounds to
  each other. }
\label{fig:erasurebounds}
\end{figure}

The erasure channel is the simplest channel for understanding the
asymptotic tradeoffs between message rate, end-to-end delay, and
probability of error when noiseless feedback is
allowed. Figure~\ref{fig:erasurebounds} illustrates how when the
erasure probability $\beta$ is small, even moderately low rates
achieve spectacular reliabilities with respect to fixed delay.

\begin{figure}[htbp]
\begin{center}
\includegraphics[width=4in,height=6in]{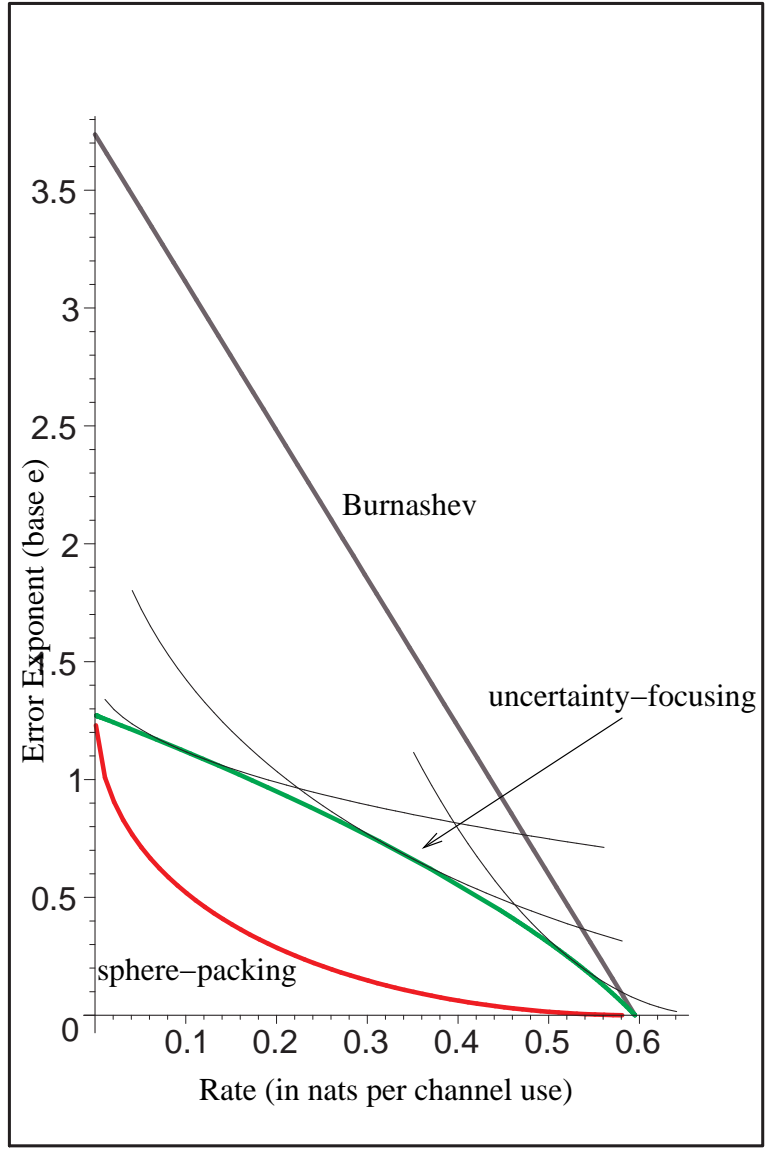}
\end{center}
\caption{The sphere-packing, uncertainty-focusing, and Burnashev
  bounds for a BSC with crossover probability $0.02$. The thin lines
  represent the parametric bounds in (\ref{eqn:generalfeedbackbound})
  setting $\lambda = \frac{1}{8}, \frac{1}{2}, \frac{7}{8}$ and the
  thick middle curve is the uncertainty-focusing bound --- the lower
  envelope of the parametric bounds over all $\lambda$. }
\label{fig:straightbsc}
\end{figure}

\begin{figure}[htbp]
\begin{center}
\includegraphics[width=4in,height=3in]{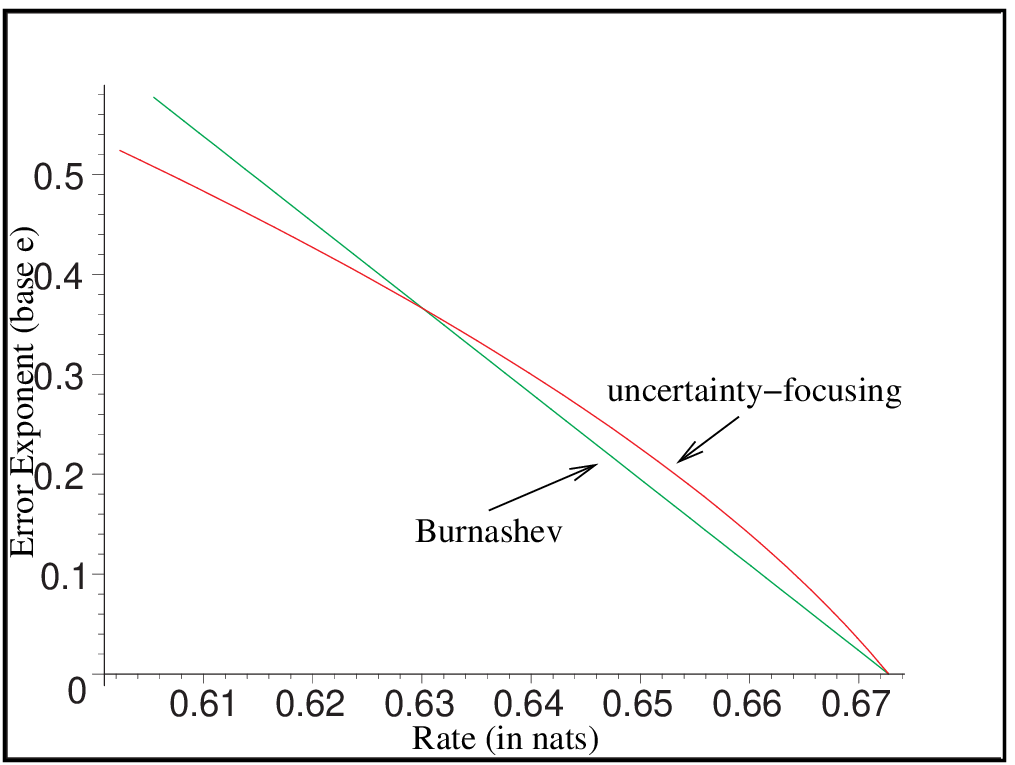}
\end{center}
\caption{The uncertainty-focusing and Burnashev bounds for a BSC with
  crossover probability $0.003$.}
\label{fig:focusvsburnashev}
\end{figure}

Now, consider a binary symmetric channel with crossover probability
$0.02$. The capacity of this channel is about $0.60$ nats per channel
use. Figure~\ref{fig:straightbsc} shows how the different choices of
$\lambda$ used in the bound (\ref{eqn:generalfeedbackbound}) kiss the
uncertainty-focusing bound for the BSC. It also shows the Burnashev
bound for variable-block-length coding for comparison. In this
particular plot, the Burnashev bound appears to always be higher than
the uncertainty-focusing bound. Figure~\ref{fig:focusvsburnashev}
illustrates that this is not always the case by plotting both bounds
in the high-rate regime for a BSC with crossover probability
$0.003$. It is unknown whether any scheme can actually achieve
fixed-delay reliabilities above the Burnashev bound since the scheme
of Theorem~\ref{thm:genericachieve} does not do so.

\begin{figure}[htbp]
\begin{center}
\includegraphics[width=4in,height=3in]{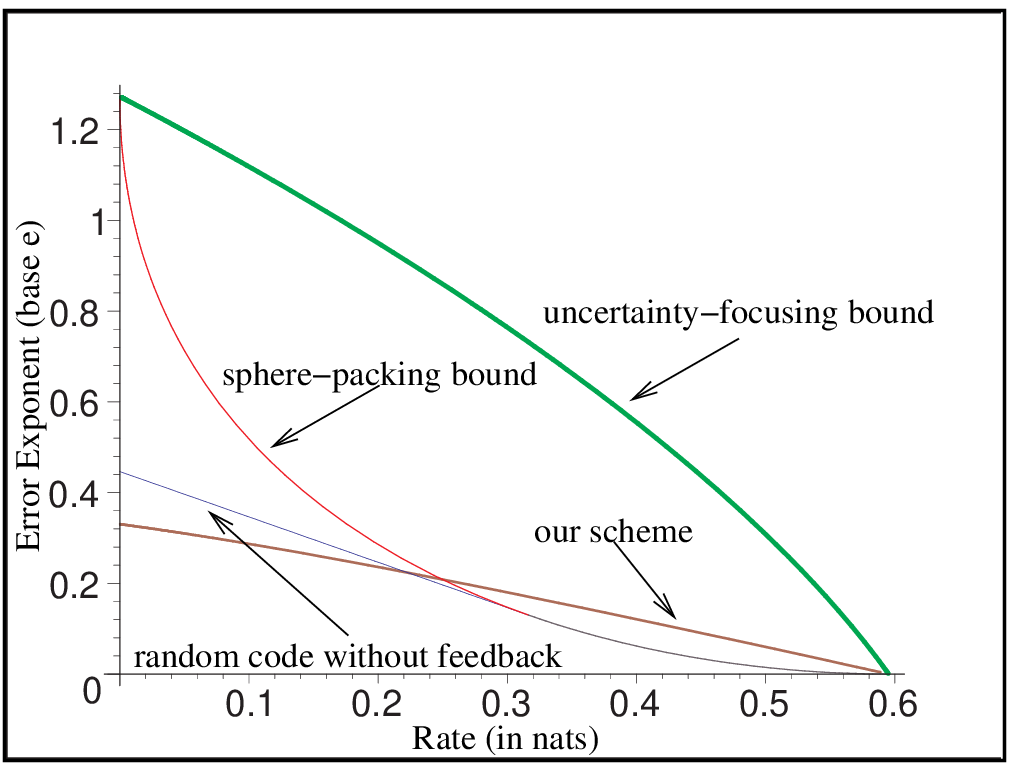}
\end{center}
\caption{The delay-reliability bounds for the BSC used with noiseless
  feedback. The sphere-packing bound approaches capacity in a
  quadratically flat manner while the uncertainty-focusing bound and
  the scheme of Theorem~\ref{thm:genericachieve} both approach the
  capacity point linearly, albeit with different slopes. The
  random-coding error exponent is also plotted for convenience since
  it beats the scheme of Theorem~\ref{thm:genericachieve} at low rates
  and can be attained without feedback.}
\label{fig:beatspherebsc}
\end{figure}

The gap between the uncertainty-focusing bound and the scheme of
Theorem~\ref{thm:genericachieve} is illustrated in
Figure~\ref{fig:beatspherebsc} for the BSC. This also shows how the
sphere-packing bound is significantly beaten at high rates even when
the channel has no zero-error capacity and thus provides an explicit
counterexample to Pinsker's Theorem~8 in
\cite{PinskerNoFeedback}. Examples showing how the
uncertainty-focusing bound is met for communication systems with
strictly positive zero-error capacity are deferred to
Section~\ref{sec:fortifiedexamples}.

\afterpage{\clearpage}

\section{Upper-bounding the fixed-delay reliability function without
  feedback} \label{sec:nofeedback}

This section proves Theorem~\ref{thm:nofeedbackbound} giving a
generalization of Pinsker's BSC argument from \cite{PinskerNoFeedback}
to the case of general DMCs. The Haroutunian exponent $E^+(R)$ from
(\ref{eqn:Haroutunianbound}) is shown to upper bound the reliability
function with delay if feedback is not available. For output-symmetric DMCs,
this is the same as the sphere-packing bound $E_{sp}(R)$. Furthermore,
we discuss why this proof {\em does not} go through when feedback is
present.

The complete proof spans the next few sections with some technical
details in the Appendices.

\subsection{Feedforward decoders and their equivalent forms}
For notational convenience, assume that $R' < 1$ so that at least one
channel use comes between each message bit's arrival. If $R' > 1$, the
same argument will work (at the cost of uglier notation) by
considering the incoming bits to arrive in pairs, triples, etc.
Theorem~\ref{thm:nofeedbackbound} is proven by considering a more
powerful class of decoders that have access to extra information that
can only improve their performance.
\begin{definition} \label{def:feedforwarddecoder}
A {\em delay-$d$ rate-$R$ decoder ${\cal D}$ with feedforward
  information} is a decoder ${\cal D}_i: \{0,1\}^{i-1}\times {\cal
  Y}^{\lceil \frac{i}{R'} \rceil +
    d} \rightarrow \{0,1\} $ that has noiseless
access to the past message bits $B_1^{i-1}$ in addition to the available
channel outputs $Y_1^{\lceil \frac{i}{R'} \rceil + d}$.
\end{definition}
\vspace{0.1in}

The first property is that with access to the feedforward information,
it suffices to ignore very old channel outputs.
\begin{lemma} \label{lem:windowdependence} For a memoryless channel,
  given a rate-$R$ encoder ${\cal E}$ without feedback and a delay-$d$
  rate-$R$ decoder ${\cal D}_i$ with feedforward for bit $i$, there
  exists a decoder ${\cal D}^f_i: \{0,1\}^{i-1}\times {\cal Y}^{d+1}
  \rightarrow \{0,1\}$ for bit $i$ that only depends
  on all the past message bits $B_1^{i-1}$ and the recent channel
  outputs $Y_{\lceil \frac{i}{R'} \rceil}^{\lceil \frac{i}{R'} \rceil
    + d}$. The bit error probability ${\cal P}(B_i \neq {\cal
    D}^f_i(B_1^{i-1},Y_{\lceil \frac{i}{R'} \rceil}^{\lceil
    \frac{i}{R'} \rceil + d})) \leq {\cal P}(B_i \neq {\cal
    D}_i(B_1^{i-1},Y_1^{\lceil \frac{i}{R'} \rceil + d}))$ assuming
  that the message bits $B$ are all iid fair coin tosses.
\end{lemma}
{\em Proof:} The result follows immediately from the following Markov
chain that holds since there is no feedback. 
\begin{equation} \label{eqn:basicmarkovchain}
Y_1^{\lceil \frac{i}{R'} \rceil -
  1}~-~B_1^{i-1}X_1^{\lceil \frac{i}{R'} \rceil - 1}~-~B_i X_{\lceil \frac{i}{R'}
  \rceil}^{\lceil \frac{i}{R'} \rceil + d} Y_{\lceil
  \frac{i}{R'} \rceil}^{\lceil \frac{i}{R'} \rceil + d}.
\end{equation}

To see the result explicitly, let ${\cal D}^{map}_i$ be the MAP
decoder for bit $i$ based on feedforward information $B_1^{i-1}$ and
observations $Y_1^{\lceil \frac{i}{R'} \rceil + d}$. 

\begin{eqnarray*}
&  & {\cal D}^{map}_i(b_1^{i-1},y_1^{\lceil \frac{i}{R'} \rceil + d}) \\
& = & 
\argmax_{b_i} {\cal P}(B_i = b_i|B_1^{i-1} = b_1^{i-1}, Y_1^{\lceil \frac{i}{R'}
  \rceil + d} = y_1^{\lceil \frac{i}{R'} \rceil + d}) \\
& = & 
\argmax_{b_i} {\cal P}(B_i = b_i, B_1^{i-1} = b_1^{i-1}, Y_1^{\lceil \frac{i}{R'}
  \rceil + d} = y_1^{\lceil \frac{i}{R'} \rceil + d}) \\
& = & 
\argmax_{b_i} \sum_{b_{i+1}^{i + \lceil dR' \rceil}} {\cal
  P}(B_i = b_i, B_1^{i-1} = b_1^{i-1}, B_{i+1}^{i + \lceil dR' \rceil} = b_{i+1}^{i + \lceil dR' \rceil}, Y_1^{\lceil \frac{i}{R'}
  \rceil + d} = y_1^{\lceil \frac{i}{R'} \rceil + d}) \\
& = & 
\argmax_{b_i} {\cal P}(B_1^{i-1} = b_1^{i-1}, Y_1^{\lceil \frac{i}{R'} \rceil - 1} = y_1^{\lceil
    \frac{i}{R'} \rceil - 1})  \\ 
& & ~~~~ \cdot  \sum_{b_{i+1}^{i + \lceil dR' \rceil}} 
{\cal P}(B_{i}^{i + \lceil dR' \rceil} = b_{i}^{i + \lceil dR' \rceil})
{\cal P}(Y_{\lceil \frac{i}{R'} \rceil}^{\lceil \frac{i}{R'}
  \rceil + d} = y_{\lceil \frac{i}{R'} \rceil}^{\lceil \frac{i}{R'} \rceil + d} 
\big| B_1^{i + \lceil dR' \rceil} = b_1^{i + \lceil dR' \rceil},
Y_1^{\lceil \frac{i}{R'} \rceil - 1} = y_1^{\lceil \frac{i}{R'} \rceil
  - 1}) \\ 
& =_{(a)} & 
\argmax_{b_i} \sum_{b_{i+1}^{i + \lceil dR' \rceil}} 
{\cal P}(Y_{\lceil \frac{i}{R'} \rceil}^{\lceil \frac{i}{R'}
  \rceil + d} = y_{\lceil \frac{i}{R'} \rceil}^{\lceil \frac{i}{R'} \rceil + d} 
\big| B_1^{i + \lceil dR' \rceil} = b_1^{i + \lceil dR' \rceil},
Y_1^{\lceil \frac{i}{R'} \rceil - 1} = y_1^{\lceil \frac{i}{R'} \rceil - 1}) \\
& =_{(b)} & 
\argmax_{b_i} \sum_{b_{i+1}^{i + \lceil dR' \rceil}} 
{\cal P}(Y_{\lceil \frac{i}{R'} \rceil}^{\lceil \frac{i}{R'}
  \rceil + d} = y_{\lceil \frac{i}{R'} \rceil}^{\lceil \frac{i}{R'} \rceil + d} 
\big| X_1^{\lceil \frac{i}{R'} \rceil + d} = {\cal E}(b_1^{i + \lceil dR' \rceil}), 
B_1^{i + \lceil dR' \rceil} = b_1^{i + \lceil dR' \rceil},
Y_1^{\lceil \frac{i}{R'} \rceil - 1} = y_1^{\lceil \frac{i}{R'} \rceil - 1}) \\
& =_{(c)} & 
\argmax_{b_i} \sum_{b_{i+1}^{i + \lceil dR' \rceil}} 
{\cal P}(Y_{\lceil \frac{i}{R'} \rceil}^{\lceil \frac{i}{R'}
  \rceil + d} = y_{\lceil \frac{i}{R'} \rceil}^{\lceil \frac{i}{R'} \rceil + d} 
\big| X_1^{\lceil \frac{i}{R'} \rceil + d} = {\cal E}(b_1^{i + \lceil dR' \rceil})). 
\end{eqnarray*}
The first few lines above are standard expansions of probability in
the MAP context and use the fact that the message bits are drawn
iid. (a) holds by dropping terms that do not depend on the exact
values for $b_i^{i + \lceil dR' \rceil}$ and thus do not impact the
$\argmax$. (b) uses the fact that the channel input $X$ is entirely
determined\footnote{Note that the same argument would also work if the
  encoder and decoder are allowed to share common randomness.} by the
message bits $B$ for an encoder without feedback. (c) is due to the
memoryless nature of the channel.

Define ${\cal D}^f_i$ directly as 
\begin{eqnarray*}
&  & {\cal D}^f_i(b_1^{i-1},y_{\lceil \frac{i}{R'} \rceil}^{\lceil \frac{i}{R'} \rceil + d}) \\
& = & 
\argmax_{b_i} \sum_{b_{i+1}^{i + \lceil dR' \rceil}} 
{\cal P}(Y_{\lceil \frac{i}{R'} \rceil}^{\lceil \frac{i}{R'}
  \rceil + d} = y_{\lceil \frac{i}{R'} \rceil}^{\lceil \frac{i}{R'} \rceil + d} 
\big| X_1^{\lceil \frac{i}{R'} \rceil + d} = {\cal E}(b_1^{i + \lceil dR' \rceil})). 
\end{eqnarray*}

This decoder only depends on the recent channel outputs in addition to
the feedforward information and achieves MAP performance. Since
MAP is optimal, the probability of bit error would be the same or
better than any other decoder. \hfill $\Diamond$ 
\vspace{0.15in} 

The second property is that it suffices to feedforward the error
sequence $\widetilde{B}_i = \widehat{B}_i + B_i \bmod 2$ rather than 
the past message bits themselves.

\begin{lemma} \label{lem:forwarderrors} Given a rate-$R$ encoder
  ${\cal E}$ and delay-$d$ rate-$R$ decoder ${\cal D}_i$ for bit $i$
  with feedforward. There exists another decoder $\widetilde{\cal
    D}_i: \{0,1\}^{i-1}\times {\cal Y}^{\lceil \frac{i}{R'} \rceil +
    d} \rightarrow \{0,1\}$ that only depends on the error sequence
  $\widetilde{B}_1^{i-1}$ in addition to the channel outputs
  $Y_{1}^{\lceil \frac{i}{R'} \rceil + d}$. If $\widetilde{B}_i =
  \widehat{B}_i + B_i \bmod 2$, then the outputs of the two
  decoders are identical $\widetilde{\cal
    D}_i(\widetilde{B}_1^{i-1},Y_{1}^{\lceil \frac{i}{R'} \rceil + d})
  = {\cal D}_i(B_1^{i-1},Y_{1}^{\lceil \frac{i}{R'} \rceil + d})$. 
\end{lemma}
{\em Proof:} This holds very generally by induction. Neither
memorylessness nor even the absence of feedback is required. It
clearly holds for $i=1$ since there are no prior bits and so the same
$\widehat{B}_1$ results. Assume now that
it holds for all $j<k$ and consider $i=k$. By the induction
hypothesis, the action of all the prior decoders $j$ can be simulated
since the decoder has access to $\widetilde{B}_1^{j-1}$ and
$Y_{1}^{\lceil \frac{j}{R'} \rceil + d}$. The resulting estimates
$\widehat{B}_j$ for $j<k$ can be XORed with $\widetilde{B}_j$ to
recover $B_j$ itself. Since $B_1^{k-1}$ can be recovered from the
given information, the original ${\cal D}_k$ decoder can be run as a
subroutine to give $\widehat{B}_k$. \hfill $\Diamond$ \vspace{0.15in} 

Lemmas~\ref{lem:windowdependence} and \ref{lem:forwarderrors} tell us
that  feedforward decoders can be thought in three ways: having access
to all past message bits and all past channel outputs, having access to
all past message bits and only a recent window of past channel outputs,
or having access to all past decoding errors and all past channel
outputs. 

\subsection{Constructing a rate-$(R - \delta_1)$ block code}
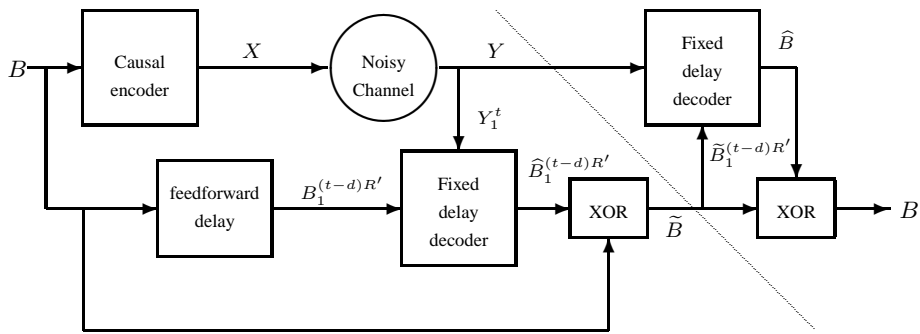
\begin{figure}[htbp]
\begin{center}
\setlength{\unitlength}{3100sp}%
\begingroup\makeatletter\ifx\SetFigFont\undefined%
\gdef\SetFigFont#1#2#3#4#5{%
  \reset@font\fontsize{#1}{#2pt}%
  \fontfamily{#3}\fontseries{#4}\fontshape{#5}%
  \selectfont}%
\fi\endgroup%
\begin{picture}(6997,2594)(1929,-3683)
\thicklines
{\color[rgb]{0,0,0}\put(3001,-3061){\framebox(900,750){}}
}%
\put(3451,-2611){\makebox(0,0)[b]{\smash{\SetFigFont{7}{6}{\rmdefault}{\mddefault}{\updefault}{\color[rgb]{0,0,0}feedforward}%
}}}
\put(3451,-2836){\makebox(0,0)[b]{\smash{\SetFigFont{7}{6}{\rmdefault}{\mddefault}{\updefault}{\color[rgb]{0,0,0}delay}%
}}}
{\color[rgb]{0,0,0}\put(4951,-3136){\framebox(900,900){}}
}%
\put(5401,-2536){\makebox(0,0)[b]{\smash{\SetFigFont{7}{6}{\rmdefault}{\mddefault}{\updefault}{\color[rgb]{0,0,0}Fixed}%
}}}
\put(5401,-2761){\makebox(0,0)[b]{\smash{\SetFigFont{7}{6}{\rmdefault}{\mddefault}{\updefault}{\color[rgb]{0,0,0}delay}%
}}}
\put(5401,-2986){\makebox(0,0)[b]{\smash{\SetFigFont{7}{6}{\rmdefault}{\mddefault}{\updefault}{\color[rgb]{0,0,0}decoder}%
}}}
{\color[rgb]{0,0,0}\put(6901,-2011){\framebox(900,900){}}
}%
\put(7351,-1411){\makebox(0,0)[b]{\smash{\SetFigFont{7}{6}{\rmdefault}{\mddefault}{\updefault}{\color[rgb]{0,0,0}Fixed}%
}}}
\put(7351,-1636){\makebox(0,0)[b]{\smash{\SetFigFont{7}{6}{\rmdefault}{\mddefault}{\updefault}{\color[rgb]{0,0,0}delay}%
}}}
\put(7351,-1861){\makebox(0,0)[b]{\smash{\SetFigFont{7}{6}{\rmdefault}{\mddefault}{\updefault}{\color[rgb]{0,0,0}decoder}%
}}}
{\color[rgb]{0,0,0}\put(4801,-1561){\circle{848}}
}%
{\color[rgb]{0,0,0}\put(2401,-2011){\framebox(900,900){}}
}%
{\color[rgb]{0,0,0}\put(1951,-1561){\vector( 1, 0){450}}
}%
{\color[rgb]{0,0,0}\put(3301,-1561){\vector( 1, 0){1050}}
}%
{\color[rgb]{0,0,0}\put(5251,-1561){\vector( 1, 0){1650}}
}%
{\color[rgb]{0,0,0}\put(2101,-1561){\line( 0,-1){1125}}
\put(2101,-2686){\vector( 1, 0){900}}
}%
{\color[rgb]{0,0,0}\put(3901,-2686){\vector( 1, 0){1050}}
}%
{\color[rgb]{0,0,0}\put(5401,-1561){\vector( 0,-1){675}}
}%
{\color[rgb]{0,0,0}\put(5851,-2686){\vector( 1, 0){450}}
}%
{\color[rgb]{0,0,0}\put(6301,-2911){\framebox(600,450){}}
}%
{\color[rgb]{0,0,0}\put(2401,-2686){\line( 0,-1){975}}
\put(2401,-3661){\line( 1, 0){4200}}
\put(6601,-3661){\vector( 0, 1){750}}
}%
{\color[rgb]{0,0,0}\put(6901,-2686){\line( 1, 0){450}}
\put(7351,-2686){\vector( 0, 1){675}}
}%
{\color[rgb]{0,0,0}\put(7801,-1561){\line( 1, 0){300}}
\put(8101,-1561){\vector( 0,-1){900}}
}%
{\color[rgb]{0,0,0}\put(7801,-2911){\framebox(600,450){}}
}%
{\color[rgb]{0,0,0}\put(7351,-2686){\vector( 1, 0){450}}
}%
{\color[rgb]{0,0,0}\put(8401,-2686){\vector( 1, 0){450}}
}%
\thinlines
{\color[rgb]{0,0,0}\multiput(5701,-1111)(12.718,-12.718){201}{\makebox(1.6667,11.6667){\SetFigFont{4}{6}{\rmdefault}{\mddefault}{\updefault}.}}
}%
\put(4801,-1561){\makebox(0,0)[b]{\smash{\SetFigFont{7}{6}{\rmdefault}{\mddefault}{\updefault}{\color[rgb]{0,0,0}Noisy}%
}}}
\put(4801,-1786){\makebox(0,0)[b]{\smash{\SetFigFont{7}{6}{\rmdefault}{\mddefault}{\updefault}{\color[rgb]{0,0,0}Channel}%
}}}
\put(2851,-1561){\makebox(0,0)[b]{\smash{\SetFigFont{7}{6}{\rmdefault}{\mddefault}{\updefault}{\color[rgb]{0,0,0}Causal}%
}}}
\put(2851,-1786){\makebox(0,0)[b]{\smash{\SetFigFont{7}{6}{\rmdefault}{\mddefault}{\updefault}{\color[rgb]{0,0,0}encoder}%
}}}
\put(6601,-2761){\makebox(0,0)[b]{\smash{\SetFigFont{7}{6}{\rmdefault}{\mddefault}{\updefault}{\color[rgb]{0,0,0}XOR}%
}}}
\put(7951,-1411){\makebox(0,0)[lb]{\smash{\SetFigFont{7}{6}{\rmdefault}{\mddefault}{\updefault}{\color[rgb]{0,0,0}$\widehat{B}$}%
}}}
\put(8101,-2761){\makebox(0,0)[b]{\smash{\SetFigFont{7}{6}{\rmdefault}{\mddefault}{\updefault}{\color[rgb]{0,0,0}XOR}%
}}}
\put(8926,-2761){\makebox(0,0)[lb]{\smash{\SetFigFont{9}{6}{\rmdefault}{\mddefault}{\updefault}{\color[rgb]{0,0,0}$B$}%
}}}
\put(1951,-1636){\makebox(0,0)[rb]{\smash{\SetFigFont{9}{6}{\rmdefault}{\mddefault}{\updefault}{\color[rgb]{0,0,0}$B$}%
}}}
\put(3751,-1486){\makebox(0,0)[b]{\smash{\SetFigFont{8}{6}{\rmdefault}{\mddefault}{\updefault}{\color[rgb]{0,0,0}$X$}%
}}}
\put(5701,-1486){\makebox(0,0)[b]{\smash{\SetFigFont{8}{6}{\rmdefault}{\mddefault}{\updefault}{\color[rgb]{0,0,0}$Y$}%
}}}
\put(5551,-2011){\makebox(0,0)[lb]{\smash{\SetFigFont{7}{6}{\rmdefault}{\mddefault}{\updefault}{\color[rgb]{0,0,0}$Y_1^t$}%
}}}
\put(7051,-2911){\makebox(0,0)[lb]{\smash{\SetFigFont{8}{6}{\rmdefault}{\mddefault}{\updefault}{\color[rgb]{0,0,0}$\widetilde{B}$}%
}}}
\put(7401,-2311){\makebox(0,0)[lb]{\smash{\SetFigFont{7}{6}{\rmdefault}{\mddefault}{\updefault}{\color[rgb]{0,0,0}$\widetilde{B}_1^{(t-d)R'}$}%
}}}
\put(4476,-2611){\makebox(0,0)[b]{\smash{\SetFigFont{7}{6}{\rmdefault}{\mddefault}{\updefault}{\color[rgb]{0,0,0}$B_1^{(t-d)R'}$}%
}}}
\put(5951,-2436){\makebox(0,0)[lb]{\smash{\SetFigFont{7}{6}{\rmdefault}{\mddefault}{\updefault}{\color[rgb]{0,0,0}$\widehat{B}_1^{(t-d)R'}$}%
}}}
\end{picture}
\end{center}
\caption{The relevant ``cutset'' illustrated. For the message bits $B$ to
  pass noiselessly across the cutset, the sum of the mutual
  information between $X$ and $Y$ and the entropy of $\widetilde{B}$ must
  be larger than the entropy of the $B$. The mutual information
  between $X$ and $Y$ is bounded by the capacity of the noisy channel
  and the entropy of $\widetilde{B}$ provides a lower bound to the
  probability of bit errors.}
\label{fig:newdataprocessing}
\end{figure}

Consider the system illustrated in
Figure~\ref{fig:newdataprocessing}. The message bitstream consisting of
fair coin tosses is encoded using the given rate-$R$ encoder. The
channel outputs are decoded using the delay-$d$ rate-$R$ decoders with
feedforward, with the feedforward in the form of the error signals
$\widetilde{B}$ by Lemma~\ref{lem:forwarderrors}. These error signals
are generated by XORing the message bits with the output of an
equivalent feedforward decoder. Finally, the feedforward error signals
are used one more time and combined with the estimates $\widehat{B}$
to recover the message bits $B$ exactly. It is immediately clear that
this hypothetical system never makes an error from end to end.

Now, this system will be interpreted as a block code. Pick an
arbitrarily small $\delta_1 > 0$. To avoid
cumbersome notation, some integer effects will be neglected. For every
delay $d$, pick a block length $n = \frac{d R}{\delta_1}$. For notational
convenience, let $\delta_1'$ be such that $\frac{(R - \delta_1)}{\ln
  2} = R' - \delta_1'$ so that $n (R' - \delta_1') = n R' - dR'$.

The data processing inequality implies:
\begin{lemma} \label{lem:newdataprocessing}
Suppose $n$ is the block length, the block rate is $R - \delta_1$ nats
per channel use, the $X_1^{n}$ are the channel inputs, the $Y_1^{n}$
are the channel outputs, and the $\widetilde{B}_1^{n (R' -
\delta_1')}$ are the error signals coming from the underlying rate-$R$
delay-$d$ encoding and decoding system. Then
\begin{equation} \label{eqn:newdataprocessing}
H(\widetilde{B}_1^{n(R' - \delta_1')}) \geq n (R-\delta_1)- I(X_1^{n};
Y_1^{n}).
\end{equation}
\end{lemma}
{\em Proof:} See Appendix~\ref{app:lemdataprocessing}.

\subsection{Lower-bounding the error probability}

Now, suppose this system of Figure~\ref{fig:newdataprocessing} were to
be run over the noisy channel $G$ that minimizes
(\ref{eqn:Haroutunianbound}) at $R - 2\delta_1$ nats per channel
use. Since the capacity of $G$ is at most $R - 2\delta_1$ nats per
channel use and there is no feedback to the encoder, the mutual
information between the channel inputs and outputs is upper-bounded by
\begin{equation} \label{eqn:goofychannelinformation}
I(X_1^{n};Y_1^{n}) \leq n(R - 2\delta_1) = n(R - \delta_1) - n\delta_1.
\end{equation}
Plugging (\ref{eqn:goofychannelinformation}) into
(\ref{eqn:newdataprocessing}) from Lemma~\ref{lem:newdataprocessing} gives 
\begin{equation} \label{eqn:toomuchentropy}
H(\widetilde{B}_1^{n(R' - \delta_1')}) \geq n \delta_1.
\end{equation}
Since the sum of marginal entropies $\sum_{i=1}^{n(R' -
  \delta_1')}H(\widetilde{B}_i) \geq H(\widetilde{B}_1^{n(R' -
  \delta_1')})$, the average entropy of the error bits
$\widetilde{B}_i$ is at least $\frac{\delta_1}{R' - \delta_1'} >
0$. Consider $i^*$ whose individual entropy
$H(\widetilde{B}_{i^*}) \geq \frac{\delta_1}{R' - \delta_1'}$.

By the strict monotonicity of the binary entropy function for
probabilities less than $\frac{1}{2}$, there exists a $\delta_2 > 0$
so that the probability of bit error ${\cal P}(\widetilde{B}_{i^*} = 1) =
{\cal P}(\widehat{B}_{i^*} \neq B_{i^*}) \geq \delta_2$. While the specific
positions $i^*$ might vary for different delays $d$, the lower bound
$\delta_2$ on minimum error probability does not vary. 

At this point, Lemma~\ref{lem:windowdependence} implies that even if
the channel $G$ were used only for the $d + 1$ time steps from
$[\lceil\frac{i^*}{R'} \rceil, \lceil \frac{i^*}{R'} \rceil + d]$,
the same minimum error probability $\delta_2$ must hold, regardless of
how large $d$ is. For each possible message prefix $b_1^{i^*}$,
there is an error event $A(b_1^{i^*})$ corresponding to the channel
outputs that would cause erroneous decoding of the $i^*$-th
bit. Formally, $A(b_1^{i^*}) := \{y_{\lceil\frac{i^*}{R'}\rceil}^{\lceil
  \frac{i^*}{R'} \rceil + d} | {\cal
  D}^f_{i^*}(b_1^{i^*-1},y_{\lceil\frac{i^*}{R'}\rceil}^{\lceil
  \frac{i^*}{R'} \rceil + d}) \neq b_{i^*}\}$. 

Averaging out the probability of error over message prefixes gives
\begin{eqnarray*}
\delta_2 
& \leq &\sum_{b_1^{i^*}} \frac{1}{2^{i^*}}
{\cal P}(A(b_1^{i^*})|B_1^{i^*} = b_1^{i^*}) \\
& = & \sum_{b_1^{\lfloor (\lceil \frac{i^*}{R'} \rceil + d_j)R'
\rfloor}} 2^{-\lfloor (\lceil \frac{i^*}{R'} \rceil + d_j)R' \rfloor} 
{\cal P}(A(b_1^{i^*})|\vec{X} = {\cal E}(b_1^{\lfloor (\lceil
\frac{i^*}{R'} \rceil + d_j)R'\rfloor})).
\end{eqnarray*}
Since the average over messages $b_1^{\lfloor (\lceil \frac{i^*}{R'}
  \rceil + d_j)R' \rfloor}$ is at least $\delta_2$, and
the probabilities can be no bigger than $1$ and no smaller than $0$,
at least a $\frac{\delta_2}{2}$ proportion of messages result in the
$A(b_1^{i^*})$ having a conditional probability of at least
$\frac{\delta_2}{2}$ if channel $G$ is used. 

All that remains is to show that the probability of this event under
the true channel $P$ cannot be too small. To distinguish between the
probability of an event when using channel $P$ or channel $G$,
subscripts are used with ${\cal P}_P$ used to refer to the probability
of an event when the channel is $P$ and ${\cal P}_G$ used for when the
channel is $G$.

This simple lemma is useful:
\begin{lemma} \label{lem:measurechange}
If under channel $G$ and input sequence $\vec{x}$, the probability 
${\cal P}_G(Y_1^d \in A|X_1^d = \vec{x}) \geq \delta > 0$, then for
any $\epsilon > 0$, there exists $d_0(\epsilon, \delta, G, P)$ so that 
as long as $d > d_0(\epsilon, \delta, G, P)$, the $A$ 
event's conditional probability using channel $P$ must satisfy ${\cal
  P}_P(Y_1^d \in A|X_1^d = \vec{x}) \geq \frac{\delta}{2} \exp(-d
(D(G||P | \vec{r}) + \epsilon)$ where $\vec{r}$ is the type of $\vec{x}$.  
\end{lemma}
{\em Proof: } See Appendix~\ref{app:lemmeasurechange}.

\vspace{0.1in}

Given $\frac{\delta_2}{2}, G$ and an arbitrary $\epsilon > 0$, apply
Lemma~\ref{lem:measurechange} to consider delays $d > d_0$. This
reveals that 
\begin{eqnarray*}
{\cal P}_P(B_{i^*} \neq \widehat{B}_{i^*}) 
& = & 
\sum_{b_1^{\lfloor (\lceil \frac{i^*}{R'} \rceil + d_j)R' \rfloor}} 
2^{-\lfloor (\lceil \frac{i^*}{R'} \rceil + d_j)R' \rfloor} 
{\cal P}_P\left(A(b_1^{i^*})|\vec{X} = 
{\cal E}(b_1^{\lfloor (\lceil \frac{i^*}{R'} \rceil + d_j)R'\rfloor})\right) \\
& \geq_{(a)} & \frac{(\delta_2)^2}{8} \exp(-(d_j+1)(\max_{\vec{r}}
D(G||P | \vec{r}) + \epsilon)) \\
& =_{(b)} & \frac{(\delta_2)^2}{8} \exp\left(-(d_j+1)(E^+(R - 2\delta_1) + \epsilon)\right).
\end{eqnarray*}
(a) follows from the fact that a proportion $\frac{\delta_2}{2}$ of the
messages must have probability of bit error of at least
$\frac{\delta_2}{2}$ with the final factor of $2$ coming from
Lemma~\ref{lem:measurechange}. Since the local type of the channel
input is unknown, the maximum is taken over the channel input type
$\vec{r}$. (b) is using the definition of $G$ and
the Haroutunian bound.

Since $\epsilon > 0$ is an arbitrary choice and $\delta_2$ does not
depend on the delay $d$, taking logs quickly reveals that the error
exponent with delay cannot be any larger than $E^+(R -
2\delta_1)$. For any $\alpha > E^+(R)$, it is always possible to pick
a $0 < \delta_1 < \frac{R - C_{0,f}}{2}$ so that $\alpha > E^+(R -
2\delta_1)$ as well since the Haroutunian bound $E^+$ is continuous in
the rate for all rates strictly below Shannon capacity and above the
feedback zero-error capacity $C_{0,f}$. Thus, no exponent $\alpha >
E^+(R)$ can be asymptotically achieved and
Theorem~\ref{thm:nofeedbackbound} is proved. \hfill $\Box$
\vspace{0.20in}

\subsection{Comments}
For output-symmetric channels, $E^+(R) = E_{sp}(R)$ and so the usual
sphere-packing bound is recovered in the fixed-delay context. Since
$E_{sp}(R)$ is achieved universally with delay at high rates by using
infinite-length random time-varying convolutional codes, this means
that such codes achieve the best possible asymptotic tradeoff between
probability of bit error and end-to-end delay. However, the proof in
the previous section does not get to the sphere-packing bound for
asymmetric channels like the Z-channel plotted in
Figure~\ref{fig:zchannelbounds}. 

We could apply the sphere-packing bound to the $n$-length block-code
by trying the $G$ channel that optimizes $E_{sp}(R - 2\delta_1 -
\gamma)$ for one of the block codeword compositions $\vec{r}$ that
contains at least $\exp(n (R - \delta_1 - \gamma))$ codewords for some
small $\gamma > 0$ that can be chosen after $\delta_1$. As a result,
there would be weak bits whose probabilities of error are at least
$\delta_2$ when used with the $G$ channel. The problem arises when we
attempt to translate this back to the original channel $P$. Because
the local $(d+1)$-length input-type is unknown in the vicinity of these
weak bits, we would only be able to prove an exponent of
\begin{equation} \label{eqn:betterHaroutunianbound}
\widetilde{E}^+(R) = \inf_{G: \max_{\vec{r}: I(\vec{r},P) \geq R}
  I(\vec{r},G) < R}  \max_x D(G(\cdot|x)||P(\cdot|x)).
\end{equation}
This is formally better than (\ref{eqn:Haroutunianbound}) since
there is slightly more flexibility in choosing the mimicking channel
$G$. It now just has to have a mutual information across it lower than
$R$ when driven with an input distribution that is good enough for the
original channel. But, it seems unlikely that
(\ref{eqn:betterHaroutunianbound}) is tight the way that $E_{sp}$ is
since for the Z-channel, it can evaluate to the same thing as
(\ref{eqn:Haroutunianbound}). 

It is more interesting to reflect upon why this proof does not go
through when feedback is available. This reveals why Pinsker's
assertion of Theorem~8 in \cite{PinskerNoFeedback} is
incorrect. Although the lack of feedback was used in many places, the
most critical point is Lemma~\ref{lem:windowdependence} which
corresponds to \cite[Eqn.~(39)]{PinskerNoFeedback}. When feedback is
present, the current channel inputs can depend on the past channel
{\em outputs}, even if we condition on the past channel {\em
  inputs}. Thus it is not possible to take a block error and then
focus attention on the channel behavior only during the delay
period. It could be that the atypical channel behavior has to begin
well before the bit in question even arrived at the encoder. This is
seen clearly in the BEC case with feedback discussed in
Section~\ref{sec:becexample} --- the most common failure mode is
for a bit to enter finding a large queue of senior bits already
waiting and then finding that service continues to be so slow that the
senior bits are not all able to leave the queue before the bit's own
deadline expires.

\section{Upper-bounding the fixed-delay reliability function with
  feedback} \label{sec:feedbackbound}

To prove Theorem~\ref{thm:generalfeedbackbound} and get a proper upper
bound to the fixed-delay reliability function when feedback is
allowed, we need to account for the fact that the dominant error event
might begin before the bit in question even arrives at the
encoder. To do this, Viterbi's argument from \cite{viterbi} is
repurposed to address delay rather than constraint length. We call
this upper bound the ``uncertainty-focusing bound'' because it is
based on the idea of focusing the decoder's uncertainty about the
message bits given the channel outputs onto bits whose deadlines are
not pending. 

To bound what is possible, a fixed-delay code is translated into a
fixed-block-length code. A lower bound on error probability for block
codes is then pulled back to give a bound on the probability of error
for the original fixed-delay code. The key difference from the
previous section is that the block-length $n$ is not automatically
made large compared to the delay. Rather, each different block length
provides its own bound at all rates, with the final bound at any given
rate and delay coming from optimizing over the block length.

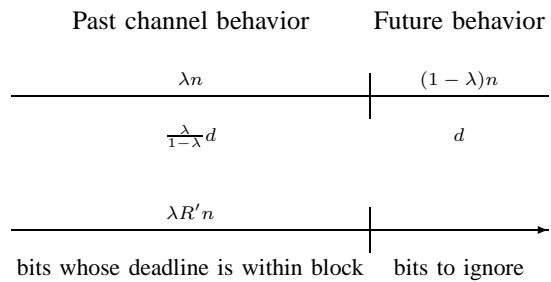
\begin{figure}[htbp]
\begin{center}
\setlength{\unitlength}{3700sp}%
\begingroup\makeatletter\ifx\SetFigFont\undefined%
\gdef\SetFigFont#1#2#3#4#5{%
  \reset@font\fontsize{#1}{#2pt}%
  \fontfamily{#3}\fontseries{#4}\fontshape{#5}%
  \selectfont}%
\fi\endgroup%
\begin{picture}(3624,2547)(589,-2023)
\thinlines
{\color[rgb]{0,0,0}\put(601,-61){\line( 1, 0){3600}}
}%
{\color[rgb]{0,0,0}\put(3001, 89){\line( 0,-1){300}}
}%
{\color[rgb]{0,0,0}\put(601,-961){\vector( 1, 0){3600}}
}%
{\color[rgb]{0,0,0}\put(3001,-811){\line( 0,-1){300}}
}%
\put(3601,-361){\makebox(0,0)[b]{\smash{\SetFigFont{7}{6}{\rmdefault}{\mddefault}{\updefault}{\color[rgb]{0,0,0}$d$}%
}}}
\put(1801,-361){\makebox(0,0)[b]{\smash{\SetFigFont{7}{6}{\rmdefault}{\mddefault}{\updefault}{\color[rgb]{0,0,0}$\frac{\lambda}{1-\lambda}d$}%
}}}
\put(1801,-1261){\makebox(0,0)[b]{\smash{\SetFigFont{9}{6}{\rmdefault}{\mddefault}{\updefault}{\color[rgb]{0,0,0}bits whose deadline is within block}%
}}}
\put(3601,-1261){\makebox(0,0)[b]{\smash{\SetFigFont{9}{6}{\rmdefault}{\mddefault}{\updefault}{\color[rgb]{0,0,0}bits to ignore}%
}}}
\put(1801, 14){\makebox(0,0)[b]{\smash{\SetFigFont{7}{6}{\rmdefault}{\mddefault}{\updefault}{\color[rgb]{0,0,0}$\lambda n$}%
}}}
\put(3601, 14){\makebox(0,0)[b]{\smash{\SetFigFont{7}{6}{\rmdefault}{\mddefault}{\updefault}{\color[rgb]{0,0,0}$(1-\lambda)n$}%
}}}
\put(1801,389){\makebox(0,0)[b]{\smash{\SetFigFont{10}{6}{\rmdefault}{\mddefault}{\updefault}{\color[rgb]{0,0,0}Past channel behavior}%
}}}
\put(3601,389){\makebox(0,0)[b]{\smash{\SetFigFont{10}{6}{\rmdefault}{\mddefault}{\updefault}{\color[rgb]{0,0,0}Future behavior}%
}}}
\put(1801,-886){\makebox(0,0)[b]{\smash{\SetFigFont{7}{6}{\rmdefault}{\mddefault}{\updefault}{\color[rgb]{0,0,0}$\lambda R'n$}%
}}}
\end{picture}
\end{center}
\caption{Using the fixed-delay code to make a block code of length
$n$: only the first $\lambda R' n$ bits are decoded by the end of the
block and so the rate is cut by a factor of $\lambda$. The error
exponent with block length $n$ is $1-\lambda$ of the exponent with the
delay $d$.}
\label{fig:lambdameaning}
\end{figure}

{\em Proof: } Given a code with fixed delay $d$, pick an arbitrary $0
< \lambda < 1$ and set the block length $n = \frac{d}{1-\lambda}$. As
illustrated in Figure~\ref{fig:lambdameaning}, this implies that $n =
\lambda n + d = \frac{\lambda}{1 - \lambda} d + d$. To avoid
cumbersome notation, integer effects are ignored here. When $d$ is
small, the fact that the block length must be an integer limits our
choices for $\lambda$ in an insignificant way.

The block decoder operates by running the delay-$d$ decoder. This
decodes the first $\lambda nR'$ bits, thus making the effective rate
for the block code $\lambda R'$ bits per channel use or $\lambda R$
nats per channel use. The encoder just applies the given causal
encoders with feedback using the actual message bits as the first
$\lambda nR'$ bits. Random coin tosses can be used for the final
$(1-\lambda) n R'$ inputs to the encoders since these will not be
decoded anyway.

Let $B_1^{\lambda nR'}$ be the original message consisting entirely of
independent fair coin tosses. The Haroutunian bound reveals that given
any $\delta_1,\epsilon > 0$ there exists a sufficiently large block
length $n_1$ and a constant $K$, so that as long as $n > n_1$, this
fixed-block-length code with feedback must have a probability of block
error that is lower bounded by \cite{Haroutunian}
\begin{equation} \label{eqn:haroutunianerror}
{\cal P}_P(B_{1}^{\lambda nR'} \neq \widehat{B}_{1}^{\lambda nR'}) \geq 
K \exp\left(-n[E^+(\lambda R - \delta_1) + \epsilon] \right).
\end{equation}

Substitute in $n = \frac{d}{1 - \lambda}$ and then notice that there
must be at least one message bit position $i^*$ whose probability of
bit error is no worse than $\frac{1}{n \lambda R'}$ times the
probability of block error. This gives
$${\cal P}_P(B_{i^*} \neq \widehat{B}_{i^*}) \geq
\frac{(1-\lambda)K}{\lambda R' d} 
\exp\left(-d[\frac{E^+(\lambda R - \delta_1)}{1-\lambda} +
\frac{\epsilon}{1-\lambda}] \right).$$

Since the $\frac{1}{d}$ term in front is dominated by the exponential
and $\delta_1,\epsilon$ are arbitrarily small and $\lambda$ was
arbitrary, taking logs and the limit $d \rightarrow \infty$ proves
(\ref{eqn:generalfeedbackbound}). 

Whenever $E^+(R) = E_{sp}(R)$, by using (\ref{eqn:rhospherepack}) and
following arguments identical to those used in the analysis of
convolutional codes, (\ref{eqn:generalfeedbackbound}) turns into
(\ref{eqn:symmetricfeedbackbound}). These arguments are given
in Appendix~\ref{app:focussymmetric} for completeness. 

Expanding (\ref{eqn:symmetricfeedbackbound}) by Taylor expansion in
the vicinity of $\eta=0$, noticing that the first derivative of $E_0$
there is the capacity $C$, and applying simple algebra leads to the
negative slope of $2C/\frac{\partial^2 E_0(0)}{\partial \eta^2}$ in
the vicinity of the $(C,0)$ point. When the second derivative term is
equal to zero, then \cite{gallager} reveals that the channel's
sphere-packing bound hits $(C,0)$ at a positive slope of at least $-1$
and thus (\ref{eqn:symmetricfeedbackbound}) evaluated at $\eta=1$
already has hit the capacity. There is no need to consider lower
values of $\eta$. The uncertainty-focusing bound in such cases jumps
discontinuously down to zero at rates above capacity. \hfill $\Box$
\vspace{0.20in}


It is also important to notice that the core idea driving the proof is
the inverse-concatenation construction from \cite{viterbi} and
\cite{ForneyML}. This allows us to map an upper bound on the
fixed-block-length reliability function into an upper bound on the
fixed-delay reliability. As a result, the uncertainty-focusing bound
can also be used for channels {\em without} feedback. 

\begin{corollary} 
For a DMC, no fixed-delay
  exponent greater than the expurgated bound at rate 0 ($\alpha >
  E_{ex}(0)$ from \cite{gallager}) is asymptotically achievable without
  feedback.
\end{corollary}
{\em Proof:} Because the straight-line bound \cite{gallager} can
tighten the low-rate exponent for block-codes without feedback, this
means that it can also be used to tighten the bound for fixed-delay
codes in the low-rate regime. The inverse concatenation construction
immediately turns the straight-line bound for fixed-block-length codes
turns into a horizontal line at $E_{ex}(0)$ for fixed-delay
codes. \hfill $\Box$

\vspace{0.1in}
Thus the best upper bound we have for the reliability function for
end-to-end delay in a system without feedback is $\min(E_{ex}(0),
E^+(R))$.

For the case of output-symmetric channels with feedback (or whenever the
$E_{a,s}$ bound is tight), it is also possible to explicitly calculate
the worst case $\lambda^*$ in parametric form using the arguments of
Appendix~\ref{app:focussymmetric}:
\begin{equation} \label{eqn:optimumlambda}
\lambda^* = 
\frac{(\frac{\partial E_0(\rho = \eta)}{\partial \rho})}{R(\eta)} 
= 
\frac{\eta}{E_0(\eta)}
(\frac{\partial E_0(\rho = \eta)}{\partial \rho}).
\end{equation}
The exponentially dominating error event involves $\frac{\lambda^*}{1
  - \lambda^*} d $ of the past channel outputs as well as the $d$ time
steps in the future --- for a error event length of $\frac{d}{1 -
  \lambda^*}$. Thus $\lambda^*$ captures the critical balance between
how badly the channel must misbehave and how long it must misbehave
for. In general, when $R$ is near $C$, the $\eta$ will be near
zero. Since $\frac{\partial E_0(\rho = 0)}{\partial \rho} = C$, this
implies $\lambda^*$ there will be near $1$, and the dominant error
events will be much longer than the desired end-to-end delay.



\section{Achievability of the fixed-delay reliability with
feedback for erasure channels} \label{sec:erasurefeedback}

This section proves Theorem~\ref{thm:erasurecase} and thereby
demonstrates the asymptotic achievability of $E_a(R) = E_{a,s}(R)$
everywhere for erasure channels with noiseless feedback. 

\subsection{The optimal code and its reliability}

The optimal scheme for the binary erasure channel with
instantaneous\footnote{If the feedback is not instantaneous, then
  there is no obvious scheme. Asymptotically optimal schemes for such
  cases are given in \cite{BalancedForwardFeedback}.}  causal
noiseless feedback is intuitively obvious --- buffer up message bits
as they arrive and attempt to transmit the oldest message bit that has
not yet been received correctly by the receiver. What is not
immediately obvious is how well this scheme actually performs with
end-to-end delay.

The Markov-chain analysis in Section~\ref{sec:becexample} becomes
unwieldy at rates that are not simple rational numbers like
$\frac{1}{2}$.  In \cite{ACC00Paper, SahaiThesis}, an analysis of this
scheme is given by translating the communication problem into a
problem of stabilization of an unstable scalar plant over a noisy
feedback link using techniques from \cite{ControlPartI}. The
stabilization problem can then be studied explicitly in terms of its
$\eta$-th moments, which can be understood using certain infinite
sums. The dominant terms in these sums are found using heuristic
arguments (rigorous only for $\eta=2,3$) and the convergence of those
reveals which $\eta$-moments are finite. This in turn implicitly gives
a lower bound to the reliability function with delay. It turns out
that this calculation agrees with the uncertainty-focusing bound. In
the following section, a direct and rigorous proof is given for
Theorem~\ref{thm:erasurecase} at all rates.

A BEC with erasure probability $\beta$ is output-symmetric and so the
Haroutunian bound and the sphere-packing bound are
identical. Evaluating the symmetric uncertainty-focusing bound
(\ref{eqn:symmetricfeedbackbound}) gives the following parametric
expression: (in units of bits and power of two reliability exponents
since the computation is simpler in that base)
\begin{equation} \label{eqn:becreliability}
E_{a}^{bec}(R') = \eta-\log_2 (1+\beta(2^\eta-1)) ~~\,,\,~~ R' =
\frac{\eta-\log_2(1+\beta(2^\eta-1))}{\eta}
\end{equation}
where $\eta$ ranges from $0$ to $\infty$. 

Simple algebraic manipulation allows the parameter $\eta$ to be
eliminated and this results in the rate-reliability tradeoff of
(\ref{eqn:becanytimecapacity}). The calculations for this and the
simple low-rate bound are in Appendix~\ref{app:erasurelowrate}.  

\subsection{Direct proof of achievability} \label{sec:becproof}
In this section, the asymptotic achievability of the BEC's
fixed-delay reliability function (\ref{eqn:becreliability}) is proven
directly using a technique that parallels the bounding technique used
for Theorem~\ref{thm:generalfeedbackbound}. 

The key idea is to use the first-in-first-out property of the ``repeat
until received'' strategy, treating the system as a D/M/1 queue.  The
only way the $i$-th bit would not be received by the deadline is if
there were too few successes. It is easy to see that this could happen
if there were zero successes after it enters the system. But it could
also happen if there were only one success since the previous bit
entered the system, and so on. This is captured in the following:
\begin{lemma} \label{lem:notthereyet}
The probability that bit $i$ is unable to meet deadline $\lceil
\frac{i}{R'} \rceil + d$ can be upper-bounded by: 
\begin{equation} \label{eqn:sumbound}
{\cal P}(\widehat{B}_i(\lceil \frac{i}{R'} \rceil + d) \neq B_i)
\leq 
\sum_{k=1}^i {\cal P}\left(\frac{1}{d + \lceil \frac{i}{R'} \rceil - \lceil
\frac{k}{R'} \rceil} \sum_{t=\lceil \frac{k}{R'} \rceil}^{\lceil \frac{i}{R'}
\rceil + d} Z_t \leq \frac{i-k}{d + \lceil \frac{i}{R'} \rceil - \lceil
\frac{k}{R'} \rceil}
\right)
\end{equation}
where the $\{Z_t\}$ are the iid random variables that are $1$ if the
$t$-th channel use is successful and $0$ if it is erased.
\end{lemma}
{\em Proof:} See Appendix~\ref{app:lemnotthereyet}.

\begin{figure}[hbtp]
\begin{center}
\setlength{\unitlength}{3700sp}%
\begingroup\makeatletter\ifx\SetFigFont\undefined%
\gdef\SetFigFont#1#2#3#4#5{%
  \reset@font\fontsize{#1}{#2pt}%
  \fontfamily{#3}\fontseries{#4}\fontshape{#5}%
  \selectfont}%
\fi\endgroup%
\begin{picture}(5274,1935)(139,-1090)
\thinlines
{\color[rgb]{0,0,0}\put(5401,389){\vector(-1, 0){5250}}
}%
{\color[rgb]{0,0,0}\put(4951,539){\line( 0,-1){300}}
}%
{\color[rgb]{0,0,0}\put(4201,614){\line( 0,-1){450}}
}%
{\color[rgb]{0,0,0}\put(4201,-61){\line( 1, 0){1200}}
}%
{\color[rgb]{0,0,0}\multiput(1951,-511)(4.49775,0.00000){668}{\makebox(1.6667,11.6667){\SetFigFont{5}{6}{\rmdefault}{\mddefault}{\updefault}.}}
}%
{\color[rgb]{0,0,0}\multiput(1951,614)(0.00000,-4.50000){101}{\makebox(1.6667,11.6667){\SetFigFont{5}{6}{\rmdefault}{\mddefault}{\updefault}.}}
}%
\put(5176,164){\makebox(0,0)[b]{\smash{\SetFigFont{7}{6}{\rmdefault}{\mddefault}{\updefault}{\color[rgb]{0,0,0}$d$}%
}}}
\put(4576,164){\makebox(0,0)[b]{\smash{\SetFigFont{7}{6}{\rmdefault}{\mddefault}{\updefault}{\color[rgb]{0,0,0}$\frac{\lambda^*}{1-\lambda^*}d$}%
}}}
\put(4801,-211){\makebox(0,0)[b]{\smash{\SetFigFont{7}{6}{\rmdefault}{\mddefault}{\updefault}{\color[rgb]{0,0,0}dominating }%
}}}
\put(4801,-361){\makebox(0,0)[b]{\smash{\SetFigFont{7}{6}{\rmdefault}{\mddefault}{\updefault}{\color[rgb]{0,0,0}error event}%
}}}
\put(1951,689){\makebox(0,0)[b]{\smash{\SetFigFont{7}{6}{\rmdefault}{\mddefault}{\updefault}{\color[rgb]{0,0,0}$\bar{k}$}%
}}}
\put(4951,689){\makebox(0,0)[b]{\smash{\SetFigFont{7}{6}{\rmdefault}{\mddefault}{\updefault}{\color[rgb]{0,0,0}$k=i$}%
}}}
\put(4201,689){\makebox(0,0)[b]{\smash{\SetFigFont{7}{6}{\rmdefault}{\mddefault}{\updefault}{\color[rgb]{0,0,0}$k^*$}%
}}}
\put(3451,-736){\makebox(0,0)[b]{\smash{\SetFigFont{7}{6}{\rmdefault}{\mddefault}{\updefault}{\color[rgb]{0,0,0}$k > \bar{k}$ type events individually}%
}}}
\put(1951,-211){\makebox(0,0)[rb]{\smash{\SetFigFont{7}{6}{\rmdefault}{\mddefault}{\updefault}{\color[rgb]{0,0,0}have probabilities that die}%
}}}
\put(1951,-361){\makebox(0,0)[rb]{\smash{\SetFigFont{7}{6}{\rmdefault}{\mddefault}{\updefault}{\color[rgb]{0,0,0}and are summed away}%
}}}
\put(3451,-886){\makebox(0,0)[b]{\smash{\SetFigFont{7}{6}{\rmdefault}{\mddefault}{\updefault}{\color[rgb]{0,0,0}bounded by dominating one }%
}}}
\put(3451,-1036){\makebox(0,0)[b]{\smash{\SetFigFont{7}{6}{\rmdefault}{\mddefault}{\updefault}{\color[rgb]{0,0,0}and union bounded collectively}%
}}}
\put(1951,-61){\makebox(0,0)[rb]{\smash{\SetFigFont{7}{6}{\rmdefault}{\mddefault}{\updefault}{\color[rgb]{0,0,0}Smaller $k\leq\bar{k}$ type events}%
}}}
\end{picture}
\end{center}
\caption{Error events beginning with message bits from earlier than
  $\bar{k}$ are those whose probabilities are getting exponentially
  small and are all less than the dominating event. The shorter events
  number linear in $d$ and are all individually smaller than the
  dominating error event. This shows that the dominating event's
  exponent governs the probability of error as a whole.}
\label{fig:erasurebounding}
\end{figure}
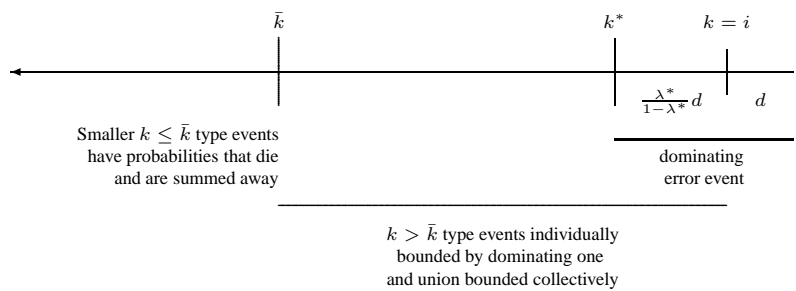

The next idea is to isolate the dominant term in the sum
(\ref{eqn:sumbound}) and to bound the whole sum explicitly in terms of
this. The idea is depicted in Figure~\ref{fig:erasurebounding}. The
potentially unbounded-length sum (since $i$ is arbitrary) is broken
into two parts. One part has a finite number of terms and each term is
upper-bounded by the dominant term. The other part has an
unbounded number of terms but that sum is bounded using a convergent
geometric series. This is done explicitly rather than relying on
asymptotic large-deviations theorems so that the resulting constants are
available to us to calculate plots for finite delays. The details are
in Appendix~\ref{app:erasuredetails}, but result in
\begin{eqnarray}
& & {\cal P}(\widehat{B}_i(\lceil \frac{i}{R'} \rceil + d) \neq B_i)
\nonumber \\
& \leq &
\exp(-d \frac{D(\lambda^* R' || 1 - \beta)}{1-\lambda^*} 
(\frac{D(R' + \frac{2}{\bar{n}} || 1 - \beta) - \epsilon_1}
      {D(R' || 1 - \beta)}
))
 \left[ \lceil \frac{1}{R'} \rceil 
 \sum_{l=0}^{\infty} 
 \exp(-l (D(R' + 2\bar{n}^{-1} || 1 - \beta) - \epsilon_1)) \right]
\nonumber \\ 
& & 
+ d(\frac{D(\lambda^*R' || 1 - \beta)}
         {(1-\lambda^*)D(R' || 1 - \beta)})
  \exp(-d \bigg((1-\epsilon_2) \frac{D(\lambda^*R' || 1 - \beta)}
                               {1 - \lambda^*} 
       - \frac{\epsilon_1}{1-\lambda^*} \bigg)) \label{eqn:erasureintermediate}
\end{eqnarray}
where $\lambda^*$ is coming from (\ref{eqn:optimumlambda}), $\bar{n} =
d \frac{D(\lambda^*R' || 1 - \beta)}{(1-\lambda^*)D(R' || 1 -
  \beta)}$, and $\epsilon_1, \epsilon_2$ are constants that can be
made arbitrarily small as $d$ gets large. The term in the brackets $[
\cdots ]$ is a convergent geometric series while $(\frac{D(R' +
  2\bar{n}^{-1} || 1 - \beta)}{D(R' || 1 - \beta)})$ approaches $1$ as
$d$ and hence $\bar{n}$ gets large.

Since $\lambda^*$ does not depend on $d$, just notice that $E_a(R) =
\frac{D(\lambda^*R' || 1 - \beta)}{1-\lambda^*}$ for the binary
erasure channel to get:
\begin{equation} \label{eqn:becachieve}
{\cal P}(\widehat{B}_i(\lceil \frac{i}{R'} \rceil + d) \neq B_i)
\leq 
(\gamma + \xi d)\exp(-d (1-\epsilon_3) E_a(R))
\end{equation}
for all $d > d_3(\epsilon_3, \beta, R)$ where $\epsilon_3 > 0$ is
arbitrary and $\gamma, \xi > 0$ are constants depending on
$\epsilon_3, \beta$ and $R$. Since the linear term is dominated by the
exponential, it is clear that the bound of
(\ref{eqn:symmetricfeedbackbound}) is asymptotically achievable for
the BEC with large delays at all rates $R$.  For non-binary erasure
channels, it is obvious that the same proof holds.

Furthermore, since the FIFO-based encoder does not need to know what
the target delay is, the code is clearly delay-universal or anytime in
nature. \hfill $\Box$ \vspace{0.20in}

\subsection{The transmission delay view}
An alternative view of the communication problem over the binary
erasure channel is useful when considering more general cases. Each
bit's delay can be viewed as the sum of a queuing delay (that can be
correlated across different bits) and its transmission delay $T_j$
which is a geometric $(1-\beta)$ random variable that is iid over
different bits $j$. The event $\{\frac{1}{d + \lceil \frac{i}{R'} \rceil
- \lceil \frac{k}{R'} \rceil} \sum_{t=\lceil \frac{k}{R'}
\rceil}^{\lceil \frac{i}{R'} \rceil + d} Z_t \leq \frac{i-k}{d +
\lceil \frac{i}{R'} \rceil - \lceil \frac{k}{R'} \rceil}\}$ from
(\ref{eqn:sumbound}) can alternatively be expressed in this language
as: $\{\sum_{j=k}^i T_j > d + \lceil \frac{i}{R'} \rceil - \lceil
\frac{k}{R'} \rceil\}$. This expresses the event that even if any
backlog before $k$ is ignored, the unlucky transmission delays alone
are too much for the bit $i$ to meet its deadline. (\ref{eqn:sumbound})
then becomes
\begin{equation} \label{eqn:sumbound2}
{\cal P}(\widehat{B}_i(\lceil \frac{i}{R'} \rceil + d) \neq B_i)
\leq 
\sum_{k=1}^i 
{\cal P}(\sum_{j=k}^i T_j > d + \lceil \frac{i}{R'} \rceil - \lceil \frac{k}{R'} \rceil).
\end{equation}
With this interpretation, Theorem~\ref{thm:erasurecase} about the
binary erasure channel implies the following result about large delays
in certain D/G/1 queues:
\begin{corollary} \label{cor:delaybasedbound} Consider a communication
  system in which point messages arrive deterministically at a steady
  rate of $R'$ messages per unit time, are FIFO queued up until ready
  to be served, and are then independently served using geometric
  $(1-\beta)$ service times $T_j$. Given any $\epsilon_4 > 0$, there
  exists a $d_4(\epsilon_4, \beta, R)$ so that for all $d \geq d_4$,
  the probability that point message $i$ has not completed service by
  time $d+\lceil \frac{i}{R'} \rceil$ is upper bounded by $\exp(-d
  (1-\epsilon_4) E_a^{bec}(R')\ln 2)$ where $E_a^{bec}$ from
  (\ref{eqn:becreliability}) is the fixed-delay error exponent for the
  binary erasure channel with erasure probability $\beta$ and rate
  $R'$ in bits per channel use. This fixed delay exponent is attained
  universally over all sufficiently long delays $d \geq d_3$.

Furthermore, this result continues to hold even if the independent
service times $T_j$ merely have complementary CDFs that are bounded
by: ${\cal P}(T_j > k) \leq \beta^{k}$. The service times do not need
to be identically distributed.

Finally, suppose the point message rate $R' = \frac{1}{m}$ where $m >
\widetilde{m}$ is a positive integer and the independent service times
$T_j$ satisfy ${\cal P}(T_j > \widetilde{m} + k) \leq \beta^{k}$. Then
the probability that point message $i$ (which arrived at time $im$)
has not completed service by time $d+ i m$ is upper bounded by
$\exp(-d (1-\epsilon_4) E_a^{bec}(R'') \ln 2)$ where $R'' = (m -
\widetilde{m})^{-1}$. This fixed delay exponent is also attained
universally over all sufficiently long delays $d \geq d_4 + m$.
\end{corollary}
{\em Proof: }In place of bits, there are messages. The geometric
random variables can be interpreted as the interarrival times for
the Bernoulli process of successful transmissions. The rate $R'$ bits
per channel use turns into $R' \ln 2$ nats per channel use. Finally,
the $\gamma + \xi d$ polynomial from (\ref{eqn:becachieve}) can be
absorbed into the exponential by just making $d_4$ and $\epsilon_4$ a
little bigger than the original $d_3$ and $\epsilon_3$. This
establishes the result for independent geometric service times.

For the case of general service times whose complementary CDF is
bounded by the geometric's complementary CDF, the reason is that the
errors all come from large deviations events of the form $\sum_{j=k}^i
T_j \geq l$. For each $j$, start with an independent continuous
uniform$[0,1]$ random variable $V_j$ and obtain both $T_j$ and $T'_j$
from $V_j$ through the inverse of their respective CDFs. This way,
each of the $T_j$ can be paired with a geometric $T'_j$ random
variable such that $\forall \omega, T_j(\omega) \leq T'_j(\omega)$
where $\omega$ represents an element from the sample space. Since
\begin{eqnarray*}
\{\omega | \sum_{j=k}^i T_j(\omega) \geq l \}
& = &
\bigcup_{\vec{l}| l_n \in {\cal N}, \sum_n l_n = l} 
\bigcap_{n} \{\omega | T_n(\omega) \geq l_n \} \\
& \subseteq &
\bigcup_{\vec{l}| l_n \in {\cal N}, \sum_n l_n = l} 
\bigcap_{n} \{\omega | T'_n(\omega) \geq l_n \} \\
& = &
\{\omega | \sum_{j=k}^i T'_j(\omega) \geq l \},
\end{eqnarray*}
it is clear that ${\cal P}(\sum_{j=k}^i T_j(\omega) \geq l) \leq
{\cal P}(\sum_{j=k}^i T'_j(\omega) \geq l)$ and so the
same error probability bounds can be achieved. 

Finally, consider the case of $R' =
\frac{1}{m}$ and independent service times bounded by those of a
constant plus geometrics. (\ref{eqn:sumbound2}) simplifies to
\begin{eqnarray} \label{eqn:sumbound3}
{\cal P}(\widehat{M}_i(im + d) \neq M_i)
& \leq &
\sum_{k=1}^i {\cal P}\left(\sum_{j=k}^i T_j > d + im -
km\right) \label{eqn:sumbound3} \\
& = & 
\sum_{k=1}^i {\cal P}\left(\sum_{j=k}^i T_j > d + (i-k)m \right) \nonumber \\
& = & 
\sum_{k=1}^i {\cal P}\left(\sum_{j=k}^i (T_j - \widetilde{m}) > d +
(i-k)(m-\widetilde{m}) \right) \nonumber \\
& = & 
\sum_{k=1}^i {\cal P}\left(\sum_{j=k}^i (T_j - \widetilde{m}) > d +
i(m-\widetilde{m}) -k(m-\widetilde{m}) \right). \label{eqn:sumbound4}
\end{eqnarray}
Notice that in (\ref{eqn:sumbound4}), the random variable $\tilde{T}_j
= T_j - \widetilde{m}$ has a complementary CDF bounded by a geometric
and corresponds to (\ref{eqn:sumbound2}) with a point-message rate of
$R'' = (m-\widetilde{m})^{-1}$. Thus, the error exponent with delay is
at least as good as $E_a^{bec}(R'') \ln 2$ for the point
messages. \hfill $\Box$ \vspace{0.20in}

\section{Achievability for general channels} \label{sec:fortifiedfeedback} 

The goal of this section is to prove Theorems \ref{thm:withzeroerror}
and \ref{thm:genericachieve}. Rather than starting with channels with
strictly positive zero-error capacity, it is conceptually easier to
start with generic DMCs but add a low-rate error-free side channel
that can be used to carry ``control'' information. This information is
interpreted as a kind of punctuation used to make the channel output
stream unambiguously understandable to the decoder. The idea is that
the rate of this error-free control channel is much lower than the
message rate that needs to be communicated. This allows the result to
extend immediately to channels with strictly positive zero-error
capacity. For general channels, the control channel is synthesized and
its own errors must be taken into account.

\subsection{The scheme for fortified systems with noiseless feedback} \label{sec:nclscheme}
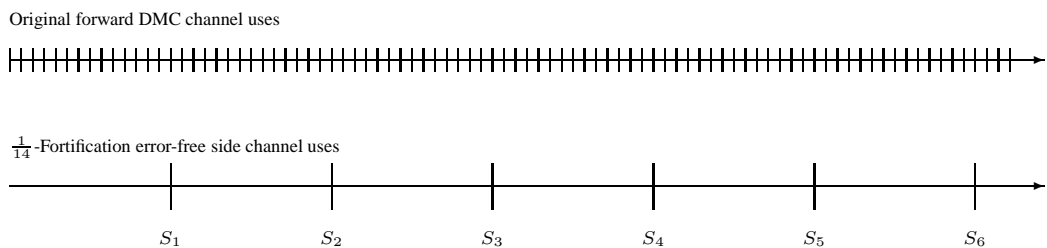
\begin{figure}[htbp]
\begin{center}
\setlength{\unitlength}{3800sp}%
\begingroup\makeatletter\ifx\SetFigFont\undefined%
\gdef\SetFigFont#1#2#3#4#5{%
  \reset@font\fontsize{#1}{#2pt}%
  \fontfamily{#3}\fontseries{#4}\fontshape{#5}%
  \selectfont}%
\fi\endgroup%
\begin{picture}(6774,1627)(-11,-1469)
\thinlines
{\color[rgb]{0,0,0}\put(  1,-136){\line( 0,-1){150}}
}%
{\color[rgb]{0,0,0}\put( 76,-136){\line( 0,-1){150}}
}%
{\color[rgb]{0,0,0}\put(151,-136){\line( 0,-1){150}}
}%
{\color[rgb]{0,0,0}\put(226,-136){\line( 0,-1){150}}
}%
{\color[rgb]{0,0,0}\put(301,-136){\line( 0,-1){150}}
}%
{\color[rgb]{0,0,0}\put(376,-136){\line( 0,-1){150}}
}%
{\color[rgb]{0,0,0}\put(451,-136){\line( 0,-1){150}}
}%
{\color[rgb]{0,0,0}\put(526,-136){\line( 0,-1){150}}
}%
{\color[rgb]{0,0,0}\put(601,-136){\line( 0,-1){150}}
}%
{\color[rgb]{0,0,0}\put(676,-136){\line( 0,-1){150}}
}%
{\color[rgb]{0,0,0}\put(751,-136){\line( 0,-1){150}}
}%
{\color[rgb]{0,0,0}\put(826,-136){\line( 0,-1){150}}
}%
{\color[rgb]{0,0,0}\put(901,-136){\line( 0,-1){150}}
}%
{\color[rgb]{0,0,0}\put(976,-136){\line( 0,-1){150}}
}%
{\color[rgb]{0,0,0}\put(1051,-136){\line( 0,-1){150}}
}%
{\color[rgb]{0,0,0}\put(1126,-136){\line( 0,-1){150}}
}%
{\color[rgb]{0,0,0}\put(1201,-136){\line( 0,-1){150}}
}%
{\color[rgb]{0,0,0}\put(1276,-136){\line( 0,-1){150}}
}%
{\color[rgb]{0,0,0}\put(1351,-136){\line( 0,-1){150}}
}%
{\color[rgb]{0,0,0}\put(1426,-136){\line( 0,-1){150}}
}%
{\color[rgb]{0,0,0}\put(1501,-136){\line( 0,-1){150}}
}%
{\color[rgb]{0,0,0}\put(1576,-136){\line( 0,-1){150}}
}%
{\color[rgb]{0,0,0}\put(1651,-136){\line( 0,-1){150}}
}%
{\color[rgb]{0,0,0}\put(1726,-136){\line( 0,-1){150}}
}%
{\color[rgb]{0,0,0}\put(1801,-136){\line( 0,-1){150}}
}%
{\color[rgb]{0,0,0}\put(1876,-136){\line( 0,-1){150}}
}%
{\color[rgb]{0,0,0}\put(1951,-136){\line( 0,-1){150}}
}%
{\color[rgb]{0,0,0}\put(2026,-136){\line( 0,-1){150}}
}%
{\color[rgb]{0,0,0}\put(2101,-136){\line( 0,-1){150}}
}%
{\color[rgb]{0,0,0}\put(2176,-136){\line( 0,-1){150}}
}%
{\color[rgb]{0,0,0}\put(2251,-136){\line( 0,-1){150}}
}%
{\color[rgb]{0,0,0}\put(2326,-136){\line( 0,-1){150}}
}%
{\color[rgb]{0,0,0}\put(2401,-136){\line( 0,-1){150}}
}%
{\color[rgb]{0,0,0}\put(2476,-136){\line( 0,-1){150}}
}%
{\color[rgb]{0,0,0}\put(2551,-136){\line( 0,-1){150}}
}%
{\color[rgb]{0,0,0}\put(2626,-136){\line( 0,-1){150}}
}%
{\color[rgb]{0,0,0}\put(2701,-136){\line( 0,-1){150}}
}%
{\color[rgb]{0,0,0}\put(2776,-136){\line( 0,-1){150}}
}%
{\color[rgb]{0,0,0}\put(2851,-136){\line( 0,-1){150}}
}%
{\color[rgb]{0,0,0}\put(2926,-136){\line( 0,-1){150}}
}%
{\color[rgb]{0,0,0}\put(3001,-136){\line( 0,-1){150}}
}%
{\color[rgb]{0,0,0}\put(3076,-136){\line( 0,-1){150}}
}%
{\color[rgb]{0,0,0}\put(3151,-136){\line( 0,-1){150}}
}%
{\color[rgb]{0,0,0}\put(3226,-136){\line( 0,-1){150}}
}%
{\color[rgb]{0,0,0}\put(3301,-136){\line( 0,-1){150}}
}%
{\color[rgb]{0,0,0}\put(3376,-136){\line( 0,-1){150}}
}%
{\color[rgb]{0,0,0}\put(3451,-136){\line( 0,-1){150}}
}%
{\color[rgb]{0,0,0}\put(3526,-136){\line( 0,-1){150}}
}%
{\color[rgb]{0,0,0}\put(3601,-136){\line( 0,-1){150}}
}%
{\color[rgb]{0,0,0}\put(3676,-136){\line( 0,-1){150}}
}%
{\color[rgb]{0,0,0}\put(3751,-136){\line( 0,-1){150}}
}%
{\color[rgb]{0,0,0}\put(3826,-136){\line( 0,-1){150}}
}%
{\color[rgb]{0,0,0}\put(3901,-136){\line( 0,-1){150}}
}%
{\color[rgb]{0,0,0}\put(3976,-136){\line( 0,-1){150}}
}%
{\color[rgb]{0,0,0}\put(4051,-136){\line( 0,-1){150}}
}%
{\color[rgb]{0,0,0}\put(4126,-136){\line( 0,-1){150}}
}%
{\color[rgb]{0,0,0}\put(4201,-136){\line( 0,-1){150}}
}%
{\color[rgb]{0,0,0}\put(4276,-136){\line( 0,-1){150}}
}%
{\color[rgb]{0,0,0}\put(4351,-136){\line( 0,-1){150}}
}%
{\color[rgb]{0,0,0}\put(4426,-136){\line( 0,-1){150}}
}%
{\color[rgb]{0,0,0}\put(4501,-136){\line( 0,-1){150}}
}%
{\color[rgb]{0,0,0}\put(4576,-136){\line( 0,-1){150}}
}%
{\color[rgb]{0,0,0}\put(4651,-136){\line( 0,-1){150}}
}%
{\color[rgb]{0,0,0}\put(4726,-136){\line( 0,-1){150}}
}%
{\color[rgb]{0,0,0}\put(4801,-136){\line( 0,-1){150}}
}%
{\color[rgb]{0,0,0}\put(4876,-136){\line( 0,-1){150}}
}%
{\color[rgb]{0,0,0}\put(4951,-136){\line( 0,-1){150}}
}%
{\color[rgb]{0,0,0}\put(5026,-136){\line( 0,-1){150}}
}%
{\color[rgb]{0,0,0}\put(5101,-136){\line( 0,-1){150}}
}%
{\color[rgb]{0,0,0}\put(5176,-136){\line( 0,-1){150}}
}%
{\color[rgb]{0,0,0}\put(5251,-136){\line( 0,-1){150}}
}%
{\color[rgb]{0,0,0}\put(5326,-136){\line( 0,-1){150}}
}%
{\color[rgb]{0,0,0}\put(5401,-136){\line( 0,-1){150}}
}%
{\color[rgb]{0,0,0}\put(5476,-136){\line( 0,-1){150}}
}%
{\color[rgb]{0,0,0}\put(5551,-136){\line( 0,-1){150}}
}%
{\color[rgb]{0,0,0}\put(5626,-136){\line( 0,-1){150}}
}%
{\color[rgb]{0,0,0}\put(5701,-136){\line( 0,-1){150}}
}%
{\color[rgb]{0,0,0}\put(5776,-136){\line( 0,-1){150}}
}%
{\color[rgb]{0,0,0}\put(5851,-136){\line( 0,-1){150}}
}%
{\color[rgb]{0,0,0}\put(5926,-136){\line( 0,-1){150}}
}%
{\color[rgb]{0,0,0}\put(6001,-136){\line( 0,-1){150}}
}%
{\color[rgb]{0,0,0}\put(6076,-136){\line( 0,-1){150}}
}%
{\color[rgb]{0,0,0}\put(6151,-136){\line( 0,-1){150}}
}%
{\color[rgb]{0,0,0}\put(6226,-136){\line( 0,-1){150}}
}%
{\color[rgb]{0,0,0}\put(6301,-136){\line( 0,-1){150}}
}%
{\color[rgb]{0,0,0}\put(6376,-136){\line( 0,-1){150}}
}%
{\color[rgb]{0,0,0}\put(6451,-136){\line( 0,-1){150}}
}%
{\color[rgb]{0,0,0}\put(6526,-136){\line( 0,-1){150}}
}%
{\color[rgb]{0,0,0}\put(  1,-1036){\vector( 1, 0){6750}}
}%
{\color[rgb]{0,0,0}\put(1051,-886){\line( 0,-1){300}}
}%
{\color[rgb]{0,0,0}\put(2101,-886){\line( 0,-1){300}}
}%
{\color[rgb]{0,0,0}\put(3151,-886){\line( 0,-1){300}}
}%
{\color[rgb]{0,0,0}\put(4201,-886){\line( 0,-1){300}}
}%
{\color[rgb]{0,0,0}\put(5251,-886){\line( 0,-1){300}}
}%
{\color[rgb]{0,0,0}\put(6301,-886){\line( 0,-1){300}}
}%
{\color[rgb]{0,0,0}\put(  1,-211){\vector( 1, 0){6750}}
}%
\put(  1, 14){\makebox(0,0)[lb]{\smash{\SetFigFont{7}{6}{\rmdefault}{\mddefault}{\updefault}{\color[rgb]{0,0,0}Original forward DMC channel uses}%
}}}
\put(1051,-1411){\makebox(0,0)[b]{\smash{\SetFigFont{7}{6}{\rmdefault}{\mddefault}{\updefault}{\color[rgb]{0,0,0}$S_1$}%
}}}
\put(2101,-1411){\makebox(0,0)[b]{\smash{\SetFigFont{7}{6}{\rmdefault}{\mddefault}{\updefault}{\color[rgb]{0,0,0}$S_2$}%
}}}
\put(3151,-1411){\makebox(0,0)[b]{\smash{\SetFigFont{7}{6}{\rmdefault}{\mddefault}{\updefault}{\color[rgb]{0,0,0}$S_3$}%
}}}
\put(4201,-1411){\makebox(0,0)[b]{\smash{\SetFigFont{7}{6}{\rmdefault}{\mddefault}{\updefault}{\color[rgb]{0,0,0}$S_4$}%
}}}
\put(5251,-1411){\makebox(0,0)[b]{\smash{\SetFigFont{7}{6}{\rmdefault}{\mddefault}{\updefault}{\color[rgb]{0,0,0}$S_5$}%
}}}
\put(6301,-1411){\makebox(0,0)[b]{\smash{\SetFigFont{7}{6}{\rmdefault}{\mddefault}{\updefault}{\color[rgb]{0,0,0}$S_6$}%
}}}
\put(
1,-811){\makebox(0,0)[lb]{\smash{\SetFigFont{7}{6}{\rmdefault}{\mddefault}{\updefault}{\color[rgb]{0,0,0}$\frac{1}{14}$-Fortification error-free side channel uses}%
}}}
\end{picture}
\end{center}
\caption{Fortification illustrated: the forward noisy channel uses are
supplemented with regular low-rate use of an error-free side channel.}
\label{fig:fortification}
\end{figure}

A ``fortified'' model is an idealization (depicted in
Figure~\ref{fig:fortification}) that makes an error-free
control channel explicit:

\begin{definition} \label{def:fortify}
Given a DMC $P$ for the forward link, a $\frac{1}{k}$-{\em fortified}
communication system built around it is one in which every $k$-th use
of $P$ is supplemented with the ability to transmit a single
error-free bit to the receiver.
\end{definition}
\vspace{0.1in} 

In comparison to the encoders with feedback from
Definition~\ref{def:nofeedbackencoder}, fortified encoders get to send
an additional error-free bit $S_{\frac{t}{k}}$ at times $t$ that are
integer multiples of $k$. The decoders are naturally modified to get
causal access to the error-free bits as well.

The idea is to generalize the repeat-until-received strategy used for
the erasure channel in Theorem~\ref{thm:erasurecase}.  A family of
schemes indexed by three parameters $(n,c,l)$ is described first, and
the asymptotic achievability of $E_{a,s}(R)$ is shown by taking an
appropriate limit over such schemes.

\begin{figure}[hbtp]
\begin{center}
\setlength{\unitlength}{3900sp}%
\begingroup\makeatletter\ifx\SetFigFont\undefined%
\gdef\SetFigFont#1#2#3#4#5{%
  \reset@font\fontsize{#1}{#2pt}%
  \fontfamily{#3}\fontseries{#4}\fontshape{#5}%
  \selectfont}%
\fi\endgroup%
\begin{picture}(7704,2310)(-41,-2161)
\thinlines
{\color[rgb]{0,0,0}\put( 38,-136){\line( 0,-1){150}}
}%
{\color[rgb]{0,0,0}\put(113,-136){\line( 0,-1){150}}
}%
{\color[rgb]{0,0,0}\put(188,-136){\line( 0,-1){150}}
}%
{\color[rgb]{0,0,0}\put(263,-136){\line( 0,-1){150}}
}%
{\color[rgb]{0,0,0}\put(338,-136){\line( 0,-1){150}}
}%
{\color[rgb]{0,0,0}\put(413,-136){\line( 0,-1){150}}
}%
{\color[rgb]{0,0,0}\put(488,-136){\line( 0,-1){150}}
}%
{\color[rgb]{0,0,0}\put(563,-136){\line( 0,-1){150}}
}%
{\color[rgb]{0,0,0}\put(638,-136){\line( 0,-1){150}}
}%
{\color[rgb]{0,0,0}\put(713,-136){\line( 0,-1){150}}
}%
{\color[rgb]{0,0,0}\put(788,-136){\line( 0,-1){150}}
}%
{\color[rgb]{0,0,0}\put(863,-136){\line( 0,-1){150}}
}%
{\color[rgb]{0,0,0}\put(938,-136){\line( 0,-1){150}}
}%
{\color[rgb]{0,0,0}\put(1013,-136){\line( 0,-1){150}}
}%
{\color[rgb]{0,0,0}\put(1088,-136){\line( 0,-1){150}}
}%
{\color[rgb]{0,0,0}\put(1163,-136){\line( 0,-1){150}}
}%
{\color[rgb]{0,0,0}\put(1238,-136){\line( 0,-1){150}}
}%
{\color[rgb]{0,0,0}\put(1313,-136){\line( 0,-1){150}}
}%
{\color[rgb]{0,0,0}\put(1388,-136){\line( 0,-1){150}}
}%
{\color[rgb]{0,0,0}\put(1463,-136){\line( 0,-1){150}}
}%
{\color[rgb]{0,0,0}\put(1538,-136){\line( 0,-1){150}}
}%
{\color[rgb]{0,0,0}\put(1613,-136){\line( 0,-1){150}}
}%
{\color[rgb]{0,0,0}\put(1688,-136){\line( 0,-1){150}}
}%
{\color[rgb]{0,0,0}\put(1763,-136){\line( 0,-1){150}}
}%
{\color[rgb]{0,0,0}\put(1838,-136){\line( 0,-1){150}}
}%
{\color[rgb]{0,0,0}\put(1913,-136){\line( 0,-1){150}}
}%
{\color[rgb]{0,0,0}\put(1988,-136){\line( 0,-1){150}}
}%
{\color[rgb]{0,0,0}\put(2063,-136){\line( 0,-1){150}}
}%
{\color[rgb]{0,0,0}\put(2138,-136){\line( 0,-1){150}}
}%
{\color[rgb]{0,0,0}\put(2213,-136){\line( 0,-1){150}}
}%
{\color[rgb]{0,0,0}\put(2288,-136){\line( 0,-1){150}}
}%
{\color[rgb]{0,0,0}\put(2363,-136){\line( 0,-1){150}}
}%
{\color[rgb]{0,0,0}\put(2438,-136){\line( 0,-1){150}}
}%
{\color[rgb]{0,0,0}\put(2513,-136){\line( 0,-1){150}}
}%
{\color[rgb]{0,0,0}\put(2588,-136){\line( 0,-1){150}}
}%
{\color[rgb]{0,0,0}\put(2663,-136){\line( 0,-1){150}}
}%
{\color[rgb]{0,0,0}\put(2738,-136){\line( 0,-1){150}}
}%
{\color[rgb]{0,0,0}\put(2813,-136){\line( 0,-1){150}}
}%
{\color[rgb]{0,0,0}\put(2888,-136){\line( 0,-1){150}}
}%
{\color[rgb]{0,0,0}\put(2963,-136){\line( 0,-1){150}}
}%
{\color[rgb]{0,0,0}\put(3038,-136){\line( 0,-1){150}}
}%
{\color[rgb]{0,0,0}\put(3113,-136){\line( 0,-1){150}}
}%
{\color[rgb]{0,0,0}\put(3188,-136){\line( 0,-1){150}}
}%
{\color[rgb]{0,0,0}\put(3263,-136){\line( 0,-1){150}}
}%
{\color[rgb]{0,0,0}\put(3338,-136){\line( 0,-1){150}}
}%
{\color[rgb]{0,0,0}\put(3413,-136){\line( 0,-1){150}}
}%
{\color[rgb]{0,0,0}\put(3488,-136){\line( 0,-1){150}}
}%
{\color[rgb]{0,0,0}\put(3563,-136){\line( 0,-1){150}}
}%
{\color[rgb]{0,0,0}\put(3638,-136){\line( 0,-1){150}}
}%
{\color[rgb]{0,0,0}\put(3713,-136){\line( 0,-1){150}}
}%
{\color[rgb]{0,0,0}\put(3788,-136){\line( 0,-1){150}}
}%
{\color[rgb]{0,0,0}\put(3863,-136){\line( 0,-1){150}}
}%
{\color[rgb]{0,0,0}\put(3938,-136){\line( 0,-1){150}}
}%
{\color[rgb]{0,0,0}\put(4013,-136){\line( 0,-1){150}}
}%
{\color[rgb]{0,0,0}\put(4088,-136){\line( 0,-1){150}}
}%
{\color[rgb]{0,0,0}\put(4163,-136){\line( 0,-1){150}}
}%
{\color[rgb]{0,0,0}\put(4238,-136){\line( 0,-1){150}}
}%
{\color[rgb]{0,0,0}\put(4313,-136){\line( 0,-1){150}}
}%
{\color[rgb]{0,0,0}\put(4388,-136){\line( 0,-1){150}}
}%
{\color[rgb]{0,0,0}\put(4463,-136){\line( 0,-1){150}}
}%
{\color[rgb]{0,0,0}\put(4538,-136){\line( 0,-1){150}}
}%
{\color[rgb]{0,0,0}\put(4613,-136){\line( 0,-1){150}}
}%
{\color[rgb]{0,0,0}\put(4688,-136){\line( 0,-1){150}}
}%
{\color[rgb]{0,0,0}\put(4763,-136){\line( 0,-1){150}}
}%
{\color[rgb]{0,0,0}\put(4838,-136){\line( 0,-1){150}}
}%
{\color[rgb]{0,0,0}\put(4913,-136){\line( 0,-1){150}}
}%
{\color[rgb]{0,0,0}\put(4988,-136){\line( 0,-1){150}}
}%
{\color[rgb]{0,0,0}\put(5063,-136){\line( 0,-1){150}}
}%
{\color[rgb]{0,0,0}\put(5138,-136){\line( 0,-1){150}}
}%
{\color[rgb]{0,0,0}\put(5213,-136){\line( 0,-1){150}}
}%
{\color[rgb]{0,0,0}\put(5288,-136){\line( 0,-1){150}}
}%
{\color[rgb]{0,0,0}\put(5363,-136){\line( 0,-1){150}}
}%
{\color[rgb]{0,0,0}\put(5438,-136){\line( 0,-1){150}}
}%
{\color[rgb]{0,0,0}\put(5513,-136){\line( 0,-1){150}}
}%
{\color[rgb]{0,0,0}\put(5588,-136){\line( 0,-1){150}}
}%
{\color[rgb]{0,0,0}\put(5663,-136){\line( 0,-1){150}}
}%
{\color[rgb]{0,0,0}\put(5738,-136){\line( 0,-1){150}}
}%
{\color[rgb]{0,0,0}\put(5813,-136){\line( 0,-1){150}}
}%
{\color[rgb]{0,0,0}\put(5888,-136){\line( 0,-1){150}}
}%
{\color[rgb]{0,0,0}\put(5963,-136){\line( 0,-1){150}}
}%
{\color[rgb]{0,0,0}\put(6038,-136){\line( 0,-1){150}}
}%
{\color[rgb]{0,0,0}\put(6113,-136){\line( 0,-1){150}}
}%
{\color[rgb]{0,0,0}\put(6188,-136){\line( 0,-1){150}}
}%
{\color[rgb]{0,0,0}\put(6263,-136){\line( 0,-1){150}}
}%
{\color[rgb]{0,0,0}\put(4201,-586){\line( 1, 0){2100}}
}%
{\color[rgb]{0,0,0}\put(4201,-511){\line( 0,-1){150}}
}%
{\color[rgb]{0,0,0}\put(6301,-511){\line( 0,-1){150}}
}%
\put(5251,-736){\makebox(0,0)[b]{\smash{\SetFigFont{7}{6}{\rmdefault}{\mddefault}{\updefault}{\color[rgb]{0,0,0}one chunk}%
}}}
{\color[rgb]{0,0,0}\put(526,-1786){\vector( 0, 1){525}}
}%
{\color[rgb]{0,0,0}\put(526,-1486){\line( 1, 0){525}}
\put(1051,-1486){\vector( 0, 1){225}}
}%
\put(526,-1936){\makebox(0,0)[lb]{\smash{\SetFigFont{7}{6}{\rmdefault}{\mddefault}{\updefault}{\color[rgb]{0,0,0}Prior block list disambiguation}%
}}}
{\color[rgb]{0,0,0}\put(  1,-136){\line( 0,-1){150}}
}%
{\color[rgb]{0,0,0}\put( 76,-136){\line( 0,-1){150}}
}%
{\color[rgb]{0,0,0}\put(151,-136){\line( 0,-1){150}}
}%
{\color[rgb]{0,0,0}\put(226,-136){\line( 0,-1){150}}
}%
{\color[rgb]{0,0,0}\put(301,-136){\line( 0,-1){150}}
}%
{\color[rgb]{0,0,0}\put(376,-136){\line( 0,-1){150}}
}%
{\color[rgb]{0,0,0}\put(451,-136){\line( 0,-1){150}}
}%
{\color[rgb]{0,0,0}\put(526,-136){\line( 0,-1){150}}
}%
{\color[rgb]{0,0,0}\put(601,-136){\line( 0,-1){150}}
}%
{\color[rgb]{0,0,0}\put(676,-136){\line( 0,-1){150}}
}%
{\color[rgb]{0,0,0}\put(751,-136){\line( 0,-1){150}}
}%
{\color[rgb]{0,0,0}\put(826,-136){\line( 0,-1){150}}
}%
{\color[rgb]{0,0,0}\put(901,-136){\line( 0,-1){150}}
}%
{\color[rgb]{0,0,0}\put(976,-136){\line( 0,-1){150}}
}%
{\color[rgb]{0,0,0}\put(1051,-136){\line( 0,-1){150}}
}%
{\color[rgb]{0,0,0}\put(1126,-136){\line( 0,-1){150}}
}%
{\color[rgb]{0,0,0}\put(1201,-136){\line( 0,-1){150}}
}%
{\color[rgb]{0,0,0}\put(1276,-136){\line( 0,-1){150}}
}%
{\color[rgb]{0,0,0}\put(1351,-136){\line( 0,-1){150}}
}%
{\color[rgb]{0,0,0}\put(1426,-136){\line( 0,-1){150}}
}%
{\color[rgb]{0,0,0}\put(1501,-136){\line( 0,-1){150}}
}%
{\color[rgb]{0,0,0}\put(1576,-136){\line( 0,-1){150}}
}%
{\color[rgb]{0,0,0}\put(1651,-136){\line( 0,-1){150}}
}%
{\color[rgb]{0,0,0}\put(1726,-136){\line( 0,-1){150}}
}%
{\color[rgb]{0,0,0}\put(1801,-136){\line( 0,-1){150}}
}%
{\color[rgb]{0,0,0}\put(1876,-136){\line( 0,-1){150}}
}%
{\color[rgb]{0,0,0}\put(1951,-136){\line( 0,-1){150}}
}%
{\color[rgb]{0,0,0}\put(2026,-136){\line( 0,-1){150}}
}%
{\color[rgb]{0,0,0}\put(2101,-136){\line( 0,-1){150}}
}%
{\color[rgb]{0,0,0}\put(2176,-136){\line( 0,-1){150}}
}%
{\color[rgb]{0,0,0}\put(2251,-136){\line( 0,-1){150}}
}%
{\color[rgb]{0,0,0}\put(2326,-136){\line( 0,-1){150}}
}%
{\color[rgb]{0,0,0}\put(2401,-136){\line( 0,-1){150}}
}%
{\color[rgb]{0,0,0}\put(2476,-136){\line( 0,-1){150}}
}%
{\color[rgb]{0,0,0}\put(2551,-136){\line( 0,-1){150}}
}%
{\color[rgb]{0,0,0}\put(2626,-136){\line( 0,-1){150}}
}%
{\color[rgb]{0,0,0}\put(2701,-136){\line( 0,-1){150}}
}%
{\color[rgb]{0,0,0}\put(2776,-136){\line( 0,-1){150}}
}%
{\color[rgb]{0,0,0}\put(2851,-136){\line( 0,-1){150}}
}%
{\color[rgb]{0,0,0}\put(2926,-136){\line( 0,-1){150}}
}%
{\color[rgb]{0,0,0}\put(3001,-136){\line( 0,-1){150}}
}%
{\color[rgb]{0,0,0}\put(3076,-136){\line( 0,-1){150}}
}%
{\color[rgb]{0,0,0}\put(3151,-136){\line( 0,-1){150}}
}%
{\color[rgb]{0,0,0}\put(3226,-136){\line( 0,-1){150}}
}%
{\color[rgb]{0,0,0}\put(3301,-136){\line( 0,-1){150}}
}%
{\color[rgb]{0,0,0}\put(3376,-136){\line( 0,-1){150}}
}%
{\color[rgb]{0,0,0}\put(3451,-136){\line( 0,-1){150}}
}%
{\color[rgb]{0,0,0}\put(3526,-136){\line( 0,-1){150}}
}%
{\color[rgb]{0,0,0}\put(3601,-136){\line( 0,-1){150}}
}%
{\color[rgb]{0,0,0}\put(3676,-136){\line( 0,-1){150}}
}%
{\color[rgb]{0,0,0}\put(3751,-136){\line( 0,-1){150}}
}%
{\color[rgb]{0,0,0}\put(3826,-136){\line( 0,-1){150}}
}%
{\color[rgb]{0,0,0}\put(3901,-136){\line( 0,-1){150}}
}%
{\color[rgb]{0,0,0}\put(3976,-136){\line( 0,-1){150}}
}%
{\color[rgb]{0,0,0}\put(4051,-136){\line( 0,-1){150}}
}%
{\color[rgb]{0,0,0}\put(4126,-136){\line( 0,-1){150}}
}%
{\color[rgb]{0,0,0}\put(4201,-136){\line( 0,-1){150}}
}%
{\color[rgb]{0,0,0}\put(4276,-136){\line( 0,-1){150}}
}%
{\color[rgb]{0,0,0}\put(4351,-136){\line( 0,-1){150}}
}%
{\color[rgb]{0,0,0}\put(4426,-136){\line( 0,-1){150}}
}%
{\color[rgb]{0,0,0}\put(4501,-136){\line( 0,-1){150}}
}%
{\color[rgb]{0,0,0}\put(4576,-136){\line( 0,-1){150}}
}%
{\color[rgb]{0,0,0}\put(4651,-136){\line( 0,-1){150}}
}%
{\color[rgb]{0,0,0}\put(4726,-136){\line( 0,-1){150}}
}%
{\color[rgb]{0,0,0}\put(4801,-136){\line( 0,-1){150}}
}%
{\color[rgb]{0,0,0}\put(4876,-136){\line( 0,-1){150}}
}%
{\color[rgb]{0,0,0}\put(4951,-136){\line( 0,-1){150}}
}%
{\color[rgb]{0,0,0}\put(5026,-136){\line( 0,-1){150}}
}%
{\color[rgb]{0,0,0}\put(5101,-136){\line( 0,-1){150}}
}%
{\color[rgb]{0,0,0}\put(5176,-136){\line( 0,-1){150}}
}%
{\color[rgb]{0,0,0}\put(5251,-136){\line( 0,-1){150}}
}%
{\color[rgb]{0,0,0}\put(5326,-136){\line( 0,-1){150}}
}%
{\color[rgb]{0,0,0}\put(5401,-136){\line( 0,-1){150}}
}%
{\color[rgb]{0,0,0}\put(5476,-136){\line( 0,-1){150}}
}%
{\color[rgb]{0,0,0}\put(5551,-136){\line( 0,-1){150}}
}%
{\color[rgb]{0,0,0}\put(5626,-136){\line( 0,-1){150}}
}%
{\color[rgb]{0,0,0}\put(5701,-136){\line( 0,-1){150}}
}%
{\color[rgb]{0,0,0}\put(5776,-136){\line( 0,-1){150}}
}%
{\color[rgb]{0,0,0}\put(5851,-136){\line( 0,-1){150}}
}%
{\color[rgb]{0,0,0}\put(5926,-136){\line( 0,-1){150}}
}%
{\color[rgb]{0,0,0}\put(6001,-136){\line( 0,-1){150}}
}%
{\color[rgb]{0,0,0}\put(6076,-136){\line( 0,-1){150}}
}%
{\color[rgb]{0,0,0}\put(6151,-136){\line( 0,-1){150}}
}%
{\color[rgb]{0,0,0}\put(6226,-136){\line( 0,-1){150}}
}%
{\color[rgb]{0,0,0}\put(  1,-211){\vector( 1, 0){7650}}
}%
{\color[rgb]{0,0,0}\put(  1,-1036){\vector( 1, 0){7650}}
}%
{\color[rgb]{0,0,0}\put(2101,-886){\line( 0,-1){300}}
}%
{\color[rgb]{0,0,0}\put(3151,-886){\line( 0,-1){300}}
}%
{\color[rgb]{0,0,0}\put(4201,-886){\line( 0,-1){300}}
}%
{\color[rgb]{0,0,0}\put(5251,-886){\line( 0,-1){300}}
}%
{\color[rgb]{0,0,0}\put(6301,-886){\line( 0,-1){300}}
}%
{\color[rgb]{0,0,0}\put(1576,-886){\line( 0,-1){300}}
}%
{\color[rgb]{0,0,0}\put(2626,-886){\line( 0,-1){300}}
}%
{\color[rgb]{0,0,0}\put(3676,-886){\line( 0,-1){300}}
}%
{\color[rgb]{0,0,0}\put(4726,-886){\line( 0,-1){300}}
}%
{\color[rgb]{0,0,0}\put(5776,-886){\line( 0,-1){300}}
}%
{\color[rgb]{0,0,0}\put(  1,-2011){\vector( 0, 1){750}}
}%
{\color[rgb]{0,0,0}\put(3676,-1786){\vector( 0, 1){525}}
}%
{\color[rgb]{0,0,0}\put(3676,-1486){\line(-1, 0){2100}}
\put(1576,-1486){\vector( 0, 1){225}}
}%
{\color[rgb]{0,0,0}\put(2626,-1486){\vector( 0, 1){225}}
}%
{\color[rgb]{0,0,0}\put(3151,-1486){\vector( 0, 1){225}}
}%
{\color[rgb]{0,0,0}\put(3676,-1486){\line( 1, 0){2100}}
\put(5776,-1486){\vector( 0, 1){225}}
}%
{\color[rgb]{0,0,0}\put(5251,-1486){\vector( 0, 1){225}}
}%
{\color[rgb]{0,0,0}\put(4726,-1486){\vector( 0, 1){225}}
}%
{\color[rgb]{0,0,0}\put(6826,-886){\line( 0,-1){300}}
}%
{\color[rgb]{0,0,0}\put(7351,-886){\line( 0,-1){300}}
}%
{\color[rgb]{0,0,0}\put(7351,-1711){\vector( 0, 1){525}}
}%
{\color[rgb]{0,0,0}\put(7351,-1486){\line(-1, 0){525}}
\put(6826,-1486){\vector( 0, 1){225}}
}%
{\color[rgb]{0,0,0}\put(6301,-136){\line( 0,-1){150}}
}%
{\color[rgb]{0,0,0}\multiput(7463,-136)(0.00000,-4.54545){34}{\makebox(1.6667,11.6667){\SetFigFont{5}{6}{\rmdefault}{\mddefault}{\updefault}.}}
}%
{\color[rgb]{0,0,0}\multiput(7426,-136)(0.00000,-4.54545){34}{\makebox(1.6667,11.6667){\SetFigFont{5}{6}{\rmdefault}{\mddefault}{\updefault}.}}
}%
{\color[rgb]{0,0,0}\multiput(7388,-136)(0.00000,-4.54545){34}{\makebox(1.6667,11.6667){\SetFigFont{5}{6}{\rmdefault}{\mddefault}{\updefault}.}}
}%
{\color[rgb]{0,0,0}\multiput(7351,-136)(0.00000,-4.54545){34}{\makebox(1.6667,11.6667){\SetFigFont{5}{6}{\rmdefault}{\mddefault}{\updefault}.}}
}%
{\color[rgb]{0,0,0}\multiput(7313,-136)(0.00000,-4.54545){34}{\makebox(1.6667,11.6667){\SetFigFont{5}{6}{\rmdefault}{\mddefault}{\updefault}.}}
}%
{\color[rgb]{0,0,0}\multiput(7276,-136)(0.00000,-4.54545){34}{\makebox(1.6667,11.6667){\SetFigFont{5}{6}{\rmdefault}{\mddefault}{\updefault}.}}
}%
{\color[rgb]{0,0,0}\multiput(7238,-136)(0.00000,-4.54545){34}{\makebox(1.6667,11.6667){\SetFigFont{5}{6}{\rmdefault}{\mddefault}{\updefault}.}}
}%
{\color[rgb]{0,0,0}\multiput(7201,-136)(0.00000,-4.54545){34}{\makebox(1.6667,11.6667){\SetFigFont{5}{6}{\rmdefault}{\mddefault}{\updefault}.}}
}%
{\color[rgb]{0,0,0}\multiput(7163,-136)(0.00000,-4.54545){34}{\makebox(1.6667,11.6667){\SetFigFont{5}{6}{\rmdefault}{\mddefault}{\updefault}.}}
}%
{\color[rgb]{0,0,0}\multiput(7126,-136)(0.00000,-4.54545){34}{\makebox(1.6667,11.6667){\SetFigFont{5}{6}{\rmdefault}{\mddefault}{\updefault}.}}
}%
{\color[rgb]{0,0,0}\multiput(7088,-136)(0.00000,-4.54545){34}{\makebox(1.6667,11.6667){\SetFigFont{5}{6}{\rmdefault}{\mddefault}{\updefault}.}}
}%
{\color[rgb]{0,0,0}\multiput(7051,-136)(0.00000,-4.54545){34}{\makebox(1.6667,11.6667){\SetFigFont{5}{6}{\rmdefault}{\mddefault}{\updefault}.}}
}%
{\color[rgb]{0,0,0}\multiput(7013,-136)(0.00000,-4.54545){34}{\makebox(1.6667,11.6667){\SetFigFont{5}{6}{\rmdefault}{\mddefault}{\updefault}.}}
}%
{\color[rgb]{0,0,0}\multiput(6976,-136)(0.00000,-4.54545){34}{\makebox(1.6667,11.6667){\SetFigFont{5}{6}{\rmdefault}{\mddefault}{\updefault}.}}
}%
{\color[rgb]{0,0,0}\multiput(6938,-136)(0.00000,-4.54545){34}{\makebox(1.6667,11.6667){\SetFigFont{5}{6}{\rmdefault}{\mddefault}{\updefault}.}}
}%
{\color[rgb]{0,0,0}\multiput(6901,-136)(0.00000,-4.54545){34}{\makebox(1.6667,11.6667){\SetFigFont{5}{6}{\rmdefault}{\mddefault}{\updefault}.}}
}%
{\color[rgb]{0,0,0}\multiput(6863,-136)(0.00000,-4.54545){34}{\makebox(1.6667,11.6667){\SetFigFont{5}{6}{\rmdefault}{\mddefault}{\updefault}.}}
}%
{\color[rgb]{0,0,0}\multiput(6826,-136)(0.00000,-4.54545){34}{\makebox(1.6667,11.6667){\SetFigFont{5}{6}{\rmdefault}{\mddefault}{\updefault}.}}
}%
{\color[rgb]{0,0,0}\multiput(6788,-136)(0.00000,-4.54545){34}{\makebox(1.6667,11.6667){\SetFigFont{5}{6}{\rmdefault}{\mddefault}{\updefault}.}}
}%
{\color[rgb]{0,0,0}\multiput(6751,-136)(0.00000,-4.54545){34}{\makebox(1.6667,11.6667){\SetFigFont{5}{6}{\rmdefault}{\mddefault}{\updefault}.}}
}%
{\color[rgb]{0,0,0}\multiput(6713,-136)(0.00000,-4.54545){34}{\makebox(1.6667,11.6667){\SetFigFont{5}{6}{\rmdefault}{\mddefault}{\updefault}.}}
}%
{\color[rgb]{0,0,0}\multiput(6676,-136)(0.00000,-4.54545){34}{\makebox(1.6667,11.6667){\SetFigFont{5}{6}{\rmdefault}{\mddefault}{\updefault}.}}
}%
{\color[rgb]{0,0,0}\multiput(6638,-136)(0.00000,-4.54545){34}{\makebox(1.6667,11.6667){\SetFigFont{5}{6}{\rmdefault}{\mddefault}{\updefault}.}}
}%
{\color[rgb]{0,0,0}\multiput(6601,-136)(0.00000,-4.54545){34}{\makebox(1.6667,11.6667){\SetFigFont{5}{6}{\rmdefault}{\mddefault}{\updefault}.}}
}%
{\color[rgb]{0,0,0}\multiput(6563,-136)(0.00000,-4.54545){34}{\makebox(1.6667,11.6667){\SetFigFont{5}{6}{\rmdefault}{\mddefault}{\updefault}.}}
}%
{\color[rgb]{0,0,0}\multiput(6526,-136)(0.00000,-4.54545){34}{\makebox(1.6667,11.6667){\SetFigFont{5}{6}{\rmdefault}{\mddefault}{\updefault}.}}
}%
{\color[rgb]{0,0,0}\multiput(6488,-136)(0.00000,-4.54545){34}{\makebox(1.6667,11.6667){\SetFigFont{5}{6}{\rmdefault}{\mddefault}{\updefault}.}}
}%
{\color[rgb]{0,0,0}\multiput(6451,-136)(0.00000,-4.54545){34}{\makebox(1.6667,11.6667){\SetFigFont{5}{6}{\rmdefault}{\mddefault}{\updefault}.}}
}%
{\color[rgb]{0,0,0}\multiput(6413,-136)(0.00000,-4.54545){34}{\makebox(1.6667,11.6667){\SetFigFont{5}{6}{\rmdefault}{\mddefault}{\updefault}.}}
}%
{\color[rgb]{0,0,0}\multiput(6376,-136)(0.00000,-4.54545){34}{\makebox(1.6667,11.6667){\SetFigFont{5}{6}{\rmdefault}{\mddefault}{\updefault}.}}
}%
{\color[rgb]{0,0,0}\multiput(6338,-136)(0.00000,-4.54545){34}{\makebox(1.6667,11.6667){\SetFigFont{5}{6}{\rmdefault}{\mddefault}{\updefault}.}}
}%
{\color[rgb]{0,0,0}\multiput(1051,-886)(0.00000,-4.47761){68}{\makebox(1.6667,11.6667){\SetFigFont{5}{6}{\rmdefault}{\mddefault}{\updefault}.}}
}%
{\color[rgb]{0,0,0}\multiput(526,-886)(0.00000,-4.47761){68}{\makebox(1.6667,11.6667){\SetFigFont{5}{6}{\rmdefault}{\mddefault}{\updefault}.}}
}%
{\color[rgb]{0,0,0}\multiput(  1,-886)(0.00000,-4.47761){68}{\makebox(1.6667,11.6667){\SetFigFont{5}{6}{\rmdefault}{\mddefault}{\updefault}.}}
}%
\put(  1, 14){\makebox(0,0)[lb]{\smash{\SetFigFont{7}{6}{\rmdefault}{\mddefault}{\updefault}{\color[rgb]{0,0,0}Forward DMC channel uses}%
}}}
\put(  1,-811){\makebox(0,0)[lb]{\smash{\SetFigFont{7}{6}{\rmdefault}{\mddefault}{\updefault}{\color[rgb]{0,0,0}Error-free side channel uses}%
}}}
\put(  1,-2161){\makebox(0,0)[lb]{\smash{\SetFigFont{7}{6}{\rmdefault}{\mddefault}{\updefault}{\color[rgb]{0,0,0}Previous block confirmation}%
}}}
\put(3676,-1936){\makebox(0,0)[lb]{\smash{\SetFigFont{7}{6}{\rmdefault}{\mddefault}{\updefault}{\color[rgb]{0,0,0}Unused control bits}%
}}}
\put(2101,-1336){\makebox(0,0)[b]{\smash{\SetFigFont{7}{6}{\rmdefault}{\mddefault}{\updefault}{\color[rgb]{0,0,0}deny}%
}}}
\put(4201,-1336){\makebox(0,0)[b]{\smash{\SetFigFont{7}{6}{\rmdefault}{\mddefault}{\updefault}{\color[rgb]{0,0,0}deny}%
}}}
\put(6301,-1336){\makebox(0,0)[b]{\smash{\SetFigFont{7}{6}{\rmdefault}{\mddefault}{\updefault}{\color[rgb]{0,0,0}confirm}%
}}}
\put(7351,-1936){\makebox(0,0)[rb]{\smash{\SetFigFont{7}{6}{\rmdefault}{\mddefault}{\updefault}{\color[rgb]{0,0,0}List disambiguation}%
}}}
\end{picture}
\end{center}
\caption{One block's transmission in the $(n,c,l)$ scheme for a
  $\frac{1}{14}$-fortified channel. In this case $l=2$ and $c=4$, so
  there are 4 error-free bits per chunk. $n$ is not visible at this
  level since the number of chunks needed for successful transmission
  is random. Typically fewer than $n$ chunks are needed so $n$ could
  be 5 in this example.} 
\label{fig:nclcode}
\end{figure}
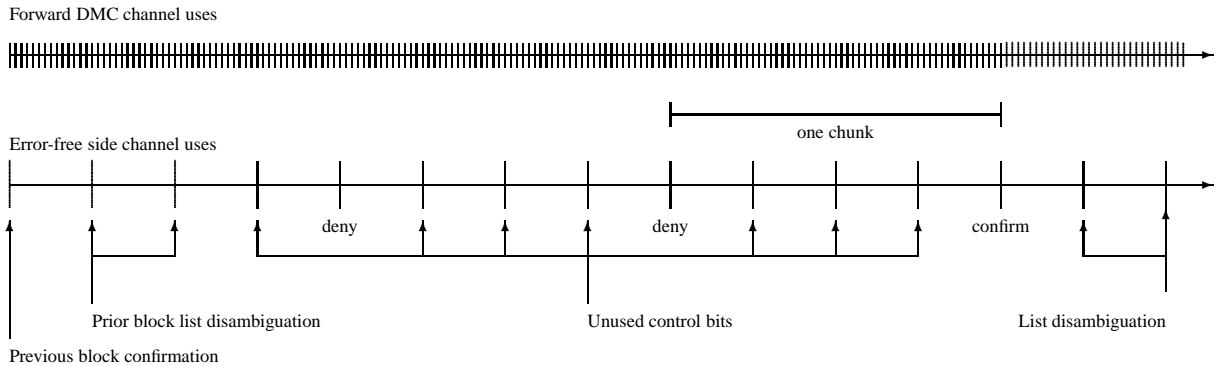

Call $c \geq 1$ the chunk length in terms of how many control bits are
associated with each chunk, $2^l$ the list length (with $l \leq c
-1$), and $n > l$ the message block length in units of chunks. The $(n,
c, l)$ randomized communication scheme (illustrated in
Figure~\ref{fig:nclcode}) is:  
\begin{enumerate} 
\item The encoder queues up incoming message bits and assembles them
  into message blocks of size $\frac{n c k R}{\ln 2}$ bits. One such
  message block arrives deterministically every $nck$ channel uses.

\item At every noisy channel use, the encoder sends the channel input
  corresponding to the next position in an infinite-length random
  codeword associated with the current message block.

  Formally, the codewords are $X_{i}(j,t)$ where $i > 0$ represents
  the current block number, $t > 0$ is the current channel-use time,
  and $0 \leq j < \exp(n c k R)$ is the value of the current message
  block. Each $X_{i}(j,t)$ is drawn iid from ${\cal X}$ using the
  $E_0(\eta)$ maximizing distribution $\vec{q}$. An $\eta$ is chosen
  such that the desired rate $R < \frac{E_0(\eta,\vec{q})}{\eta}$,
  while the target reliability is also $\alpha < E_0(\eta,\vec{q})$.

  If there is no message block to send, the encoder just idles by
  transmitting the next letter in the past message block.

\item If the time is an integer multiple of $ck$, the encoder uses the
  noiselessly fedback channel outputs to simulate the decoder's
  attempt to ML-decode the current codeword to within a list of size
  $2^l$.

  If the true codeword is one of the $2^l$ entries on the list, the
  encoder sends a $1$ ({\em confirm}) over the noiseless forward
  link. The encoder places into a control queue $l$ bits representing
  the true codeword's index within the decoder's list. The encoder
  then removes the current message block from the message queue.

  If the true block is not in the decoder's list, the encoder sends a
  $0$ ({\em deny}) over the error-free forward link.

  The $0$ can be viewed as a null punctuation mark while the $1$
  corresponds to a comma delimiting one variable-length block from
  another. When the list-disambiguation information is sent, it can be
  interpreted as a specific type of comma. There are thus just $l+1$
  different kinds of punctuation in the system.

\item If the time is an integer multiple of $k$ but not an integer
  multiple of $ck$, then the encoder looks in the control queue and
  transmits one of these bits over the error-free link, removing it
  from this second queue. If there are no control bits waiting, then
  the error-free link is ignored.

  Since $c > l$, all $l$ of the control bits will be communicated
  within one chunk.

\item At the decoder, the encoder's message queue length is known
  perfectly since it can only change by the deterministic arrival of
  message blocks or when an error-free confirm or deny bit has been
  sent over the noise-free link. Thus the decoder can correctly parse
  the received channel uses and always knows which message block a
  given channel output $Y_t$ or fortification symbol $S_t$ corresponds
  to.
  
\item If the time is an integer multiple of $ck$ and the decoder
  receives a $1$ noiselessly, then it decodes what it has seen to a
  list of the top $2^l$ possibilities for this message block. It uses
  the next $l$ error-free bits to disambiguate this list and commits
  to the result as its estimate for the message block.
\end{enumerate}

\subsection{Analysis of end-to-end delay and probability of error}

It is clear that this hybrid-ARQ scheme does not commit any errors at
the decoder. Some blocks just take longer to make it across than
others do. Furthermore, notice that the delay experienced by any
message {\em bit} can be divided into four parts:
\begin{enumerate} 
 \item {\em Assembly delay}: How long it takes before the rest of the
   message block has arrived at the encoder. This is bounded by a
   constant $nck$ channel uses.
 \item {\em Queuing delay}: How long the message block must wait
   before it begins to be transmitted. 
 \item {\em Transmission delay}: How many channel uses it takes before
   the codeword can be correctly decoded to within a list of
   $2^l$. This is a random quantity $T_j$ that must be an integer
   multiple of $ck$ channel uses. The $T_j$ are iid since the channel
   is memoryless and the random codebooks are also iid.
 \item {\em Termination delay}: How long the decoder must wait before
   the block is disambiguated by the error-free control signals. This
   is bounded by a constant $lk$ channel uses. 
\end{enumerate}

Since the assembly and termination delays are constants that do not
depend on the target end-to-end delay, they can be ignored and the
focus kept on the queuing and transmission delays. This is because our
interest is in the fixed-delay behavior for asymptotically large
delays much longer than $nck$. Since the transmission delays are iid,
the approach is to apply Corollary~\ref{cor:delaybasedbound} and this
requires a bound in terms of a constant plus a geometric. 
\begin{lemma} \label{lem:transmissiontimes}
The $(n,c,l)$ transmission scheme using input-distribution $\vec{q}$
at rate $R$ for a $\frac{1}{k}$-fortified communication system over a
DMC has iid transmission times $T_j$ satisfying
\begin{equation} \label{eqn:bettertransmissionbound}
{\cal P}(T_j - \lceil \widetilde{t}(\rho,R,n,\vec{q}) \rceil c k  > t ck) \leq 
[\exp(-ck E_0(\rho,\vec{q}) )
]^{t}
\end{equation}
for all $0 \leq \rho \leq 2^l$ and positive integer $t$ where
$\widetilde{t}(\rho,R,n,\vec{q}) = \frac{R}{\widetilde{C}(\rho,\vec{q})}n$ and
$\widetilde{C}(\rho,\vec{q}) = \frac{E_0(\rho,\vec{q})}{\rho}$.
\end{lemma}
{\em Proof: }See Appendix~\ref{app:lemtransmissiondelay}. 
\vspace{0.1in}

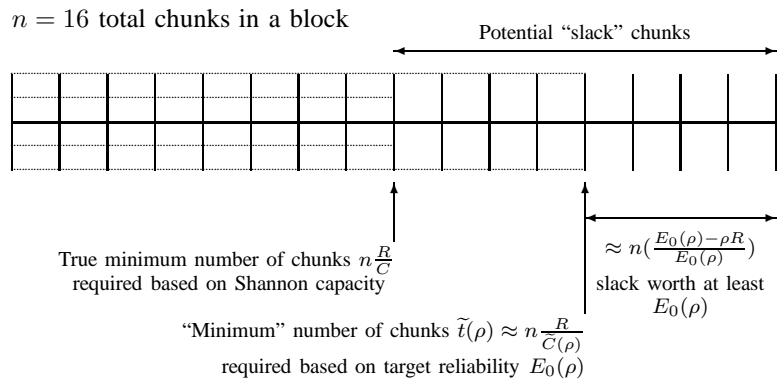
\begin{figure}[htbp]
\begin{center}
\setlength{\unitlength}{3947sp}%
\begingroup\makeatletter\ifx\SetFigFont\undefined%
\gdef\SetFigFont#1#2#3#4#5{%
  \reset@font\fontsize{#1}{#2pt}%
  \fontfamily{#3}\fontseries{#4}\fontshape{#5}%
  \selectfont}%
\fi\endgroup%
\begin{picture}(4824,2319)(1189,-1972)
\thinlines
{\color[rgb]{0,0,0}\put(1201,-361){\line( 1, 0){4800}}
}%
{\color[rgb]{0,0,0}\put(1201,-61){\line( 0,-1){600}}
}%
{\color[rgb]{0,0,0}\put(1801,-61){\line( 0,-1){600}}
}%
{\color[rgb]{0,0,0}\put(2401,-61){\line( 0,-1){600}}
}%
{\color[rgb]{0,0,0}\put(3001,-61){\line( 0,-1){600}}
}%
{\color[rgb]{0,0,0}\put(3601,-61){\line( 0,-1){600}}
}%
{\color[rgb]{0,0,0}\put(4201,-61){\line( 0,-1){600}}
}%
{\color[rgb]{0,0,0}\put(4801,-61){\line( 0,-1){600}}
}%
{\color[rgb]{0,0,0}\put(5401,-61){\line( 0,-1){600}}
}%
{\color[rgb]{0,0,0}\put(6001,-61){\line( 0,-1){600}}
}%
{\color[rgb]{0,0,0}\put(3601,-1111){\vector( 0, 1){375}}
}%
{\color[rgb]{0,0,0}\put(4801,-1561){\vector( 0, 1){825}}
}%
{\color[rgb]{0,0,0}\put(4801, 89){\vector(-1, 0){1200}}
}%
{\color[rgb]{0,0,0}\put(4801, 89){\vector( 1, 0){1200}}
}%
{\color[rgb]{0,0,0}\put(5401,-961){\vector(-1, 0){600}}
}%
{\color[rgb]{0,0,0}\put(5401,-961){\vector( 1, 0){600}}
}%
{\color[rgb]{0,0,0}\put(1501,-61){\line( 0,-1){600}}
}%
{\color[rgb]{0,0,0}\put(2101,-61){\line( 0,-1){600}}
}%
{\color[rgb]{0,0,0}\put(2701,-61){\line( 0,-1){600}}
}%
{\color[rgb]{0,0,0}\put(3301,-61){\line( 0,-1){600}}
}%
{\color[rgb]{0,0,0}\put(3901,-61){\line( 0,-1){600}}
}%
{\color[rgb]{0,0,0}\put(4501,-61){\line( 0,-1){600}}
}%
{\color[rgb]{0,0,0}\put(5101,-61){\line( 0,-1){600}}
}%
{\color[rgb]{0,0,0}\put(5701,-61){\line( 0,-1){600}}
}%
{\color[rgb]{0,0,0}\multiput(1201,-661)(18.00000,0.00000){201}{\makebox(1.6667,11.6667){\SetFigFont{5}{6}{\rmdefault}{\mddefault}{\updefault}.}}
}%
{\color[rgb]{0,0,0}\multiput(1201,-511)(18.04511,0.00000){134}{\makebox(1.6667,11.6667){\SetFigFont{5}{6}{\rmdefault}{\mddefault}{\updefault}.}}
}%
{\color[rgb]{0,0,0}\multiput(1201,-211)(18.04511,0.00000){134}{\makebox(1.6667,11.6667){\SetFigFont{5}{6}{\rmdefault}{\mddefault}{\updefault}.}}
}%
{\color[rgb]{0,0,0}\multiput(1201,-61)(18.00000,0.00000){201}{\makebox(1.6667,11.6667){\SetFigFont{5}{6}{\rmdefault}{\mddefault}{\updefault}.}}
}%
\put(4801,164){\makebox(0,0)[b]{\smash{\SetFigFont{8}{9.6}{\rmdefault}{\mddefault}{\updefault}{\color[rgb]{0,0,0}Potential ``slack'' chunks}%
}}}
\put(3601,-1411){\makebox(0,0)[rb]{\smash{\SetFigFont{8}{9.6}{\rmdefault}{\mddefault}{\updefault}{\color[rgb]{0,0,0}required based on Shannon capacity }%
}}}
\put(5401,-1186){\makebox(0,0)[b]{\smash{\SetFigFont{8}{9.6}{\rmdefault}{\mddefault}{\updefault}{\color[rgb]{0,0,0}$\approx n(\frac{E_0(\rho) -\rho R}{E_0(\rho)})$}%
}}}
\put(1351,-1261){\makebox(0,0)[b]{\smash{\SetFigFont{10}{12.0}{\rmdefault}{\mddefault}{\updefault}{\color[rgb]{0,0,0} }%
}}}
\put(3601,-1261){\makebox(0,0)[rb]{\smash{\SetFigFont{8}{9.6}{\rmdefault}{\mddefault}{\updefault}{\color[rgb]{0,0,0}True minimum number of chunks $n\frac{R}{C}$}%
}}}
\put(4801,-1711){\makebox(0,0)[rb]{\smash{\SetFigFont{8}{9.6}{\rmdefault}{\mddefault}{\updefault}{\color[rgb]{0,0,0}``Minimum'' number of chunks $\widetilde{t}(\rho) \approx n\frac{R}{\widetilde{C}(\rho)}$}%
}}}
\put(1201,239){\makebox(0,0)[lb]{\smash{\SetFigFont{10}{12.0}{\rmdefault}{\mddefault}{\updefault}{\color[rgb]{0,0,0}$n=16$ total chunks in a block}%
}}}
\put(5401,-1411){\makebox(0,0)[b]{\smash{\SetFigFont{8}{9.6}{\rmdefault}{\mddefault}{\updefault}{\color[rgb]{0,0,0}slack worth at least}%
}}}
\put(5401,-1561){\makebox(0,0)[b]{\smash{\SetFigFont{8}{9.6}{\rmdefault}{\mddefault}{\updefault}{\color[rgb]{0,0,0}$E_0(\rho)$}%
}}}
\put(4801,-1936){\makebox(0,0)[rb]{\smash{\SetFigFont{8}{9.6}{\rmdefault}{\mddefault}{\updefault}{\color[rgb]{0,0,0} required based on target reliability $E_0(\rho)$}%
}}}
\end{picture}
\end{center}
\caption{Because the message rate is less than capacity, there is some
  ``slack'' in the system. The amount of slack varies with the target
  reliability $E_0(\rho)$ and goes to zero when $R =
  \frac{E_0(\rho)}{\rho}$. The ``essential'' part of the block is
  denoted $\widetilde{t}(\rho)$ and its complement is the slack.}
\label{fig:blockslack}
\end{figure}

\begin{figure}[htbp]
\begin{center}
\includegraphics[width=4in,height=3in]{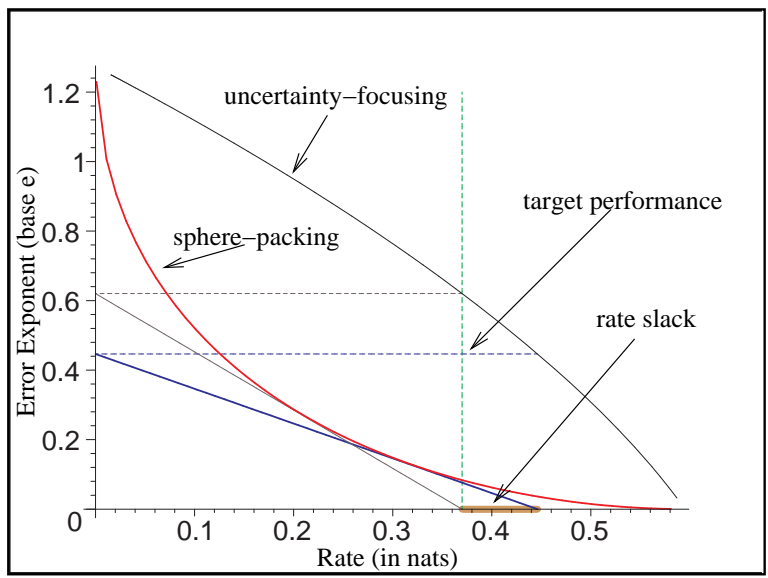}
\end{center}
\caption{Why at least a $0.44$ fixed-delay exponent is achievable at
  rate $0.37$ nats per channel use. The thick curve is the
  sphere-packing bound and the thin curve on top is the
  uncertainty-focusing bound. The thick tangent represents using a
  list size of $1$. The gap between $0.44$ and $0.37$ on the rate axis
  depicts the fraction of ``slack'' channel uses that are
  available. The thin tangent is the one used in the
  inverse-concatenation construction for the bound at rate $0.37$ nats
  per channel use.}
\label{fig:slackillustrated}
\end{figure}

\begin{figure}[hbtp]
\begin{center}
\setlength{\unitlength}{3700sp}%
\begingroup\makeatletter\ifx\SetFigFont\undefined%
\gdef\SetFigFont#1#2#3#4#5{%
  \reset@font\fontsize{#1}{#2pt}%
  \fontfamily{#3}\fontseries{#4}\fontshape{#5}%
  \selectfont}%
\fi\endgroup%
\begin{picture}(9204,7470)(289,-7111)
\thicklines
{\color[rgb]{0,0,0}\put(901,-2311){\line( 0,-1){450}}
}%
\thinlines
{\color[rgb]{0,0,0}\put(1051,-2611){\line( 0,-1){300}}
}%
{\color[rgb]{0,0,0}\put(1201,-2611){\line( 0,-1){300}}
}%
{\color[rgb]{0,0,0}\put(1351,-2611){\line( 0,-1){300}}
}%
{\color[rgb]{0,0,0}\put(1501,-2611){\line( 0,-1){300}}
}%
{\color[rgb]{0,0,0}\put(1651,-2611){\line( 0,-1){300}}
}%
{\color[rgb]{0,0,0}\put(1801,-2611){\line( 0,-1){300}}
}%
{\color[rgb]{0,0,0}\put(1951,-2611){\line( 0,-1){300}}
}%
{\color[rgb]{0,0,0}\put(2251,-2611){\line( 0,-1){300}}
}%
{\color[rgb]{0,0,0}\put(2401,-2611){\line( 0,-1){300}}
}%
{\color[rgb]{0,0,0}\put(2551,-2611){\line( 0,-1){300}}
}%
{\color[rgb]{0,0,0}\put(2701,-2611){\line( 0,-1){300}}
}%
{\color[rgb]{0,0,0}\put(2851,-2611){\line( 0,-1){300}}
}%
{\color[rgb]{0,0,0}\put(3001,-2611){\line( 0,-1){300}}
}%
{\color[rgb]{0,0,0}\put(3151,-2611){\line( 0,-1){300}}
}%
{\color[rgb]{0,0,0}\put(3451,-2611){\line( 0,-1){300}}
}%
{\color[rgb]{0,0,0}\put(3601,-2611){\line( 0,-1){300}}
}%
{\color[rgb]{0,0,0}\put(3751,-2611){\line( 0,-1){300}}
}%
{\color[rgb]{0,0,0}\put(3901,-2611){\line( 0,-1){300}}
}%
{\color[rgb]{0,0,0}\put(4051,-2611){\line( 0,-1){300}}
}%
{\color[rgb]{0,0,0}\put(4201,-2611){\line( 0,-1){300}}
}%
{\color[rgb]{0,0,0}\put(4351,-2611){\line( 0,-1){300}}
}%
{\color[rgb]{0,0,0}\put(4501,-2611){\line( 0,-1){300}}
}%
{\color[rgb]{0,0,0}\put(4651,-2611){\line( 0,-1){300}}
}%
{\color[rgb]{0,0,0}\put(4801,-2611){\line( 0,-1){300}}
}%
{\color[rgb]{0,0,0}\put(4951,-2611){\line( 0,-1){300}}
}%
{\color[rgb]{0,0,0}\put(5101,-2611){\line( 0,-1){300}}
}%
{\color[rgb]{0,0,0}\put(5251,-2611){\line( 0,-1){300}}
}%
{\color[rgb]{0,0,0}\put(5401,-2611){\line( 0,-1){300}}
}%
{\color[rgb]{0,0,0}\put(5551,-2611){\line( 0,-1){300}}
}%
{\color[rgb]{0,0,0}\put(5851,-2611){\line( 0,-1){300}}
}%
{\color[rgb]{0,0,0}\put(6001,-2611){\line( 0,-1){300}}
}%
{\color[rgb]{0,0,0}\put(6151,-2611){\line( 0,-1){300}}
}%
{\color[rgb]{0,0,0}\put(6301,-2611){\line( 0,-1){300}}
}%
{\color[rgb]{0,0,0}\put(6451,-2611){\line( 0,-1){300}}
}%
{\color[rgb]{0,0,0}\put(6601,-2611){\line( 0,-1){300}}
}%
{\color[rgb]{0,0,0}\put(6751,-2611){\line( 0,-1){300}}
}%
{\color[rgb]{0,0,0}\put(6901,-2611){\line( 0,-1){300}}
}%
{\color[rgb]{0,0,0}\put(7051,-2611){\line( 0,-1){300}}
}%
{\color[rgb]{0,0,0}\put(7201,-2611){\line( 0,-1){300}}
}%
{\color[rgb]{0,0,0}\put(7351,-2611){\line( 0,-1){300}}
}%
{\color[rgb]{0,0,0}\put(7501,-2611){\line( 0,-1){300}}
}%
{\color[rgb]{0,0,0}\put(7651,-2611){\line( 0,-1){300}}
}%
{\color[rgb]{0,0,0}\put(7801,-2611){\line( 0,-1){300}}
}%
{\color[rgb]{0,0,0}\put(7951,-2611){\line( 0,-1){300}}
}%
{\color[rgb]{0,0,0}\put(2101,-2611){\line( 0,-1){300}}
}%
\thicklines
{\color[rgb]{0,0,0}\put(3301,-2311){\line( 0,-1){450}}
}%
{\color[rgb]{0,0,0}\put(5701,-2311){\line( 0,-1){450}}
}%
{\color[rgb]{0,0,0}\put(8101,-2311){\line( 0,-1){300}}
}%
\thinlines
{\color[rgb]{0,0,0}\put(8251,-2611){\line( 0,-1){300}}
}%
{\color[rgb]{0,0,0}\put(8401,-2611){\line( 0,-1){300}}
}%
{\color[rgb]{0,0,0}\put(8701,-2611){\line( 0,-1){300}}
}%
{\color[rgb]{0,0,0}\put(8851,-2611){\line( 0,-1){300}}
}%
{\color[rgb]{0,0,0}\put(9001,-2611){\line( 0,-1){300}}
}%
{\color[rgb]{0,0,0}\put(9301,-2611){\line( 0,-1){300}}
}%
{\color[rgb]{0,0,0}\put(8551,-2611){\line( 0,-1){300}}
}%
{\color[rgb]{0,0,0}\put(9151,-2611){\line( 0,-1){300}}
}%
{\color[rgb]{0,0,0}\put(9451,-2761){\vector(-1, 0){9150}}
}%
{\color[rgb]{0,0,0}\put(751,-2611){\line( 0,-1){300}}
}%
{\color[rgb]{0,0,0}\put(601,-2611){\line( 0,-1){300}}
}%
\thicklines
{\color[rgb]{0,0,0}\put(901,-361){\line( 0,-1){450}}
}%
\thinlines
{\color[rgb]{0,0,0}\put(1051,-661){\line( 0,-1){300}}
}%
{\color[rgb]{0,0,0}\put(1201,-661){\line( 0,-1){300}}
}%
{\color[rgb]{0,0,0}\put(1351,-661){\line( 0,-1){300}}
}%
{\color[rgb]{0,0,0}\put(1501,-661){\line( 0,-1){300}}
}%
{\color[rgb]{0,0,0}\put(1651,-661){\line( 0,-1){300}}
}%
{\color[rgb]{0,0,0}\put(1801,-661){\line( 0,-1){300}}
}%
{\color[rgb]{0,0,0}\put(1951,-661){\line( 0,-1){300}}
}%
{\color[rgb]{0,0,0}\put(2251,-661){\line( 0,-1){300}}
}%
{\color[rgb]{0,0,0}\put(2401,-661){\line( 0,-1){300}}
}%
{\color[rgb]{0,0,0}\put(2551,-661){\line( 0,-1){300}}
}%
{\color[rgb]{0,0,0}\put(2701,-661){\line( 0,-1){300}}
}%
{\color[rgb]{0,0,0}\put(2851,-661){\line( 0,-1){300}}
}%
{\color[rgb]{0,0,0}\put(3001,-661){\line( 0,-1){300}}
}%
{\color[rgb]{0,0,0}\put(3151,-661){\line( 0,-1){300}}
}%
{\color[rgb]{0,0,0}\put(3451,-661){\line( 0,-1){300}}
}%
{\color[rgb]{0,0,0}\put(3601,-661){\line( 0,-1){300}}
}%
{\color[rgb]{0,0,0}\put(3751,-661){\line( 0,-1){300}}
}%
{\color[rgb]{0,0,0}\put(3901,-661){\line( 0,-1){300}}
}%
{\color[rgb]{0,0,0}\put(4051,-661){\line( 0,-1){300}}
}%
{\color[rgb]{0,0,0}\put(4201,-661){\line( 0,-1){300}}
}%
{\color[rgb]{0,0,0}\put(4351,-661){\line( 0,-1){300}}
}%
{\color[rgb]{0,0,0}\put(4501,-661){\line( 0,-1){300}}
}%
{\color[rgb]{0,0,0}\put(4651,-661){\line( 0,-1){300}}
}%
{\color[rgb]{0,0,0}\put(4801,-661){\line( 0,-1){300}}
}%
{\color[rgb]{0,0,0}\put(4951,-661){\line( 0,-1){300}}
}%
{\color[rgb]{0,0,0}\put(5101,-661){\line( 0,-1){300}}
}%
{\color[rgb]{0,0,0}\put(5251,-661){\line( 0,-1){300}}
}%
{\color[rgb]{0,0,0}\put(5401,-661){\line( 0,-1){300}}
}%
{\color[rgb]{0,0,0}\put(5551,-661){\line( 0,-1){300}}
}%
{\color[rgb]{0,0,0}\put(5851,-661){\line( 0,-1){300}}
}%
{\color[rgb]{0,0,0}\put(6001,-661){\line( 0,-1){300}}
}%
{\color[rgb]{0,0,0}\put(6151,-661){\line( 0,-1){300}}
}%
{\color[rgb]{0,0,0}\put(6301,-661){\line( 0,-1){300}}
}%
{\color[rgb]{0,0,0}\put(6451,-661){\line( 0,-1){300}}
}%
{\color[rgb]{0,0,0}\put(6601,-661){\line( 0,-1){300}}
}%
{\color[rgb]{0,0,0}\put(6751,-661){\line( 0,-1){300}}
}%
{\color[rgb]{0,0,0}\put(6901,-661){\line( 0,-1){300}}
}%
{\color[rgb]{0,0,0}\put(7051,-661){\line( 0,-1){300}}
}%
{\color[rgb]{0,0,0}\put(7201,-661){\line( 0,-1){300}}
}%
{\color[rgb]{0,0,0}\put(7351,-661){\line( 0,-1){300}}
}%
{\color[rgb]{0,0,0}\put(7501,-661){\line( 0,-1){300}}
}%
{\color[rgb]{0,0,0}\put(7651,-661){\line( 0,-1){300}}
}%
{\color[rgb]{0,0,0}\put(7801,-661){\line( 0,-1){300}}
}%
{\color[rgb]{0,0,0}\put(7951,-661){\line( 0,-1){300}}
}%
{\color[rgb]{0,0,0}\put(2101,-661){\line( 0,-1){300}}
}%
\thicklines
{\color[rgb]{0,0,0}\put(3301,-361){\line( 0,-1){450}}
}%
{\color[rgb]{0,0,0}\put(5701,-361){\line( 0,-1){450}}
}%
{\color[rgb]{0,0,0}\put(8101,-361){\line( 0,-1){450}}
}%
\thinlines
{\color[rgb]{0,0,0}\put(8251,-661){\line( 0,-1){300}}
}%
{\color[rgb]{0,0,0}\put(8401,-661){\line( 0,-1){300}}
}%
{\color[rgb]{0,0,0}\put(8701,-661){\line( 0,-1){300}}
}%
{\color[rgb]{0,0,0}\put(8851,-661){\line( 0,-1){300}}
}%
{\color[rgb]{0,0,0}\put(9001,-661){\line( 0,-1){300}}
}%
{\color[rgb]{0,0,0}\put(9301,-661){\line( 0,-1){300}}
}%
{\color[rgb]{0,0,0}\put(8551,-661){\line( 0,-1){300}}
}%
{\color[rgb]{0,0,0}\put(9151,-661){\line( 0,-1){300}}
}%
{\color[rgb]{0,0,0}\put(9301,-811){\vector(-1, 0){9000}}
}%
{\color[rgb]{0,0,0}\put(751,-661){\line( 0,-1){300}}
}%
{\color[rgb]{0,0,0}\put(601,-661){\line( 0,-1){300}}
}%
{\color[rgb]{0,0,0}\put(826,-361){\makebox(1.6667,11.6667){\SetFigFont{5}{6}{\rmdefault}{\mddefault}{\updefault}.}}
}%
{\color[rgb]{0,0,0}\put(3301,-61){\vector( 0,-1){250}}
}%
{\color[rgb]{0,0,0}\put(3601,-61){\line(-1, 0){2700}}
\put(901,-61){\vector( 0,-1){250}}
}%
{\color[rgb]{0,0,0}\put(5701,-61){\line( 1, 0){2400}}
\put(8101,-61){\vector( 0,-1){250}}
}%
{\color[rgb]{0,0,0}\put(5701,-1561){\vector( 1, 0){3600}}
}%
{\color[rgb]{0,0,0}\put(901,-661){\line( 0,-1){300}}
}%
{\color[rgb]{0,0,0}\put(3301,-661){\line( 0,-1){300}}
}%
{\color[rgb]{0,0,0}\put(5701,-661){\line( 0,-1){300}}
}%
{\color[rgb]{0,0,0}\put(8101,-661){\line( 0,-1){300}}
}%
\thicklines
{\color[rgb]{0,0,0}\put(601,-2761){\line( 0,-1){300}}
}%
\thinlines
{\color[rgb]{0,0,0}\put(901,-2611){\line( 0,-1){300}}
}%
{\color[rgb]{0,0,0}\put(3301,-2611){\line( 0,-1){300}}
}%
{\color[rgb]{0,0,0}\put(5701,-2611){\line( 0,-1){300}}
}%
{\color[rgb]{0,0,0}\put(8101,-2611){\line( 0,-1){300}}
}%
{\color[rgb]{0,0,0}\put(5251,-3811){\line( 0, 1){300}}
\put(5251,-3511){\line(-1, 0){4650}}
\put(601,-3511){\vector( 0, 1){375}}
}%
\thicklines
{\color[rgb]{0,0,0}\put(3601,-2761){\line( 0,-1){450}}
}%
{\color[rgb]{0,0,0}\put(6151,-2761){\line( 0,-1){450}}
}%
{\color[rgb]{0,0,0}\put(9451,-2761){\line( 0,-1){450}}
}%
\thinlines
{\color[rgb]{0,0,0}\put(5251,-3511){\line( 1, 0){4200}}
\put(9451,-3511){\vector( 0, 1){250}}
}%
{\color[rgb]{0,0,0}\put(6151,-3511){\vector( 0, 1){250}}
}%
{\color[rgb]{0,0,0}\put(3601,-3511){\vector( 0, 1){250}}
}%
{\color[rgb]{0,0,0}\put(3601,164){\line( 0,-1){225}}
\put(3601,-61){\line( 1, 0){2100}}
\put(5701,-61){\vector( 0,-1){250}}
}%
{\color[rgb]{0,0,0}\multiput(901,-4111)(0.00000,9.00000){126}{\makebox(1.6667,11.6667){\SetFigFont{5}{6}{\rmdefault}{\mddefault}{\updefault}.}}
}%
{\color[rgb]{0,0,0}\put(5701,-1861){\vector( 1, 0){1775}}
}%
{\color[rgb]{0,0,0}\put(7501,-1861){\vector( 1, 0){1800}}
}%
{\color[rgb]{0,0,0}\put(7501,-5086){\vector( 1, 0){1800}}
}%
\thicklines
{\color[rgb]{0,0,0}\put(7501,-5386){\line( 0,-1){450}}
}%
{\color[rgb]{0,0,0}\put(6901,-5386){\line( 0,-1){450}}
}%
{\color[rgb]{0,0,0}\put(6301,-5386){\line( 0,-1){450}}
}%
\thinlines
{\color[rgb]{0,0,0}\put(7051,-5686){\line( 0,-1){300}}
}%
{\color[rgb]{0,0,0}\put(6901,-5686){\line( 0,-1){300}}
}%
{\color[rgb]{0,0,0}\put(6751,-5686){\line( 0,-1){300}}
}%
{\color[rgb]{0,0,0}\put(6601,-5686){\line( 0,-1){300}}
}%
{\color[rgb]{0,0,0}\put(6451,-5686){\line( 0,-1){300}}
}%
{\color[rgb]{0,0,0}\put(6301,-5686){\line( 0,-1){300}}
}%
{\color[rgb]{0,0,0}\put(6151,-5686){\line( 0,-1){300}}
}%
{\color[rgb]{0,0,0}\put(8101,-5686){\line( 0,-1){300}}
}%
{\color[rgb]{0,0,0}\put(7951,-5686){\line( 0,-1){300}}
}%
{\color[rgb]{0,0,0}\put(7801,-5686){\line( 0,-1){300}}
}%
{\color[rgb]{0,0,0}\put(7651,-5686){\line( 0,-1){300}}
}%
{\color[rgb]{0,0,0}\put(7501,-5686){\line( 0,-1){300}}
}%
{\color[rgb]{0,0,0}\put(7351,-5686){\line( 0,-1){300}}
}%
{\color[rgb]{0,0,0}\put(7201,-5686){\line( 0,-1){300}}
}%
{\color[rgb]{0,0,0}\put(9151,-5686){\line( 0,-1){300}}
}%
{\color[rgb]{0,0,0}\put(9001,-5686){\line( 0,-1){300}}
}%
{\color[rgb]{0,0,0}\put(8851,-5686){\line( 0,-1){300}}
}%
{\color[rgb]{0,0,0}\put(8701,-5686){\line( 0,-1){300}}
}%
{\color[rgb]{0,0,0}\put(8551,-5686){\line( 0,-1){300}}
}%
{\color[rgb]{0,0,0}\put(8401,-5686){\line( 0,-1){300}}
}%
{\color[rgb]{0,0,0}\put(8251,-5686){\line( 0,-1){300}}
}%
{\color[rgb]{0,0,0}\put(6001,-5686){\line( 0,-1){300}}
}%
{\color[rgb]{0,0,0}\put(5851,-5686){\line( 0,-1){300}}
}%
{\color[rgb]{0,0,0}\put(5701,-5686){\line( 0,-1){300}}
}%
{\color[rgb]{0,0,0}\put(5551,-5686){\line( 0,-1){300}}
}%
{\color[rgb]{0,0,0}\put(5401,-5686){\line( 0,-1){300}}
}%
{\color[rgb]{0,0,0}\put(9451,-5836){\vector(-1, 0){4350}}
}%
\thicklines
{\color[rgb]{0,0,0}\put(5701,-5536){\line( 0,-1){300}}
}%
{\color[rgb]{0,0,0}\put(6001,-5836){\line( 0,-1){300}}
}%
{\color[rgb]{0,0,0}\put(7201,-5836){\line( 0,-1){450}}
}%
{\color[rgb]{0,0,0}\put(8101,-5386){\line( 0,-1){300}}
}%
{\color[rgb]{0,0,0}\put(8701,-5386){\line( 0,-1){300}}
}%
{\color[rgb]{0,0,0}\put(7951,-5836){\line( 0,-1){450}}
}%
\thinlines
{\color[rgb]{0,0,0}\put(9301,-5686){\line( 0,-1){300}}
}%
\thicklines
{\color[rgb]{0,0,0}\put(9451,-5836){\line( 0,-1){450}}
}%
\thinlines
{\color[rgb]{0,0,0}\multiput(6301,-6961)(0.00000,9.00000){126}{\makebox(1.6667,11.6667){\SetFigFont{5}{6}{\rmdefault}{\mddefault}{\updefault}.}}
}%
\thicklines
{\color[rgb]{0,0,0}\put(3376,-4861){\line( 0,-1){1275}}
\put(3376,-6136){\vector( 1, 0){1725}}
}%
\thinlines
{\color[rgb]{0,0,0}\put(901,-3061){\vector( 1, 0){2675}}
}%
{\color[rgb]{0,0,0}\put(3601,-3061){\vector( 1, 0){2525}}
}%
{\color[rgb]{0,0,0}\put(6151,-3061){\vector( 1, 0){3275}}
}%
{\color[rgb]{0,0,0}\put(6301,-6136){\vector( 1, 0){875}}
}%
{\color[rgb]{0,0,0}\put(7201,-6136){\vector( 1, 0){725}}
}%
{\color[rgb]{0,0,0}\put(7951,-6136){\vector( 1, 0){1475}}
}%
{\color[rgb]{0,0,0}\multiput(2701,-2911)(-18.00000,0.00000){101}{\makebox(1.6667,11.6667){\SetFigFont{5}{6}{\rmdefault}{\mddefault}{\updefault}.}}
}%
{\color[rgb]{0,0,0}\multiput(901,-2611)(18.00000,0.00000){101}{\makebox(1.6667,11.6667){\SetFigFont{5}{6}{\rmdefault}{\mddefault}{\updefault}.}}
}%
{\color[rgb]{0,0,0}\multiput(3301,-2611)(18.00000,0.00000){101}{\makebox(1.6667,11.6667){\SetFigFont{5}{6}{\rmdefault}{\mddefault}{\updefault}.}}
}%
{\color[rgb]{0,0,0}\multiput(3601,-2911)(18.00000,0.00000){101}{\makebox(1.6667,11.6667){\SetFigFont{5}{6}{\rmdefault}{\mddefault}{\updefault}.}}
}%
{\color[rgb]{0,0,0}\multiput(5701,-2611)(18.00000,0.00000){101}{\makebox(1.6667,11.6667){\SetFigFont{5}{6}{\rmdefault}{\mddefault}{\updefault}.}}
}%
{\color[rgb]{0,0,0}\multiput(6151,-2911)(18.00000,0.00000){101}{\makebox(1.6667,11.6667){\SetFigFont{5}{6}{\rmdefault}{\mddefault}{\updefault}.}}
}%
\put(7501,-1411){\makebox(0,0)[b]{\smash{\SetFigFont{10}{12.0}{\rmdefault}{\mddefault}{\updefault}{\color[rgb]{0,0,0}Target delay $d$ chunks after block arrival}%
}}}
\put(5701,-2236){\makebox(0,0)[b]{\smash{\SetFigFont{8}{9.6}{\rmdefault}{\mddefault}{\updefault}{\color[rgb]{0,0,0}$i$}%
}}}
\put(8101,-2236){\makebox(0,0)[b]{\smash{\SetFigFont{8}{9.6}{\rmdefault}{\mddefault}{\updefault}{\color[rgb]{0,0,0}$i+1$}%
}}}
\put(3301,-2236){\makebox(0,0)[b]{\smash{\SetFigFont{8}{9.6}{\rmdefault}{\mddefault}{\updefault}{\color[rgb]{0,0,0}$i-1$}%
}}}
\put(5701,-1186){\makebox(0,0)[b]{\smash{\SetFigFont{8}{9.6}{\rmdefault}{\mddefault}{\updefault}{\color[rgb]{0,0,0}$i$}%
}}}
\put(8101,-1186){\makebox(0,0)[b]{\smash{\SetFigFont{8}{9.6}{\rmdefault}{\mddefault}{\updefault}{\color[rgb]{0,0,0}$i+1$}%
}}}
\put(3301,-1186){\makebox(0,0)[b]{\smash{\SetFigFont{8}{9.6}{\rmdefault}{\mddefault}{\updefault}{\color[rgb]{0,0,0}$i-1$}%
}}}
\put(901,-1186){\makebox(0,0)[b]{\smash{\SetFigFont{8}{9.6}{\rmdefault}{\mddefault}{\updefault}{\color[rgb]{0,0,0}$i-2$}%
}}}
\put(3601,-3661){\makebox(0,0)[b]{\smash{\SetFigFont{8}{9.6}{\rmdefault}{\mddefault}{\updefault}{\color[rgb]{0,0,0}$i-2$}%
}}}
\put(901,-2236){\makebox(0,0)[b]{\smash{\SetFigFont{8}{9.6}{\rmdefault}{\mddefault}{\updefault}{\color[rgb]{0,0,0}$i-2$}%
}}}
\put(9451,-3661){\makebox(0,0)[b]{\smash{\SetFigFont{8}{9.6}{\rmdefault}{\mddefault}{\updefault}{\color[rgb]{0,0,0}$i$}%
}}}
\put(6151,-3661){\makebox(0,0)[b]{\smash{\SetFigFont{8}{9.6}{\rmdefault}{\mddefault}{\updefault}{\color[rgb]{0,0,0}$i-1$}%
}}}
\put(601,-3661){\makebox(0,0)[b]{\smash{\SetFigFont{8}{9.6}{\rmdefault}{\mddefault}{\updefault}{\color[rgb]{0,0,0}$i-3$}%
}}}
\put(3601,239){\makebox(0,0)[b]{\smash{\SetFigFont{10}{12.0}{\rmdefault}{\mddefault}{\updefault}{\color[rgb]{0,0,0}$nR$ bit message block arrival times}%
}}}
\put(8401,-2041){\makebox(0,0)[b]{\smash{\SetFigFont{10}{12.0}{\rmdefault}{\mddefault}{\updefault}{\color[rgb]{0,0,0}Extra delay $d-\widetilde{t}(\rho)$}%
}}}
\put(6601,-1786){\makebox(0,0)[b]{\smash{\SetFigFont{10}{12.0}{\rmdefault}{\mddefault}{\updefault}{\color[rgb]{0,0,0}``Essential'' delay $\widetilde{t}(\rho)$}%
}}}
\put(901,-4261){\makebox(0,0)[lb]{\smash{\SetFigFont{10}{12.0}{\rmdefault}{\mddefault}{\updefault}{\color[rgb]{0,0,0}Assumed renewal time --- new block enters an empty queue}%
}}}
\put(8401,-5011){\makebox(0,0)[b]{\smash{\SetFigFont{10}{12.0}{\rmdefault}{\mddefault}{\updefault}{\color[rgb]{0,0,0}Extra delay $d-\widetilde{t}(\rho)$}%
}}}
\put(7501,-5311){\makebox(0,0)[b]{\smash{\SetFigFont{8}{9.6}{\rmdefault}{\mddefault}{\updefault}{\color[rgb]{0,0,0}$i$}%
}}}
\put(6901,-5311){\makebox(0,0)[b]{\smash{\SetFigFont{8}{9.6}{\rmdefault}{\mddefault}{\updefault}{\color[rgb]{0,0,0}$i-1$}%
}}}
\put(6301,-5311){\makebox(0,0)[b]{\smash{\SetFigFont{8}{9.6}{\rmdefault}{\mddefault}{\updefault}{\color[rgb]{0,0,0}$i-2$}%
}}}
\put(5701,-5311){\makebox(0,0)[b]{\smash{\SetFigFont{8}{9.6}{\rmdefault}{\mddefault}{\updefault}{\color[rgb]{0,0,0}$i-3$}%
}}}
\put(8101,-5311){\makebox(0,0)[b]{\smash{\SetFigFont{8}{9.6}{\rmdefault}{\mddefault}{\updefault}{\color[rgb]{0,0,0}$i+1$}%
}}}
\put(8701,-5311){\makebox(0,0)[b]{\smash{\SetFigFont{8}{9.6}{\rmdefault}{\mddefault}{\updefault}{\color[rgb]{0,0,0}$i+2$}%
}}}
\put(3151,-5536){\makebox(0,0)[rb]{\smash{\SetFigFont{12}{14.4}{\rmdefault}{\mddefault}{\updefault}{\color[rgb]{0,0,0}View at the level of point messages}%
}}}
\put(3151,-5761){\makebox(0,0)[rb]{\smash{\SetFigFont{12}{14.4}{\rmdefault}{\mddefault}{\updefault}{\color[rgb]{0,0,0}as compared to underlying ``slack''}%
}}}
\put(5251,-3961){\makebox(0,0)[b]{\smash{\SetFigFont{10}{12.0}{\rmdefault}{\mddefault}{\updefault}{\color[rgb]{0,0,0} possible  message block decoding times leading to an error}%
}}}
\put(2251,-3211){\makebox(0,0)[b]{\smash{\SetFigFont{8}{9.6}{\rmdefault}{\mddefault}{\updefault}{\color[rgb]{0,0,0}$T_{i-2}$}%
}}}
\put(4876,-3211){\makebox(0,0)[b]{\smash{\SetFigFont{8}{9.6}{\rmdefault}{\mddefault}{\updefault}{\color[rgb]{0,0,0}$T_{i-1}$}%
}}}
\put(7801,-3211){\makebox(0,0)[b]{\smash{\SetFigFont{8}{9.6}{\rmdefault}{\mddefault}{\updefault}{\color[rgb]{0,0,0}$T_i$}%
}}}
\put(6751,-6361){\makebox(0,0)[b]{\smash{\SetFigFont{8}{9.6}{\rmdefault}{\mddefault}{\updefault}{\color[rgb]{0,0,0}$\widetilde{T}_{i-2}(\rho)$}%
}}}
\put(7576,-6361){\makebox(0,0)[b]{\smash{\SetFigFont{8}{9.6}{\rmdefault}{\mddefault}{\updefault}{\color[rgb]{0,0,0}$\widetilde{T}_{i-1}(\rho)$}%
}}}
\put(8701,-6361){\makebox(0,0)[b]{\smash{\SetFigFont{8}{9.6}{\rmdefault}{\mddefault}{\updefault}{\color[rgb]{0,0,0}$\widetilde{T}_{i}(\rho)$}%
}}}
\put(7201,-6511){\makebox(0,0)[b]{\smash{\SetFigFont{8}{9.6}{\rmdefault}{\mddefault}{\updefault}{\color[rgb]{0,0,0}$i-2$}%
}}}
\put(7951,-6511){\makebox(0,0)[b]{\smash{\SetFigFont{8}{9.6}{\rmdefault}{\mddefault}{\updefault}{\color[rgb]{0,0,0}$i-1$}%
}}}
\put(6301,-7111){\makebox(0,0)[lb]{\smash{\SetFigFont{10}{12.0}{\rmdefault}{\mddefault}{\updefault}{\color[rgb]{0,0,0}Assumed renewal time}%
}}}
\put(6001,-6286){\makebox(0,0)[b]{\smash{\SetFigFont{8}{9.6}{\rmdefault}{\mddefault}{\updefault}{\color[rgb]{0,0,0}$i-3$}%
}}}
\put(9451,-6511){\makebox(0,0)[b]{\smash{\SetFigFont{8}{9.6}{\rmdefault}{\mddefault}{\updefault}{\color[rgb]{0,0,0}$i$}%
}}}
\end{picture}

\end{center}
\caption{At the top, the original timeline depicts the arrival of
  message blocks and the target delay. The ``essential'' component
  $\widetilde{t}(\rho)$ is also shown. In the middle, a particular
  realization is shown illustrating how an error can happen when the
  service times $T_j$ of the blocks become too large. At the
  bottom, the essential components of the service times are removed
  and the performance of the system is shown to be bounded by that of
  a queue with iid geometric service times serving the low-rate
  deterministic arrival of point-messages corresponding to the message
  blocks.} 
\label{fig:chunkslackerror}
\end{figure}
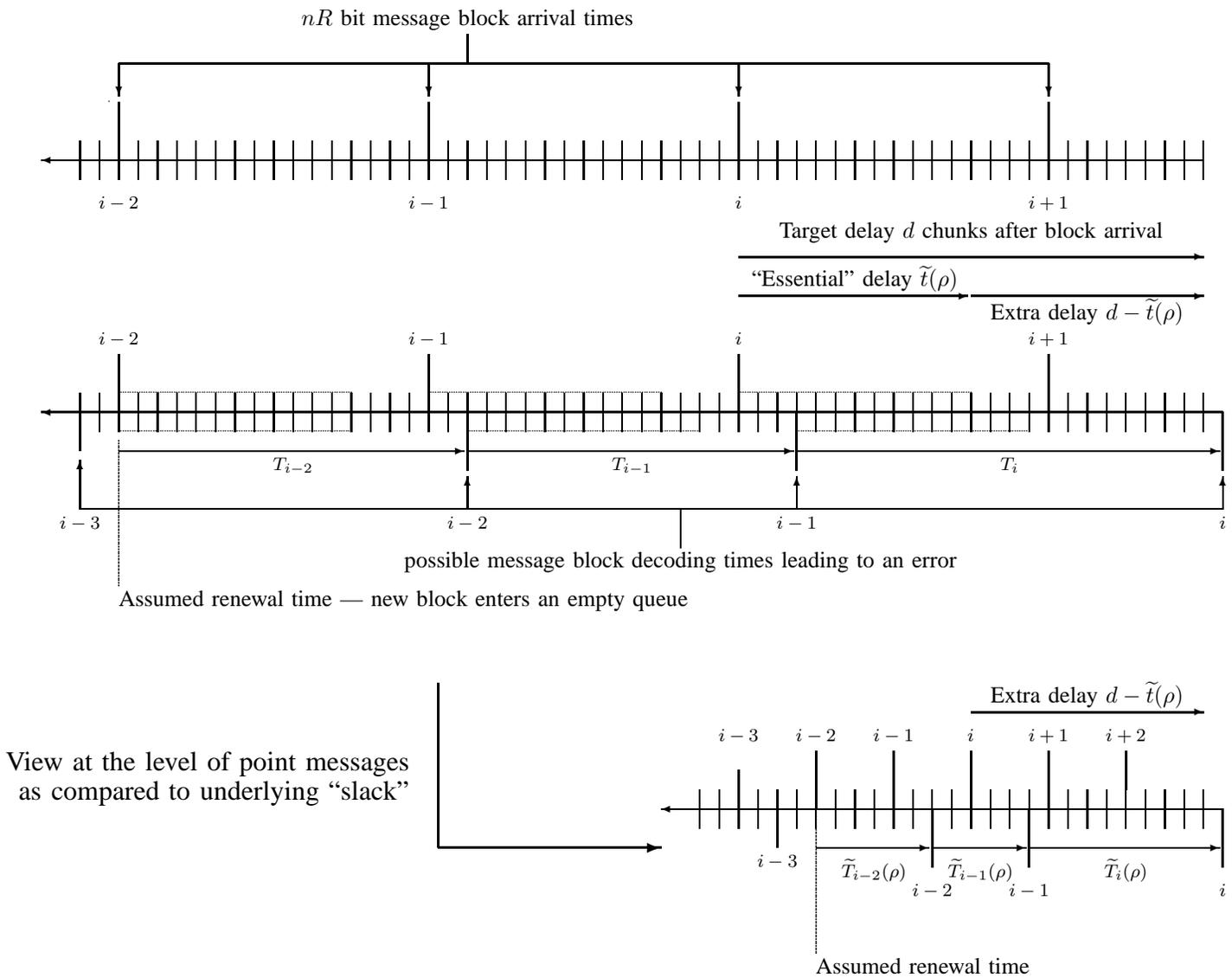

Lemma~\ref{lem:transmissiontimes} is illustrated in
Figure~\ref{fig:blockslack} and then the application of
Corollary~\ref{cor:delaybasedbound} is illustrated in Figures
\ref{fig:slackillustrated} and \ref{fig:chunkslackerror}. These
illustrate the achievability of the fixed-delay exponent $0.44$ at a
rate of $0.37$ nats. The gap between $0.44$ and $0.37$ on the rate
axis in Figure~\ref{fig:slackillustrated} depicts the fraction of
``slack'' channel uses that are available to communicate a message
block with reliability $0.44$ while still draining the queue faster
than it is being filled. The block length $n$ must be long enough so
that the slack represents at least a few channel uses. As the block
length $n$ becomes longer, it is possible to move up to the
reliability limit illustrated by the inverse concatenation
construction.

Consider time in $ck$ units. Let $R' = \frac{1}{n}$ be the rate at
which message blocks are generated in terms of blocks generated per
$ck$ channel uses. $R'' = \frac{1}{n - \lceil
  \widetilde{t}(\rho,R,n,\vec{q}) \rceil}$ is the rate at which we
evaluate the BEC's fixed-delay reliability in the application of
Corollary~\ref{cor:delaybasedbound}. The effective ``erasure
probability'' is $\beta = \exp(-ck E_0(\rho,\vec{q}))$.

Recall that $R < \frac{E_0(\rho)}{\rho}$ where the $\vec{q}$
distribution is chosen as the $E_0(\rho)$ achieving distribution. The
quantity $n - \lceil \widetilde{t}(\rho,R,n,\vec{q}) \rceil$ has a
special significance since it captures the amount of slack in the
system when viewed with parameter $\rho$. This slack term is positive
for large enough $n >
\frac{\widetilde{C}(\rho,\vec{q})}{\widetilde{C}(\rho,\vec{q}) - R}$ since

\begin{eqnarray}
n - \lceil \widetilde{t}(\rho,R,n,\vec{q}) \rceil
& \geq &  
\left(\frac{\widetilde{C}(\rho,\vec{q}) -
    R}{\widetilde{C}(\rho,\vec{q})} \right) n  - 1. \label{eqn:slackbound}
\end{eqnarray}
Thus
\begin{eqnarray*}
R'' & = & \frac{1}{n - \lceil \widetilde{t}(\rho,R,n,\vec{q}) \rceil} \\
& \leq & (\left(\frac{\widetilde{C}(\rho,\vec{q}) - R}{\widetilde{C}(\rho,\vec{q})} \right) n - 1 )^{-1}.
\end{eqnarray*}
Notice that $R''$ can be made as small as desired by choosing $n$ large
while $\beta$ can be made extremely small by choosing $c$
large. Applying Theorem~\ref{thm:erasurecase} tells us to set
\begin{eqnarray}
1 + 2r & = &  
\frac{\widetilde{C}(\rho,\vec{q}) - R}{\widetilde{C}(\rho,\vec{q})} n 
- 1 \nonumber \\
n(\rho, c, k, l, r)
& = & \frac{\widetilde{C}(\rho,\vec{q})}{\widetilde{C}(\rho,\vec{q}) -
  R}(2 + 2r)  
\label{eqn:ndefinition}
\end{eqnarray}
in order to get to within $(2 \ln 2)\exp(-rck E_0(\rho, \vec{q}))$ of
the exponent $ck E_0(\rho, \vec{q})$ in terms of delays measured in
$ck$ time units, or to within $\frac{2 \ln 2}{ck}\exp(-rck E_0(\rho,
\vec{q}))$ of the exponent of $E_0(\rho,\vec{q})$ in terms of delays
measured in channel uses.  

Putting it all together, for any small $\Delta > 0$, and $\rho \geq 0$
such that $R < \frac{E_0(\rho)}{\rho}$ a delay-exponent of $E_0(\rho)
- (3 \ln 2) \Delta$ is clearly achievable by setting $l = \max(0, \lceil
\log_2 \rho \rceil)$, choosing chunk size
\begin{equation} \label{eqn:cselection}
c = \max(l+1, \lceil \frac{\ln 16}{k E_0(\rho)} \rceil)
\end{equation}
and then choosing $r$ big enough using
\begin{equation} \label{eqn:rselection}
  r \geq \max(0,\frac{\ln(\Delta c k)}{ck E_0(\rho)}).
\end{equation}
With $c$ and $r$ defined, $n$ can be obtained from
(\ref{eqn:ndefinition}).

Notice that $k$ is arbitrary here and can thus be made as large as
desired. This corresponds to the fact that the amount of
``punctuation'' information can be made as small as desired, assuming
that the target end-to-end delay is large enough.\footnote{The
  target end-to-end delay must at least be large enough to absorb the
  roughly $2nck$ channel uses corresponding to the sum of assembly
  delay and essential service time for the message block. It is
  beyond that point that the delay exponent analysis here kicks in.}

Each $(n,c,l)$ code is also delay universal since it is not designed
with a maximum $d$ in mind. The longer the decoder is willing to wait,
the lower the probability of error becomes. This property is inherited
from the repeat-until-success code for the erasure channel through
Corollary~\ref{cor:delaybasedbound}. \hfill $\Box$ \vspace{0.20in}

\subsection{Channels with strictly positive zero-error capacity} \label{sec:strictzero}

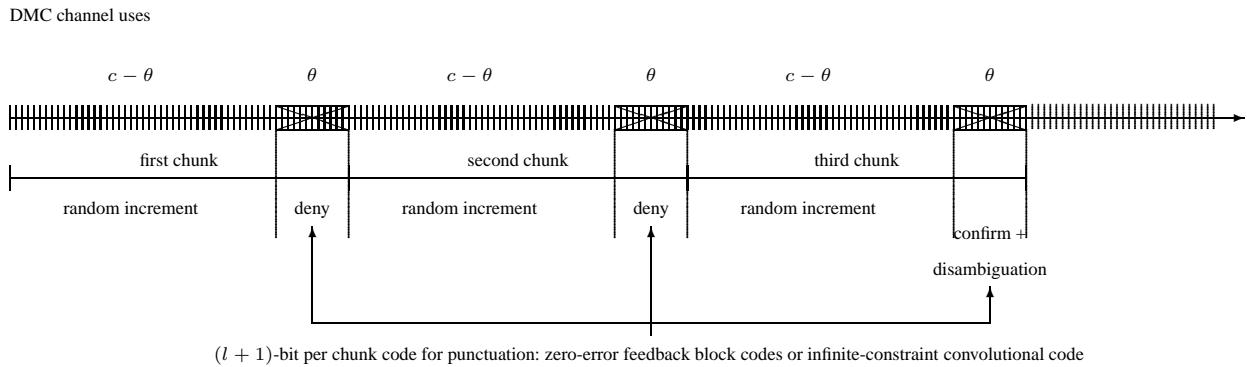
\begin{figure}[htbp]
\begin{center}
\setlength{\unitlength}{4000sp}%
\begingroup\makeatletter\ifx\SetFigFont\undefined%
\gdef\SetFigFont#1#2#3#4#5{%
  \reset@font\fontsize{#1}{#2pt}%
  \fontfamily{#3}\fontseries{#4}\fontshape{#5}%
  \selectfont}%
\fi\endgroup%
\begin{picture}(7674,2284)(-11,-1760)
\thinlines
{\color[rgb]{0,0,0}\put(1651,-136){\line( 1, 0){450}}
}%
{\color[rgb]{0,0,0}\put(1651,-286){\line( 1, 0){450}}
}%
{\color[rgb]{0,0,0}\put(  1,-211){\vector( 1, 0){7650}}
}%
{\color[rgb]{0,0,0}\put(6301,-136){\line( 0,-1){150}}
}%
{\color[rgb]{0,0,0}\multiput(7463,-136)(0.00000,-4.54545){34}{\makebox(1.6667,11.6667){\SetFigFont{5}{6}{\rmdefault}{\mddefault}{\updefault}.}}
}%
{\color[rgb]{0,0,0}\multiput(7426,-136)(0.00000,-4.54545){34}{\makebox(1.6667,11.6667){\SetFigFont{5}{6}{\rmdefault}{\mddefault}{\updefault}.}}
}%
{\color[rgb]{0,0,0}\multiput(7388,-136)(0.00000,-4.54545){34}{\makebox(1.6667,11.6667){\SetFigFont{5}{6}{\rmdefault}{\mddefault}{\updefault}.}}
}%
{\color[rgb]{0,0,0}\multiput(7351,-136)(0.00000,-4.54545){34}{\makebox(1.6667,11.6667){\SetFigFont{5}{6}{\rmdefault}{\mddefault}{\updefault}.}}
}%
{\color[rgb]{0,0,0}\multiput(7313,-136)(0.00000,-4.54545){34}{\makebox(1.6667,11.6667){\SetFigFont{5}{6}{\rmdefault}{\mddefault}{\updefault}.}}
}%
{\color[rgb]{0,0,0}\multiput(7276,-136)(0.00000,-4.54545){34}{\makebox(1.6667,11.6667){\SetFigFont{5}{6}{\rmdefault}{\mddefault}{\updefault}.}}
}%
{\color[rgb]{0,0,0}\multiput(7238,-136)(0.00000,-4.54545){34}{\makebox(1.6667,11.6667){\SetFigFont{5}{6}{\rmdefault}{\mddefault}{\updefault}.}}
}%
{\color[rgb]{0,0,0}\multiput(7201,-136)(0.00000,-4.54545){34}{\makebox(1.6667,11.6667){\SetFigFont{5}{6}{\rmdefault}{\mddefault}{\updefault}.}}
}%
{\color[rgb]{0,0,0}\multiput(7163,-136)(0.00000,-4.54545){34}{\makebox(1.6667,11.6667){\SetFigFont{5}{6}{\rmdefault}{\mddefault}{\updefault}.}}
}%
{\color[rgb]{0,0,0}\multiput(7126,-136)(0.00000,-4.54545){34}{\makebox(1.6667,11.6667){\SetFigFont{5}{6}{\rmdefault}{\mddefault}{\updefault}.}}
}%
{\color[rgb]{0,0,0}\multiput(7088,-136)(0.00000,-4.54545){34}{\makebox(1.6667,11.6667){\SetFigFont{5}{6}{\rmdefault}{\mddefault}{\updefault}.}}
}%
{\color[rgb]{0,0,0}\multiput(7051,-136)(0.00000,-4.54545){34}{\makebox(1.6667,11.6667){\SetFigFont{5}{6}{\rmdefault}{\mddefault}{\updefault}.}}
}%
{\color[rgb]{0,0,0}\multiput(7013,-136)(0.00000,-4.54545){34}{\makebox(1.6667,11.6667){\SetFigFont{5}{6}{\rmdefault}{\mddefault}{\updefault}.}}
}%
{\color[rgb]{0,0,0}\multiput(6976,-136)(0.00000,-4.54545){34}{\makebox(1.6667,11.6667){\SetFigFont{5}{6}{\rmdefault}{\mddefault}{\updefault}.}}
}%
{\color[rgb]{0,0,0}\multiput(6938,-136)(0.00000,-4.54545){34}{\makebox(1.6667,11.6667){\SetFigFont{5}{6}{\rmdefault}{\mddefault}{\updefault}.}}
}%
{\color[rgb]{0,0,0}\multiput(6901,-136)(0.00000,-4.54545){34}{\makebox(1.6667,11.6667){\SetFigFont{5}{6}{\rmdefault}{\mddefault}{\updefault}.}}
}%
{\color[rgb]{0,0,0}\multiput(6863,-136)(0.00000,-4.54545){34}{\makebox(1.6667,11.6667){\SetFigFont{5}{6}{\rmdefault}{\mddefault}{\updefault}.}}
}%
{\color[rgb]{0,0,0}\multiput(6826,-136)(0.00000,-4.54545){34}{\makebox(1.6667,11.6667){\SetFigFont{5}{6}{\rmdefault}{\mddefault}{\updefault}.}}
}%
{\color[rgb]{0,0,0}\multiput(6788,-136)(0.00000,-4.54545){34}{\makebox(1.6667,11.6667){\SetFigFont{5}{6}{\rmdefault}{\mddefault}{\updefault}.}}
}%
{\color[rgb]{0,0,0}\multiput(6751,-136)(0.00000,-4.54545){34}{\makebox(1.6667,11.6667){\SetFigFont{5}{6}{\rmdefault}{\mddefault}{\updefault}.}}
}%
{\color[rgb]{0,0,0}\multiput(6713,-136)(0.00000,-4.54545){34}{\makebox(1.6667,11.6667){\SetFigFont{5}{6}{\rmdefault}{\mddefault}{\updefault}.}}
}%
{\color[rgb]{0,0,0}\multiput(6676,-136)(0.00000,-4.54545){34}{\makebox(1.6667,11.6667){\SetFigFont{5}{6}{\rmdefault}{\mddefault}{\updefault}.}}
}%
{\color[rgb]{0,0,0}\multiput(6638,-136)(0.00000,-4.54545){34}{\makebox(1.6667,11.6667){\SetFigFont{5}{6}{\rmdefault}{\mddefault}{\updefault}.}}
}%
{\color[rgb]{0,0,0}\multiput(6601,-136)(0.00000,-4.54545){34}{\makebox(1.6667,11.6667){\SetFigFont{5}{6}{\rmdefault}{\mddefault}{\updefault}.}}
}%
{\color[rgb]{0,0,0}\multiput(6563,-136)(0.00000,-4.54545){34}{\makebox(1.6667,11.6667){\SetFigFont{5}{6}{\rmdefault}{\mddefault}{\updefault}.}}
}%
{\color[rgb]{0,0,0}\multiput(6526,-136)(0.00000,-4.54545){34}{\makebox(1.6667,11.6667){\SetFigFont{5}{6}{\rmdefault}{\mddefault}{\updefault}.}}
}%
{\color[rgb]{0,0,0}\multiput(6488,-136)(0.00000,-4.54545){34}{\makebox(1.6667,11.6667){\SetFigFont{5}{6}{\rmdefault}{\mddefault}{\updefault}.}}
}%
{\color[rgb]{0,0,0}\multiput(6451,-136)(0.00000,-4.54545){34}{\makebox(1.6667,11.6667){\SetFigFont{5}{6}{\rmdefault}{\mddefault}{\updefault}.}}
}%
{\color[rgb]{0,0,0}\multiput(6413,-136)(0.00000,-4.54545){34}{\makebox(1.6667,11.6667){\SetFigFont{5}{6}{\rmdefault}{\mddefault}{\updefault}.}}
}%
{\color[rgb]{0,0,0}\multiput(6376,-136)(0.00000,-4.54545){34}{\makebox(1.6667,11.6667){\SetFigFont{5}{6}{\rmdefault}{\mddefault}{\updefault}.}}
}%
{\color[rgb]{0,0,0}\multiput(6338,-136)(0.00000,-4.54545){34}{\makebox(1.6667,11.6667){\SetFigFont{5}{6}{\rmdefault}{\mddefault}{\updefault}.}}
}%
{\color[rgb]{0,0,0}\put(451,-136){\line( 0,-1){150}}
}%
{\color[rgb]{0,0,0}\put(526,-136){\line( 0,-1){150}}
}%
{\color[rgb]{0,0,0}\put(301,-136){\line( 0,-1){150}}
}%
{\color[rgb]{0,0,0}\put(376,-136){\line( 0,-1){150}}
}%
{\color[rgb]{0,0,0}\put(  1,-136){\line( 0,-1){150}}
}%
{\color[rgb]{0,0,0}\put( 76,-136){\line( 0,-1){150}}
}%
{\color[rgb]{0,0,0}\put(151,-136){\line( 0,-1){150}}
}%
{\color[rgb]{0,0,0}\put(226,-136){\line( 0,-1){150}}
}%
{\color[rgb]{0,0,0}\put(188,-136){\line( 0,-1){150}}
}%
{\color[rgb]{0,0,0}\put(263,-136){\line( 0,-1){150}}
}%
{\color[rgb]{0,0,0}\put( 38,-136){\line( 0,-1){150}}
}%
{\color[rgb]{0,0,0}\put(113,-136){\line( 0,-1){150}}
}%
{\color[rgb]{0,0,0}\put(338,-136){\line( 0,-1){150}}
}%
{\color[rgb]{0,0,0}\put(413,-136){\line( 0,-1){150}}
}%
{\color[rgb]{0,0,0}\put(488,-136){\line( 0,-1){150}}
}%
{\color[rgb]{0,0,0}\put(563,-136){\line( 0,-1){150}}
}%
{\color[rgb]{0,0,0}\put(601,-136){\line( 0,-1){150}}
}%
{\color[rgb]{0,0,0}\put(676,-136){\line( 0,-1){150}}
}%
{\color[rgb]{0,0,0}\put(638,-136){\line( 0,-1){150}}
}%
{\color[rgb]{0,0,0}\put(713,-136){\line( 0,-1){150}}
}%
{\color[rgb]{0,0,0}\put(751,-136){\line( 0,-1){150}}
}%
{\color[rgb]{0,0,0}\put(826,-136){\line( 0,-1){150}}
}%
{\color[rgb]{0,0,0}\put(788,-136){\line( 0,-1){150}}
}%
{\color[rgb]{0,0,0}\put(863,-136){\line( 0,-1){150}}
}%
{\color[rgb]{0,0,0}\put(901,-136){\line( 0,-1){150}}
}%
{\color[rgb]{0,0,0}\put(976,-136){\line( 0,-1){150}}
}%
{\color[rgb]{0,0,0}\put(938,-136){\line( 0,-1){150}}
}%
{\color[rgb]{0,0,0}\put(1013,-136){\line( 0,-1){150}}
}%
{\color[rgb]{0,0,0}\put(1088,-136){\line( 0,-1){150}}
}%
{\color[rgb]{0,0,0}\put(1163,-136){\line( 0,-1){150}}
}%
{\color[rgb]{0,0,0}\put(1051,-136){\line( 0,-1){150}}
}%
{\color[rgb]{0,0,0}\put(1126,-136){\line( 0,-1){150}}
}%
{\color[rgb]{0,0,0}\put(1238,-136){\line( 0,-1){150}}
}%
{\color[rgb]{0,0,0}\put(1313,-136){\line( 0,-1){150}}
}%
{\color[rgb]{0,0,0}\put(1388,-136){\line( 0,-1){150}}
}%
{\color[rgb]{0,0,0}\put(1463,-136){\line( 0,-1){150}}
}%
{\color[rgb]{0,0,0}\put(1201,-136){\line( 0,-1){150}}
}%
{\color[rgb]{0,0,0}\put(1276,-136){\line( 0,-1){150}}
}%
{\color[rgb]{0,0,0}\put(1351,-136){\line( 0,-1){150}}
}%
{\color[rgb]{0,0,0}\put(1426,-136){\line( 0,-1){150}}
}%
{\color[rgb]{0,0,0}\put(1501,-136){\line( 0,-1){150}}
}%
{\color[rgb]{0,0,0}\put(1576,-136){\line( 0,-1){150}}
}%
{\color[rgb]{0,0,0}\put(1651,-136){\line( 0,-1){150}}
}%
{\color[rgb]{0,0,0}\put(1726,-136){\line( 0,-1){150}}
}%
{\color[rgb]{0,0,0}\put(1688,-136){\line( 0,-1){150}}
}%
{\color[rgb]{0,0,0}\put(1763,-136){\line( 0,-1){150}}
}%
{\color[rgb]{0,0,0}\put(1538,-136){\line( 0,-1){150}}
}%
{\color[rgb]{0,0,0}\put(1613,-136){\line( 0,-1){150}}
}%
{\color[rgb]{0,0,0}\put(1838,-136){\line( 0,-1){150}}
}%
{\color[rgb]{0,0,0}\put(1913,-136){\line( 0,-1){150}}
}%
{\color[rgb]{0,0,0}\put(1988,-136){\line( 0,-1){150}}
}%
{\color[rgb]{0,0,0}\put(2063,-136){\line( 0,-1){150}}
}%
{\color[rgb]{0,0,0}\put(1801,-136){\line( 0,-1){150}}
}%
{\color[rgb]{0,0,0}\put(1876,-136){\line( 0,-1){150}}
}%
{\color[rgb]{0,0,0}\put(1951,-136){\line( 0,-1){150}}
}%
{\color[rgb]{0,0,0}\put(2026,-136){\line( 0,-1){150}}
}%
{\color[rgb]{0,0,0}\put(2101,-136){\line( 0,-1){150}}
}%
{\color[rgb]{0,0,0}\put(2176,-136){\line( 0,-1){150}}
}%
{\color[rgb]{0,0,0}\put(2251,-136){\line( 0,-1){150}}
}%
{\color[rgb]{0,0,0}\put(2326,-136){\line( 0,-1){150}}
}%
{\color[rgb]{0,0,0}\put(2138,-136){\line( 0,-1){150}}
}%
{\color[rgb]{0,0,0}\put(2213,-136){\line( 0,-1){150}}
}%
{\color[rgb]{0,0,0}\put(2288,-136){\line( 0,-1){150}}
}%
{\color[rgb]{0,0,0}\put(2363,-136){\line( 0,-1){150}}
}%
{\color[rgb]{0,0,0}\put(2438,-136){\line( 0,-1){150}}
}%
{\color[rgb]{0,0,0}\put(2513,-136){\line( 0,-1){150}}
}%
{\color[rgb]{0,0,0}\put(2588,-136){\line( 0,-1){150}}
}%
{\color[rgb]{0,0,0}\put(2663,-136){\line( 0,-1){150}}
}%
{\color[rgb]{0,0,0}\put(2738,-136){\line( 0,-1){150}}
}%
{\color[rgb]{0,0,0}\put(2813,-136){\line( 0,-1){150}}
}%
{\color[rgb]{0,0,0}\put(2888,-136){\line( 0,-1){150}}
}%
{\color[rgb]{0,0,0}\put(2963,-136){\line( 0,-1){150}}
}%
{\color[rgb]{0,0,0}\put(2401,-136){\line( 0,-1){150}}
}%
{\color[rgb]{0,0,0}\put(2476,-136){\line( 0,-1){150}}
}%
{\color[rgb]{0,0,0}\put(2551,-136){\line( 0,-1){150}}
}%
{\color[rgb]{0,0,0}\put(2626,-136){\line( 0,-1){150}}
}%
{\color[rgb]{0,0,0}\put(2701,-136){\line( 0,-1){150}}
}%
{\color[rgb]{0,0,0}\put(2776,-136){\line( 0,-1){150}}
}%
{\color[rgb]{0,0,0}\put(2851,-136){\line( 0,-1){150}}
}%
{\color[rgb]{0,0,0}\put(2926,-136){\line( 0,-1){150}}
}%
{\color[rgb]{0,0,0}\put(3151,-136){\line( 0,-1){150}}
}%
{\color[rgb]{0,0,0}\put(3226,-136){\line( 0,-1){150}}
}%
{\color[rgb]{0,0,0}\put(3038,-136){\line( 0,-1){150}}
}%
{\color[rgb]{0,0,0}\put(3113,-136){\line( 0,-1){150}}
}%
{\color[rgb]{0,0,0}\put(3188,-136){\line( 0,-1){150}}
}%
{\color[rgb]{0,0,0}\put(3263,-136){\line( 0,-1){150}}
}%
{\color[rgb]{0,0,0}\put(3301,-136){\line( 0,-1){150}}
}%
{\color[rgb]{0,0,0}\put(3376,-136){\line( 0,-1){150}}
}%
{\color[rgb]{0,0,0}\put(3001,-136){\line( 0,-1){150}}
}%
{\color[rgb]{0,0,0}\put(3076,-136){\line( 0,-1){150}}
}%
{\color[rgb]{0,0,0}\put(3338,-136){\line( 0,-1){150}}
}%
{\color[rgb]{0,0,0}\put(3413,-136){\line( 0,-1){150}}
}%
{\color[rgb]{0,0,0}\put(3451,-136){\line( 0,-1){150}}
}%
{\color[rgb]{0,0,0}\put(3526,-136){\line( 0,-1){150}}
}%
{\color[rgb]{0,0,0}\put(3488,-136){\line( 0,-1){150}}
}%
{\color[rgb]{0,0,0}\put(3563,-136){\line( 0,-1){150}}
}%
{\color[rgb]{0,0,0}\put(3751,-136){\line( 0,-1){150}}
}%
{\color[rgb]{0,0,0}\put(3826,-136){\line( 0,-1){150}}
}%
{\color[rgb]{0,0,0}\put(3601,-136){\line( 0,-1){150}}
}%
{\color[rgb]{0,0,0}\put(3676,-136){\line( 0,-1){150}}
}%
{\color[rgb]{0,0,0}\put(4088,-136){\line( 0,-1){150}}
}%
{\color[rgb]{0,0,0}\put(4163,-136){\line( 0,-1){150}}
}%
{\color[rgb]{0,0,0}\put(4238,-136){\line( 0,-1){150}}
}%
{\color[rgb]{0,0,0}\put(4313,-136){\line( 0,-1){150}}
}%
{\color[rgb]{0,0,0}\put(4201,-136){\line( 0,-1){150}}
}%
{\color[rgb]{0,0,0}\put(4276,-136){\line( 0,-1){150}}
}%
{\color[rgb]{0,0,0}\put(3938,-136){\line( 0,-1){150}}
}%
{\color[rgb]{0,0,0}\put(4013,-136){\line( 0,-1){150}}
}%
{\color[rgb]{0,0,0}\put(3638,-136){\line( 0,-1){150}}
}%
{\color[rgb]{0,0,0}\put(3713,-136){\line( 0,-1){150}}
}%
{\color[rgb]{0,0,0}\put(3788,-136){\line( 0,-1){150}}
}%
{\color[rgb]{0,0,0}\put(3863,-136){\line( 0,-1){150}}
}%
{\color[rgb]{0,0,0}\put(3901,-136){\line( 0,-1){150}}
}%
{\color[rgb]{0,0,0}\put(3976,-136){\line( 0,-1){150}}
}%
{\color[rgb]{0,0,0}\put(4051,-136){\line( 0,-1){150}}
}%
{\color[rgb]{0,0,0}\put(4126,-136){\line( 0,-1){150}}
}%
{\color[rgb]{0,0,0}\put(4351,-136){\line( 0,-1){150}}
}%
{\color[rgb]{0,0,0}\put(4426,-136){\line( 0,-1){150}}
}%
{\color[rgb]{0,0,0}\put(4388,-136){\line( 0,-1){150}}
}%
{\color[rgb]{0,0,0}\put(4463,-136){\line( 0,-1){150}}
}%
{\color[rgb]{0,0,0}\put(4501,-136){\line( 0,-1){150}}
}%
{\color[rgb]{0,0,0}\put(4576,-136){\line( 0,-1){150}}
}%
{\color[rgb]{0,0,0}\put(4651,-136){\line( 0,-1){150}}
}%
{\color[rgb]{0,0,0}\put(4726,-136){\line( 0,-1){150}}
}%
{\color[rgb]{0,0,0}\put(4538,-136){\line( 0,-1){150}}
}%
{\color[rgb]{0,0,0}\put(4613,-136){\line( 0,-1){150}}
}%
{\color[rgb]{0,0,0}\put(4688,-136){\line( 0,-1){150}}
}%
{\color[rgb]{0,0,0}\put(4763,-136){\line( 0,-1){150}}
}%
{\color[rgb]{0,0,0}\put(5138,-136){\line( 0,-1){150}}
}%
{\color[rgb]{0,0,0}\put(5213,-136){\line( 0,-1){150}}
}%
{\color[rgb]{0,0,0}\put(4988,-136){\line( 0,-1){150}}
}%
{\color[rgb]{0,0,0}\put(5063,-136){\line( 0,-1){150}}
}%
{\color[rgb]{0,0,0}\put(4838,-136){\line( 0,-1){150}}
}%
{\color[rgb]{0,0,0}\put(4913,-136){\line( 0,-1){150}}
}%
{\color[rgb]{0,0,0}\put(4951,-136){\line( 0,-1){150}}
}%
{\color[rgb]{0,0,0}\put(5026,-136){\line( 0,-1){150}}
}%
{\color[rgb]{0,0,0}\put(4801,-136){\line( 0,-1){150}}
}%
{\color[rgb]{0,0,0}\put(4876,-136){\line( 0,-1){150}}
}%
{\color[rgb]{0,0,0}\put(5101,-136){\line( 0,-1){150}}
}%
{\color[rgb]{0,0,0}\put(5176,-136){\line( 0,-1){150}}
}%
{\color[rgb]{0,0,0}\put(5251,-136){\line( 0,-1){150}}
}%
{\color[rgb]{0,0,0}\put(5326,-136){\line( 0,-1){150}}
}%
{\color[rgb]{0,0,0}\put(5288,-136){\line( 0,-1){150}}
}%
{\color[rgb]{0,0,0}\put(5363,-136){\line( 0,-1){150}}
}%
{\color[rgb]{0,0,0}\put(5438,-136){\line( 0,-1){150}}
}%
{\color[rgb]{0,0,0}\put(5513,-136){\line( 0,-1){150}}
}%
{\color[rgb]{0,0,0}\put(5551,-136){\line( 0,-1){150}}
}%
{\color[rgb]{0,0,0}\put(5626,-136){\line( 0,-1){150}}
}%
{\color[rgb]{0,0,0}\put(5588,-136){\line( 0,-1){150}}
}%
{\color[rgb]{0,0,0}\put(5663,-136){\line( 0,-1){150}}
}%
{\color[rgb]{0,0,0}\put(5401,-136){\line( 0,-1){150}}
}%
{\color[rgb]{0,0,0}\put(5476,-136){\line( 0,-1){150}}
}%
{\color[rgb]{0,0,0}\put(5701,-136){\line( 0,-1){150}}
}%
{\color[rgb]{0,0,0}\put(5776,-136){\line( 0,-1){150}}
}%
{\color[rgb]{0,0,0}\put(6151,-136){\line( 0,-1){150}}
}%
{\color[rgb]{0,0,0}\put(6226,-136){\line( 0,-1){150}}
}%
{\color[rgb]{0,0,0}\put(6001,-136){\line( 0,-1){150}}
}%
{\color[rgb]{0,0,0}\put(6076,-136){\line( 0,-1){150}}
}%
{\color[rgb]{0,0,0}\put(5851,-136){\line( 0,-1){150}}
}%
{\color[rgb]{0,0,0}\put(5926,-136){\line( 0,-1){150}}
}%
{\color[rgb]{0,0,0}\put(5738,-136){\line( 0,-1){150}}
}%
{\color[rgb]{0,0,0}\put(5813,-136){\line( 0,-1){150}}
}%
{\color[rgb]{0,0,0}\put(6188,-136){\line( 0,-1){150}}
}%
{\color[rgb]{0,0,0}\put(6263,-136){\line( 0,-1){150}}
}%
{\color[rgb]{0,0,0}\put(6038,-136){\line( 0,-1){150}}
}%
{\color[rgb]{0,0,0}\put(6113,-136){\line( 0,-1){150}}
}%
{\color[rgb]{0,0,0}\put(5888,-136){\line( 0,-1){150}}
}%
{\color[rgb]{0,0,0}\put(5963,-136){\line( 0,-1){150}}
}%
{\color[rgb]{0,0,0}\put(4201,-136){\line(-1, 0){450}}
}%
{\color[rgb]{0,0,0}\put(4201,-286){\line(-1, 0){450}}
}%
{\color[rgb]{0,0,0}\put(5851,-136){\line( 1, 0){450}}
}%
{\color[rgb]{0,0,0}\put(5851,-286){\line( 1, 0){450}}
}%
{\color[rgb]{0,0,0}\put(1651,-136){\line( 3,-1){450}}
}%
{\color[rgb]{0,0,0}\put(1651,-286){\line( 3, 1){450}}
}%
{\color[rgb]{0,0,0}\put(3751,-286){\line( 3, 1){450}}
}%
{\color[rgb]{0,0,0}\put(3751,-136){\line( 3,-1){450}}
}%
{\color[rgb]{0,0,0}\put(5851,-136){\line( 3,-1){450}}
}%
{\color[rgb]{0,0,0}\put(5851,-286){\line( 3, 1){450}}
}%
{\color[rgb]{0,0,0}\multiput(1651,-286)(0.00000,-4.50000){151}{\makebox(1.6667,11.6667){\SetFigFont{5}{6}{\rmdefault}{\mddefault}{\updefault}.}}
}%
{\color[rgb]{0,0,0}\multiput(2101,-286)(0.00000,-4.50000){151}{\makebox(1.6667,11.6667){\SetFigFont{5}{6}{\rmdefault}{\mddefault}{\updefault}.}}
}%
{\color[rgb]{0,0,0}\multiput(3751,-286)(0.00000,-4.50000){151}{\makebox(1.6667,11.6667){\SetFigFont{5}{6}{\rmdefault}{\mddefault}{\updefault}.}}
}%
{\color[rgb]{0,0,0}\multiput(4201,-286)(0.00000,-4.50000){151}{\makebox(1.6667,11.6667){\SetFigFont{5}{6}{\rmdefault}{\mddefault}{\updefault}.}}
}%
{\color[rgb]{0,0,0}\multiput(5851,-286)(0.00000,-4.50000){151}{\makebox(1.6667,11.6667){\SetFigFont{5}{6}{\rmdefault}{\mddefault}{\updefault}.}}
}%
{\color[rgb]{0,0,0}\multiput(6301,-286)(0.00000,-4.50000){151}{\makebox(1.6667,11.6667){\SetFigFont{5}{6}{\rmdefault}{\mddefault}{\updefault}.}}
}%
{\color[rgb]{0,0,0}\put(3976,-1561){\vector( 0, 1){675}}
}%
{\color[rgb]{0,0,0}\put(3976,-1486){\line( 1, 0){2100}}
\put(6076,-1486){\vector( 0, 1){225}}
}%
{\color[rgb]{0,0,0}\put(3976,-1486){\line(-1, 0){2100}}
\put(1876,-1486){\vector( 0, 1){600}}
}%
{\color[rgb]{0,0,0}\put(  1,-511){\line( 0,-1){150}}
}%
{\color[rgb]{0,0,0}\put(2101,-511){\line( 0,-1){150}}
}%
{\color[rgb]{0,0,0}\put(2101,-586){\line( 1, 0){2100}}
}%
{\color[rgb]{0,0,0}\put(2101,-511){\line( 0,-1){150}}
}%
{\color[rgb]{0,0,0}\put(4201,-511){\line( 0,-1){150}}
}%
{\color[rgb]{0,0,0}\put(4201,-586){\line( 1, 0){2100}}
}%
{\color[rgb]{0,0,0}\put(4201,-511){\line( 0,-1){150}}
}%
{\color[rgb]{0,0,0}\put(6301,-511){\line( 0,-1){150}}
}%
{\color[rgb]{0,0,0}\put(  1,-586){\line( 1, 0){2100}}
}%
\put(6076,-1186){\makebox(0,0)[b]{\smash{\SetFigFont{7}{6}{\rmdefault}{\mddefault}{\updefault}{\color[rgb]{0,0,0}disambiguation}%
}}}
\put(3976,-1711){\makebox(0,0)[b]{\smash{\SetFigFont{7}{6}{\rmdefault}{\mddefault}{\updefault}{\color[rgb]{0,0,0}$(l+1)$-bit per chunk code for punctuation: zero-error feedback block codes or infinite-constraint convolutional code }%
}}}
\put(  1,389){\makebox(0,0)[lb]{\smash{\SetFigFont{7}{6}{\rmdefault}{\mddefault}{\updefault}{\color[rgb]{0,0,0}DMC channel uses}%
}}}
\put(1876, 14){\makebox(0,0)[b]{\smash{\SetFigFont{7}{6}{\rmdefault}{\mddefault}{\updefault}{\color[rgb]{0,0,0}$\theta$}%
}}}
\put(3976, 14){\makebox(0,0)[b]{\smash{\SetFigFont{7}{6}{\rmdefault}{\mddefault}{\updefault}{\color[rgb]{0,0,0}$\theta$}%
}}}
\put(6076, 14){\makebox(0,0)[b]{\smash{\SetFigFont{7}{6}{\rmdefault}{\mddefault}{\updefault}{\color[rgb]{0,0,0}$\theta$}%
}}}
\put(1876,-811){\makebox(0,0)[b]{\smash{\SetFigFont{7}{6}{\rmdefault}{\mddefault}{\updefault}{\color[rgb]{0,0,0}deny}%
}}}
\put(3976,-811){\makebox(0,0)[b]{\smash{\SetFigFont{7}{6}{\rmdefault}{\mddefault}{\updefault}{\color[rgb]{0,0,0}deny}%
}}}
\put(6076,-961){\makebox(0,0)[b]{\smash{\SetFigFont{7}{6}{\rmdefault}{\mddefault}{\updefault}{\color[rgb]{0,0,0}confirm +}%
}}}
\put(751, 14){\makebox(0,0)[b]{\smash{\SetFigFont{7}{6}{\rmdefault}{\mddefault}{\updefault}{\color[rgb]{0,0,0}$c-\theta$}%
}}}
\put(2851, 14){\makebox(0,0)[b]{\smash{\SetFigFont{7}{6}{\rmdefault}{\mddefault}{\updefault}{\color[rgb]{0,0,0}$c-\theta$}%
}}}
\put(4951, 14){\makebox(0,0)[b]{\smash{\SetFigFont{7}{6}{\rmdefault}{\mddefault}{\updefault}{\color[rgb]{0,0,0}$c-\theta$}%
}}}
\put(1051,-511){\makebox(0,0)[b]{\smash{\SetFigFont{7}{6}{\rmdefault}{\mddefault}{\updefault}{\color[rgb]{0,0,0}first chunk}%
}}}
\put(3151,-511){\makebox(0,0)[b]{\smash{\SetFigFont{7}{6}{\rmdefault}{\mddefault}{\updefault}{\color[rgb]{0,0,0}second chunk}%
}}}
\put(5251,-511){\makebox(0,0)[b]{\smash{\SetFigFont{7}{6}{\rmdefault}{\mddefault}{\updefault}{\color[rgb]{0,0,0}third chunk}%
}}}
\put(751,-811){\makebox(0,0)[b]{\smash{\SetFigFont{7}{6}{\rmdefault}{\mddefault}{\updefault}{\color[rgb]{0,0,0}random increment}%
}}}
\put(2851,-811){\makebox(0,0)[b]{\smash{\SetFigFont{7}{6}{\rmdefault}{\mddefault}{\updefault}{\color[rgb]{0,0,0}random increment}%
}}}
\put(4951,-811){\makebox(0,0)[b]{\smash{\SetFigFont{7}{6}{\rmdefault}{\mddefault}{\updefault}{\color[rgb]{0,0,0}random increment}%
}}}
\end{picture}
\end{center}
\caption{One block's transmission in the channel code with
  time-sharing between the message code and punctuation code. Each
  chunk is terminated with a $\theta$-length segment to convey
  punctuation information. If the channel has zero-error capacity,
  then a zero-error block code can be used to tell the decoder whether
  to move on to the next block of the message or not. If it is to move
  on, the chunk terminator also tells which of the $2^l$ most likely
  messages was conveyed by this particular block. By making the chunk
  length $c$ long, the overhead of the control messages becomes
  asymptotically negligible since $\theta$ remains fixed if there
  is zero-error capacity. When there is no zero-error capacity, then
  $\theta$ stays proportional to $c$ and an infinite-constraint-length
  random convolutional code is used to carry punctuation
  information.}
\label{fig:zeroerrorcode}
\end{figure}

The above communication scheme is easily adapted to channels with
strictly positive zero-error capacity by just using a zero-error code
to carry the punctuation information. There is no $k$. Instead, let
$\theta$ be the block length required to realize feedback zero-error
transmission of at least $l+1$ bits. As illustrated in
Figure~\ref{fig:zeroerrorcode}, terminate each chunk with a
block-length-$\theta$ feedback zero-error code and use it to transmit
the punctuation information. If the chunk size is $c$ channel uses,
then it is as though we are operating with only a fraction $(1 -
\frac{\theta}{c})$ of the channel uses. This effectively increases the
rate to $R/(1 - \frac{\theta}{c})$ and reduces the achieved delay
exponent to $\alpha(1 - \frac{\theta}{c})$ as well. This overhead
becomes negligible by making the chunk size $c$ large giving us the
desired result.

\subsection{Delayed feedback}
Let $\phi$ be the delay in the noiseless feedback. So the encoders now
know only $Y_1^{t-\phi}$ in addition to the message bits. Everything
continues to work because the chunks $c$ can be made much longer than
$\phi$. The last $\phi - 1$ channel uses in a chunk can then
be discarded without any significant overhead.

Thus, Theorem~\ref{thm:withzeroerror} holds for any communication
system in the asymptotic limit of large end-to-end delays even if
there are small round-trip delays in the feedback. All that is
required is some way to provide infrequent, but unmistakable,
punctuation information from the encoder to the decoder. \hfill $\Box$ 

\vspace{0.20in}

\subsection{Channels without zero-error capacity: paying for punctuation} \label{sec:comments}

All that remains is to prove Theorem~\ref{thm:genericachieve}. When
the channel has no zero-error capacity, then it is still possible to
follow the Section-\ref{sec:strictzero} approach of allocating
$\theta$ channel uses per chunk to carry punctuation information. The
channel uses are partitioned as before into two streams assigned to
two sub-encoders. The first is exactly as it was in the
Section~\ref{sec:strictzero} and carries the message itself using a
variable-length channel code with the dynamic length chosen to ensure
correct list-decoding. This first encoder generates punctuation
messages at the end of every chunk and these are the input to the
second encoder. The second encoder's role is to convey this
punctuation information consisting of $l+1$ bits for every chunk.

Instead of using a zero-error code, the second encoder is implemented
using an infinite-constraint-length time-varying random convolutional
code. The trick of Appendix~\ref{app:feedbackconvolution} can be used
to reduce the expected computational burden for encoding/decoding by
using feedback, but essentially this sub-code operates without
feedback.


The decoder also runs with two subsystems. One subsystem is
responsible for decoding the punctuation stream. This can be
implemented using either an ML decoder or a sequential decoder from
\cite{JelinekSequential}. Either way, it is responsible for giving its
current best estimate for all punctuation so far. By the properties of
random infinite-constraint-length convolutional codes, this attains
the random-coding error exponent with respect to delay for every piece
of punctuation in the stream. The earlier punctuation marks are almost
certainly decoded correctly while more recent punctuation marks are
more likely to be subject to error.

This current estimate for all the punctuation so far is then used by
the subsystem responsible for decoding the message bits
themselves. The decoded punctuation is used to tentatively parse the
channel outputs into variable-length blocks and then tentatively
decode those blocks under the assumption that the punctuation is
correct. Any bits that have reached their deadlines are then
emitted. Although the decisions for those bits are now committed from
the destination's point of view, this does not prevent the system from
re-parsing them in the future when considering estimates for other
bits. 

\subsubsection{Analysis}
An error can occur at the decoder in two different ways.  As before,
the message-carrying stream could be delayed due to channel
atypicality in its own channel slots. The new source of errors is that
the punctuation stream could also become corrupted through atypicality
in these other channel slots. As a result, the punctuation overhead
$\theta$ must be kept proportional to the chunk length $c$ to avoid
having punctuation errors cause too many decoding errors.

Set $\theta = \psi c$ for a constant $\psi$ to be optimized. The rate
of the punctuation information is $\frac{l+1}{\theta} =
\frac{l+1}{\psi c} $ and goes to zero as $c \rightarrow \infty$.
Since the random-coding error exponent at rate $0$ approaches
$E_0(1)$, this is the relevant error exponent for the second stream
relative to the channel uses that it gets. But there are only $\psi$
punctuation-code channel uses per second and so the delay-exponent for
the punctuation stream is actually $\psi E_0(1)$ with respect to true
delay.

Meanwhile, the chunk size in the message stream is $c' = c
(1-\psi)$. The effective rate of the message stream is thereby
increased to $\frac{R}{1-\psi}$. Assuming that the punctuation
information is correct, the fixed-delay error-exponent is as close as
we would like to $E_{a,s}(\frac{R}{1-\psi})$ with respect to the delay
in terms of message-code channel uses. But there are only $(1-\psi)$
message-code channel uses per second and so the delay exponent
approaches $(1-\psi)E_{a,s}(\frac{R}{1-\psi})$ with respect to true
end-to-end delay.

Consider a large fixed delay $d$. It can be written as $d = d_f + d_m$
in $d$ different ways. Let $d_f$ be the part of the end-to-end delay
that is burned by errors in the punctuation stream. That is, with
probability exponentially small in $d_f$, this suffix of time has
possibly incorrect punctuation information and so cannot be trusted to
be interpreted correctly. If the bit did not make it out correctly in
the $d_m$ time steps (corresponding to $(1-\psi)d_m$ channel uses for
the message-code) where the punctuation is correct, we assume that
it will not come out correctly.

Since the channel uses are disjoint between the punctuation and message
streams, the two error events are independent. The probability of an
error with delay $d$ can thus be union-bounded as
\begin{eqnarray*}
& & {\cal P}(\widehat{B}_i(d) \neq B_i) \\
& \leq & \sum_{d_m = 1}^d {\cal P}(\mbox{message error with delay }d_m) 
{\cal P}(\mbox{punctuation error with delay }d-d_m) \\
& \leq & \sum_{d_m = 1}^d K_1 \exp(-d_m((1-\psi)E_a(\frac{R}{1-\psi}) +
\epsilon_1)) K_2 \exp(-(d-d_m)(\psi E_0(1) +
\epsilon_2)) \\
& \leq & d K_1 K_2 \exp(-d (\min\left(\psi
  E_0(1),(1-\psi)E_{a,s}(\frac{R}{1-\psi})\right) - \epsilon_1
-\epsilon_2)
\end{eqnarray*}
where $\epsilon_1,\epsilon_2$ are arbitrarily tiny constants and $K_1
K_2$ are large constants that together capture the nonasymptotic terms
in the earlier analysis.

Since the focus here is on the asymptotic error exponent with delay,
the polynomial and $\epsilon$ terms can be ignored and an achievable
exponent is found by choosing $\psi$ so that the two exponents are
balanced:
$$E' = \psi E_0(1) = (1-\psi)E_{a,s}(\frac{R}{1-\psi}).$$

Evaluating the parametric forms (\ref{eqn:symmetricfeedbackbound})
using $\eta = \rho$ for $E_{a,s}$, we get a pair of equations
\begin{eqnarray}
\psi E_0(1) & = & (1-\psi)E_0(\rho), \\
\frac{E_0(\rho)}{\rho} & = & \frac{R}{1-\psi}.
\end{eqnarray}
The first thing to notice is that simple substitution gives
$$R = \frac{(1-\psi)E_0(\rho)}{\rho} = \frac{\psi E_0(1)}{\rho} = \frac{E'}{\rho}.$$

Solving for $\psi$ shows (after a little algebra) that 
\begin{equation} \label{eqn:psidef}
\psi = \frac{E_0(\rho)}{E_0(1) + E_0(\rho)}.
\end{equation}
This way $1-\psi = \frac{E_0(1)}{E_0(1) + E_0(\rho)}$ and the first
equation is clearly true. Similarly $\frac{1}{1 - \psi} = 1 +
\frac{E_0(\rho)}{E_0(1)}$ and $\frac{\psi}{1 - \psi} =
\frac{E_0(\rho)}{E_0(1)}$ and thus the second equation is also true.
Evaluating,

\begin{eqnarray*}
E' & = & \psi E_0(1) \\
& = & \frac{E_0(\rho) E_0(1)}{E_0(1) + E_0(\rho)} \\
& = & \left(\frac{1}{E_0(\rho)} + \frac{1}{E_0(1)}\right)^{-1}.
\end{eqnarray*}

Simple (but mildly tedious) Taylor series expansion around the $\rho =
0$ point gives $E'(\rho) = 0 + C\rho + \frac{1}{2}(\frac{\partial^2
  E_0(0)}{\partial \rho^2} - 2\frac{C^2}{E_0(1)})\rho^2 +
o(\rho^2)$ and thus $R(\rho) = C  - (\frac{C^2}{E_0(1)} - \frac{1}{2}  
\frac{\partial^2 E_0(0)}{\partial \rho^2})\rho + o(\rho)$. Taking
the ratio of the first order terms gives the desired slope in the
vicinity of the $(C,0)$ point. The fact that this slope is strictly
negative is clear from the fact that $\frac{\partial^2
  E_0(0)}{\partial \rho^2} \leq 0$. \hfill $\Box$ 

\subsubsection{Computation}
As in the rate-$\frac{1}{2}$ erasure case discussed in
Section~\ref{sec:becexample}, the computational burden for the
$(n,c,l)$ schemes is a constant that depends only on the particular
scheme (and hence indirectly on the target rate-reliability pair) and
not on the target end-to-end delay. As described, the complexity is
exponential in the block length $nc$ since both the encoder and
decoder must do list decoding among the codewords. The computational
burden of the punctuation code is light since by
Appendix~\ref{app:feedbackconvolution} it is like running a sequential
decoder for a very-low-rate convolutional code.

\subsection{More examples} \label{sec:fortifiedexamples}

\begin{figure}[htbp]
\begin{center}
\includegraphics[width=4in,height=3in]{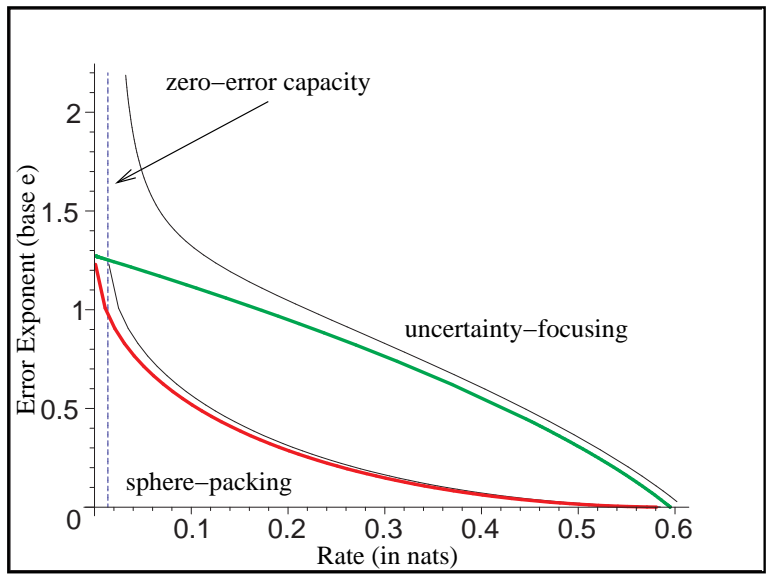}
\end{center}
\caption{The sphere-packing and uncertainty-focusing bounds, with and
  without a noiseless side-channel of rate $\frac{1}{50}$ for a
  BSC with crossover probability $0.02$. The lower curves are the
  sphere-packing bounds and the upper curves are the
  uncertainty-focusing bounds. The thin lines represent the fortified 
  cases with the added noiseless side-channel.}
\label{fig:fortifiedbsc}
\end{figure}

Rather than considering an example using a DMC with strictly positive
zero-error capacity, it is more instructive to consider a BSC with a
fortification side-channel of rate $\frac{1}{50}$ bits per channel
use. The capacity of the BSC with crossover probability $0.02$
increases to $0.61$ nats with such fortification and the Burnashev
bound becomes infinite. Figure~\ref{fig:fortifiedbsc} shows the effect
of zero-error capacity on the sphere-packing and uncertainty-focusing
bounds. At high rates, the fortified uncertainty-focusing bound looks
like it has just been shifted in rate by $0.01$ nats, just like the
fortified sphere-packing bound. However, because of the flatness of
the classical sphere-packing bound at high rates, the sphere-packing
bound visually appears unchanged by fortification on a plot. At very
low rates, the two behave differently. The fortified
uncertainty-focusing bound tends smoothly to infinity at $0.01$ nats
while the fortified sphere-packing bound jumps abruptly to infinity,
reflecting the typical behavior of the error exponent curves for
channels with strictly positive zero-error capacity.

\begin{figure}[htbp]
\begin{center}
\includegraphics[width=4in,height=3in]{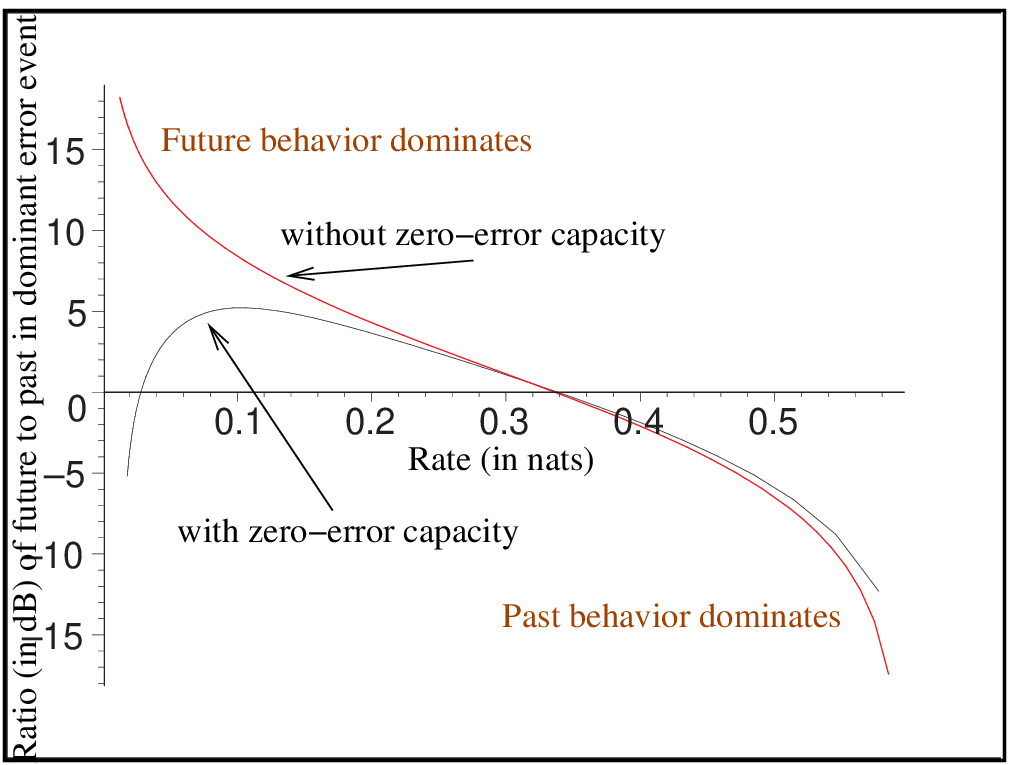}
\end{center}
\caption{The dominant error events illustrated by plotting the ratio
  of future to past in dB scale. The horizontal axis is rate and the
  vertical axis is $10\log_{10}\frac{1-\lambda^*}{\lambda^*}$ where
  $\lambda^*$ is from (\ref{eqn:optimumlambda}). The thicker red curve
  represents the unfortified channel while the thin black curve is the
  $\frac{1}{50}$-fortified system.} 
\label{fig:pastfuturedb}
\end{figure}
Looking at a deeper level of detail, Figure~\ref{fig:pastfuturedb}
illustrates the time-nature of the dominant error events at different
rates. The question is for how long does the channel behave atypically
for a bit to miss its deadline. In fixed-block-length coding, the
usual source of errors is slightly atypical behavior across the entire
block. As shown in Section~\ref{sec:nofeedback}, when feedback is not
available, the usual errors mainly involve the channel behaving
atypically {\em after} the bit in question arrived at the encoder.

By contrast, in the fixed-delay context with feedback, the dominant
error events involve more and more of the past as the rates get
large. This means that the typical way for a bit to miss its deadline
is for the channel to have been behaving atypically for some time
before the bit even arrived at the encoder, and for this atypical
behavior to continue till the deadline.  At intermediate rates, the
future behavior (after the bit has arrived at the encoder) becomes
more important since it is more likely for the channel to become very
bad for a shorter period.

At very low rates, the fortified and unfortified systems exhibit
qualitatively different behavior. For unfortified systems, the
dominant error events soon involve essentially only the future. The
dominant event approaches the channel going into complete ``outage''
(e.g.~the channel flipping half the inputs of a BSC) after the bit
arrives at the encoder. For systems with positive zero-error capacity,
such a complete outage is not possible as the message bits can always
dribble across the error-free part. For an error to occur, it is
essential to build up a large enough backlog in the queue and thus the
past behavior starts to become dominant again. The curves diverge for
the same rates at which the fortified case's uncertainty-focusing
bound is much better than the unfortified case.

\begin{figure}[htbp]
\begin{center}
\includegraphics[width=4in,height=3in]{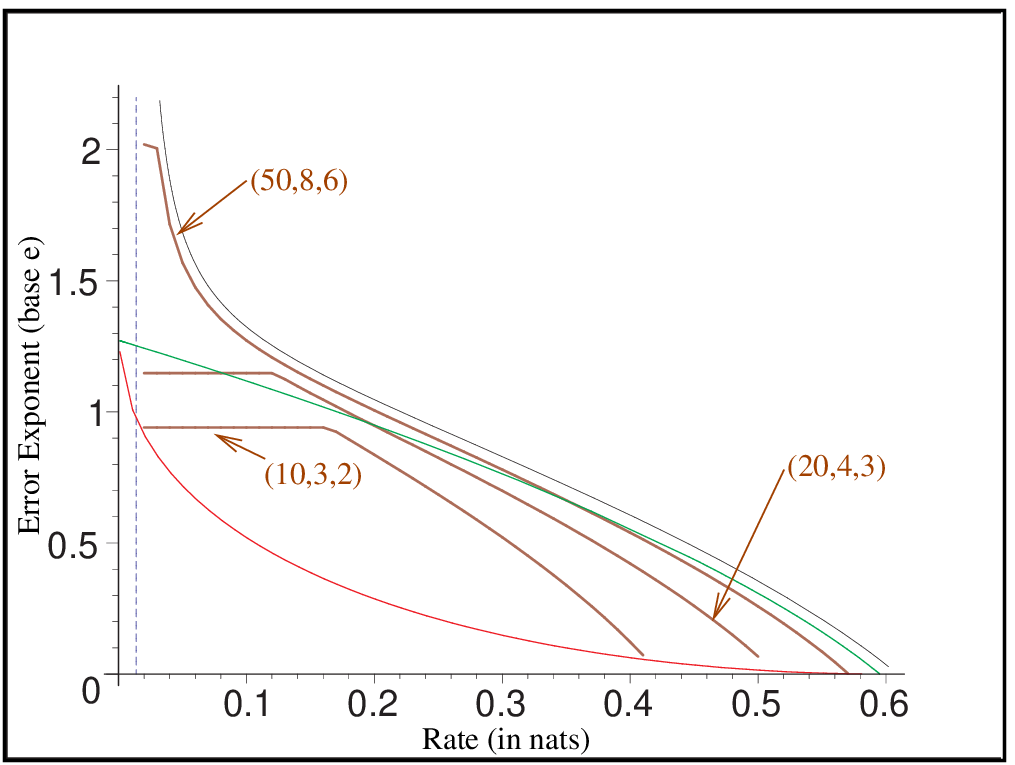}
\end{center}
\caption{The fixed-delay error exponents of different schemes for a
  $\frac{1}{50}$-fortified BSC with crossover probability $0.02$ used
  with noiseless feedback. The lowest curve is the sphere-packing
  bound limiting feedback-free performance. The three new curves
  represent what is attained by the $(10,3,2), (20,4,3), (50,8,6)$
  schemes described in Section~\ref{sec:nclscheme} and vary by
  block length, granularity, and the size of the lists used for
  list decoding. The uncertainty-focusing bound with and without
  fortification is plotted for reference.}
\label{fig:fortifiedperformance}
\end{figure}

Figure~\ref{fig:fortifiedperformance} shows the fixed-delay
reliabilities achieved by the $(n,c,l)$ schemes\footnote{The schemes
  plotted here have been slightly modified to use the noiseless
  side-channel to carry codeword information whenever it is not needed
  to carry punctuation information. This more accurately reflects the
  typical behavior of channels with strictly positive zero-error
  capacity.} of Section~\ref{sec:nclscheme} for the specific cases of
$(10,3,2)$, $(20,4,3)$, $(50,8,6)$. These are delay universal since
they hold with all sufficiently long delays. Increasing $l$ increases
the list size and helps the low rate performance while large block
lengths $n$ are needed to perform well at higher rates. It is
interesting to see how how the $(10,3,2)$ scheme is already
spectacularly better than the feedback-free case for all low to
moderate rates. In this case, there are $10*3*50 = 1500$ BSC uses and
only $10*3 = 30$ error-free control bits corresponding to a typical
message block.

\section{Conclusions}

This paper has shown that fixed-block-length and fixed-delay systems
behave very differently when feedback is allowed. While
fixed-block-length systems do not usually gain substantially in
reliability with noiseless feedback, fixed-delay systems can achieve
very substantial gains for any generic DMC. The uncertainty-focusing
bound complements the classical sphere-packing bound and gives limits
to what is possible. Furthermore, these limits can be approached in a
delay-universal fashion for erasure channels and any channel with
positive feedback zero-error capacity if the encoders have access to
noiseless channel output feedback, even if that feedback is slightly
delayed. The computational requirements in doing so do not scale with
the desired probability of error and only depend on the target rate
and delay exponent. The details of this work establish a connection
between queuing and communication over noisy channels with
feedback. For the constructions given here, the end-to-end delay is
asymptotically dominated by time spent waiting in a queue.

Given that complete noiseless feedback now has unambiguously clear
value for reliable communication, it is important for the community to
explore the required quality of feedback. This paper only shows that
slightly delayed feedback can be tolerated. The case of noisy or
rate-constrained feedback in the fixed-delay context is almost
entirely open (see \cite{BalancedForwardFeedback} for the case of
erasure channels on both the forward and feedback links). In addition,
both the upper and lower bounds here only cover the case of a single
message stream. The multistream rate/reliability region is still
unknown even for the BEC case \cite{ControlPartII}.

Stepping back, these results are also interesting because they show
how feedback changes the qualitative nature of the dominant error
events. Without feedback, errors are dominated by future channel
behavior, but when feedback is available, the dominant event involves
a mixture of the past and future. When the rate is low, the future
tends to be more important but when the rate is high, the past starts
to dominate. This brings to mind Shannon's intriguing comment at the
close of \cite{ShannonLossy}:
\begin{quotation}
 [The duality between source and channel coding] can be pursued
 further and is related to a duality between past and future and the
 notions of control and knowledge. Thus we may have knowledge of the
 past and cannot control it; we may control the future but have no
 knowledge of it. 
\end{quotation}

In \cite{OurSourceCodingBound}, we explore the source-coding analogs
of the results given here. In particular, feedback is found to be
irrelevant in point-to-point lossless source coding and the dominant
error events involve only the past! That makes precise the duality
hinted at by Shannon.

Finally, in \cite{WaterslidePaper}, the techniques developed here are
extended to lower-bound the complexity of decoding based on iterative
message-passing for general codes. The linear concept of time here is
generalized to the message-passing graph. The role of delay is thus
played by the decoding neighborhood within the graph and the
corresponding bounds reveal the complexity cost of approaching
capacity with such decoding algorithms.

\section*{Acknowledgments}
The author thanks his student Tunc Simsek for many productive
discussions. This work builds on the line of investigation that we
opened up in Tunc's doctoral thesis \cite{TuncThesis}, although the
proofs are different. I also thank Pravin Varaiya for his support,
Sanjoy Mitter and Nicola Elia for many discussions over a long period
of time which influenced this work in important ways, and the Berkeley
students in the Fall 2004 advanced information theory course who
forced me to simplify the presentation considerably. The anonymous
reviewers are also thanked for their very helpful comments.

\appendices 

\section{Feedback, convolutional codes, and
  complexity} \label{app:feedbackconvolution} The encoder is allowed
to ``look over the shoulder'' of the decoder and have access to
noiseless feedback of the channel outputs. This appendix
gives\footnote{The scheme we describe in this subsection is too
  obvious to be original to us, but we are unaware of who might have
  come up with it earlier.} the convolutional parallel to the
Burnashev problem of variable-block-length codes. For ease of
exposition, suppose the channel is binary input and that the uniform
distribution is an optimal input distribution. If another input
distribution is desired, mappings in the style of Figure~6.2.1 of
\cite{gallager} can be used to approximate the desired channel input
distribution.  Use $R' = \frac{R}{\ln 2}$ to refer to the input rate
in bits per channel use rather than nats per channel use. Apply the
``encode the error signals'' advice of \cite{ooiwornell} to get the
following simple construction of a random code:

\begin{itemize}
 \item Start with an infinite-constraint-length random time-varying 
       convolutional code. The $j$-th channel input $X_j = \sum_k
       H_k(j)B_k \bmod 2$ is generated by correlating the input
       bits $B_1^{jR'}$ with a random binary string $H_1^{jR'}(j)$.

 \item Use the noiseless feedback to run a sequential decoder at the
       encoder. This gives the encoder access to
       $\widehat{B}_1^{(j-1)R'}(j-1)$ --- the tentative estimates of
       the past input bits based on the channel outputs so far. Set
       $\widehat{B}_{jR'}(j-1) = 0$ since there is no estimate for the
       new bit, and then compute $\widetilde{B}_k(j) = B_k +
       \widehat{B}_k(j-1) \bmod 2$ to represent the current error
       sequence. Since the probability of bit error is exponentially
       decreasing in delay \cite{JelinekSequential}, only a small
       number of the $\widetilde{B}_k(j)$ are nonzero, and
       furthermore, these are all around the more recent bits. The
       expected number of nonzero error bits is therefore upper
       bounded by some constant.

 \item Run the infinite-constraint-length convolutional code using the
       error sequence rather than the input bits.  $\widetilde{X}_j =
       \sum_k H_k(j) \widetilde{B}_k(j) \bmod 2 = X_j + [\sum_k H_k(j)
       \widehat{B}_k(j-1)] \bmod 2$. Input the resulting $\widetilde{X}_j$
       into the channel. 

       Since the additional term $[\cdots]$ is entirely known at the
       receiver and modulo $2$ addition is invertible, this feedback
       code has exactly the same distance properties as the original
       code without feedback. Furthermore, since there are only a
       finite random number of nonzero error bits and the encoder
       knows where these are, the encoding complexity is a random
       variable with finite expectation.
\end{itemize}

If a block-code is desired, then pick an arbitrary length $d$ to
terminate a block, and choose an overall block length $n$ so that $d$
is insignificant in comparison.

The expected per-channel-input constraint-length used by the code is a
finite constant that only depends on the rate, while the overall
probability of block error dies exponentially with the terminator
length $d$. Consequently, the expected-constraint-length
error-exponent for variable-constraint-length convolutional codes is
infinite with noiseless feedback. If we also count the expected number
of computations required to run the encoder's copy of the decoder,
then this result holds for all rates strictly below\footnote{At depth
  $\tau$ within a false path, each node expansion for a sequential
  decoder requires $O(\tau)$ multiply-accumulate operations to
  evaluate. This polynomial-order term is insignificant when compared
  to the rate-dependent exponential increase in the number of false
  nodes with increasing search depth. Thus, the polynomial term can be
  bounded away by just treating it as slight increase in the rate.}
the computational cutoff rate $E_0(1)$. Even though noiseless feedback
is used by the encoder to generate the channel inputs, the decoding is
``sequential'' in the sense of Jacobs and Berlekamp
\cite{JacobsBerlekamp} and suffers from the resulting computational
limitation of having a search-effort distribution with certain
unbounded moments.

At rates above $E_0(1)$, the same flavor of result can be preserved in
principle by using the concatenated-coding transformations of Pinsker
\cite{PinskerComplexity} (as well as others described more recently by
Arikan \cite{Arikan05}) to bring the computational-cutoff rate
$E_0(1)$ as close to $C$ as desired. Thus, {\em the
  expected-computation error exponent for convolutional-style codes
  with noiseless output feedback can be made essentially infinite at
  all rates below capacity.} The expected complexity is a constant
that depends only on the desired rate, not on the target probability
of error.

\section{Extended Proofs} 
\subsection{Lemma \ref{lem:newdataprocessing}} \label{app:lemdataprocessing}

\begin{eqnarray*}
n (R - \delta_1) & = & H(B_1^{n (R' - \delta_1')}) \\
& = & I(B_1^{n (R' - \delta_1')} ; B_1^{n(R' - \delta_1')}) \\
& \leq_{(a)} & I(B_1^{n(R' - \delta_1')} ; \widetilde{B}_1^{n(R' - \delta_1')},
Y_1^{n}) \\
& = & H(Y_1^{n}) + H(\widetilde{B}_1^{n(R' - \delta_1')}|Y_1^{n}) 
     -H(Y_1^{n}|B_1^{n(R' - \delta_1')}) 
     -H(\widetilde{B}_1^{n(R' - \delta_1')}|Y_1^{n},B_1^{n(R' -
     \delta_1')}) \\
& =_{(b)} & H(Y_1^{n}) + H(\widetilde{B}_1^{n(R' - \delta_1')}|Y_1^{n}) 
     -H(Y_1^{n}|B_1^{n(R' - \delta_1')}) \\
& = & H(\widetilde{B}_1^{n(R' - \delta_1')}|Y_1^{n}) 
     + I(Y_1^{n};B_1^{n(R' - \delta_1')}) \\
& \leq_{(c)} & H(\widetilde{B}_1^{n(R' - \delta_1')}) + I(Y_1^{n};B_1^{n(R' -
      \delta_1')}) \\
& \leq_{(d)} & H(\widetilde{B}_1^{n(R' - \delta_1')}) + I(X_1^{n};Y_1^{n}).
\end{eqnarray*}
The first equality holds because the message bits are fair coin
tosses. (a) comes from the data processing inequality when considering
the following trivial Markov chain:
$B_1^{n(R'-\delta_1')}~-~(\widetilde{B}_1^{n(R' - \delta_1')},
Y_1^n)~-~B_1^{n(R'-\delta_1')}$ that comes from the fact that the
channel outputs and the error signals are enough to reconstruct the
original bits. After expanding in terms of entropies, the
$H(\widetilde{B}_1^{n(R' - \delta_1')}|Y_1^{n},B_1^{n(R' -
  \delta_1')})$ term can be dropped to give (b) since this conditional
entropy is zero because the error signal can be reconstructed from the
message bits and the channel outputs. (c) comes from dropping
conditioning, while the final inequality (d) comes from applying the data
processing inequality to the Markov chain $B_1^{n(R'-\delta_1')}~-~
X_1^{n} ~-~ Y_1^n$ capturing the lack of feedback in the
system. \hfill $\Diamond$ \vspace{0.15in}

\subsection{Lemma
  \ref{lem:measurechange}} \label{app:lemmeasurechange}
Before proving Lemma~\ref{lem:measurechange}, it is useful to
establish a result involving typical sets. 
\subsubsection{Typical set lemma}
\begin{lemma} \label{lem:typicalsequenceprob}
For every finite DMC $G$ and $\epsilon_1, \epsilon_2 > 0$,
there exists a constant $K$ such that for every $\vec{x}$ 
\begin{equation} \label{eqn:smallprob}
{\cal P}_G(\vec{Y} \in J_{\vec{x}}^{\epsilon_1, \epsilon_2} | \vec{X} =
\vec{x}) \geq 1 - |{\cal X}||{\cal Y}| \exp(-K d)
\end{equation}
where $d$ is the length of the vectors $\vec{x}, \vec{Y}$, and the
appropriate typical set is
\begin{equation} \label{eqn:Jdef}
J_{\vec{x}}^{\epsilon_1, \epsilon_2} = \left\{\vec{y} | \forall x \in {\cal
X} \mbox{ either }\frac{n_x(\vec{x})}{d} < \epsilon_2 \mbox{ or } \forall y \in {\cal Y},
\frac{n_{x,y}(\vec{x},\vec{y})}{n_x(\vec{x})} \in (g_{y|x} -
\epsilon_1, g_{y|x} + \epsilon_1) \right\}
\end{equation}
where $n_{x,y}(\vec{x},\vec{y})$ is the count of how many times
$(x,y)$ occurs in the sequence $(x_1, y_1), (x_2, y_2),
\ldots, (x_d, y_d)$, and $n_{x}(\vec{x})$ is the count of how many $x$
are present in the $d$ length vector $\vec{x}$.

Furthermore, for any $\vec{y} \in J_{\vec{x}}^{\epsilon_1,
\epsilon_2}$, the probability of the sequence $\vec{y}$
under a different channel $P$ satisfies
\begin{equation} \label{eqn:typicalratio}
\frac{{\cal P}_P(Y_1^d = \vec{y} | X_1^d = \vec{x})}{{\cal P}_G(Y_1^d
  = \vec{y} | X_1^d = \vec{x})} 
\geq
\exp(-d \left[D\bigg(G || P | \vec{r}(\vec{x})\bigg) + 
(2\epsilon_2 + \epsilon_1) \sum_{x,y : n_x(\vec{x}) \neq 0, g_{y|x} \neq 0}
|\ln\frac{g_{y|x}}{p_{y|x}}|\right])
\end{equation}
where $\vec{r}(\vec{x})$ is the type of $\vec{x}$. In particular,
\begin{equation} \label{eqn:typicalratioworst}
\frac{{\cal P}_P(Y_1^d = \vec{y} | X_1^d = \vec{x})}{{\cal P}_G(Y_1^d = \vec{y} | X_1^d
= \vec{x})} 
\geq
\exp(-d [\max_{\vec{r}} D(G || P | \vec{r}) + 
(2\epsilon_2 + \epsilon_1) \sum_{x,y : n_x(\vec{x}) \neq 0, g_{y|x} \neq 0}
|\ln\frac{g_{y|x}}{p_{y|x}}|]).
\end{equation}
\end{lemma} 

{\em Proof: } The first goal is to establish
(\ref{eqn:smallprob}). For every $x \in {\cal X}$, the weak law of
large numbers for iid finite random variables says that the relative
frequency of $y$'s will concentrate around $g_{y|x} \pm
\epsilon_1$. Simple Chernoff bounds for the Bernoulli random variables
representing the indicator functions tell us that this convergence is
exponentially fast in that $\forall (x,y) \exists \zeta_{x,y} > 0$ so
that if the channel input is always $x$ for a length $d$, the random
number $N_y$ of times the channel output is $y$ satisfies
$${\cal P}(|\frac{N_y}{d} - g_{y|x}| \geq \epsilon_1 |X_1^d = x_1^d)
\leq \exp(-\zeta_{x,y} d).$$ 

Let $K' = \min_{x \in {\cal X}, y \in {\cal Y}} \zeta_{x,y} > 0$. Set
$K = \epsilon_2 K'$ since there are at least $\epsilon_2 d$
occurrences of the relevant $x$ values. Finally, apply the union bound
over all $|{\cal X}||{\cal Y}|$ possible pairs to get
(\ref{eqn:smallprob}).

To show (\ref{eqn:typicalratio}), first note that those $(x,y)$ pairs
for which $g_{y|x} = 0$ can be ignored since these cannot occur in
any sequence with nonzero probability under $G$. Then
\begin{eqnarray*}
{\cal P}_G(Y_1^d = \vec{y} | X_1^d = \vec{x}) 
& = & 
\prod_{i=1}^d g_{y_i|x_i} \\
& = & 
\prod_{x \in {\cal X}, y \in {\cal Y}} g_{y|x}^{n_{x,y}(\vec{x},\vec{y})} \\
& = & 
(\prod_{x \in {\cal X}, y \in {\cal Y}} 
g_{y|x}^{\frac{n_{x,y}(\vec{x},\vec{y})}{d}})^d \\
& = & \exp(d \sum_{x \in {\cal X}, y \in {\cal Y}} \frac{n_{x,y}(\vec{x},\vec{y})}{d}
\ln g_{y|x}) \\
& = & \exp(d \sum_{x \in {\cal X}} \frac{n_x(\vec{x})}{d} \sum_{y \in {\cal Y}}
\frac{n_{x,y}(\vec{x},\vec{y})}{n_x(\vec{x})}\ln g_{y|x}).
\end{eqnarray*}
Similarly
$${\cal P}_P(Y_1^d = \vec{y} | X_1^d = \vec{x}) = 
\exp(d \sum_{x \in {\cal X}} \frac{n_x(\vec{x})}{d} \sum_{y \in {\cal Y}}
\frac{n_{x,y}(\vec{x},\vec{y})}{n_x(\vec{x})}\ln p_{y|x}).$$
The ratio of the two probabilities is thus
$$\frac{{\cal P}_P(Y_1^d = \vec{y} | X_1^d = \vec{x})}{{\cal P}_G(Y_1^d = \vec{y} |
X_1^d = \vec{x})} = \exp(-d \sum_{x \in {\cal X}} r_x(\vec{x}) \sum_{y
\in {\cal Y}} \frac{n_{x,y}(\vec{x},\vec{y})}{n_x(\vec{x})}\ln
\frac{g_{y|x}}{p_{y|x}}).$$ 

Now apply the definition of $J_{\vec{x}}^{\epsilon_1,
\epsilon_2}$ and first bound the contribution to the exponent by those
inputs $x \in {\cal X}^{rare}$ that occur too rarely:
$n_x(\vec{x}) < \epsilon_2 d$. We drop the arguments of $(\vec{x},
\vec{y})$ when they are obvious from context.
\begin{eqnarray*}
\sum_{x \in {\cal X}^{rare}} r_x \sum_{y \in {\cal Y}}
\frac{n_{x,y}}{r_x d}\ln \frac{g_{y|x}}{p_{y|x}} 
& \leq &
\sum_{x \in {\cal X}^{rare}} \left[r_x \sum_{y \in {\cal Y}}
g_{y|x} \ln \frac{g_{y|x}}{p_{y|x}} 
+ 2\epsilon_2 \sum_{y \in {\cal Y},
g_{y|x} \neq 0} |\ln\frac{g_{y|x}}{p_{y|x}}|\right] \\
& \leq &
2\epsilon_2 \sum_{x,y : r_x \neq 0, g_{y|x} \neq 0} |\ln\frac{g_{y|x}}{p_{y|x}}| + 
\sum_{x \in {\cal X}^{rare}} r_x \sum_{y \in {\cal Y}}
g_{y|x} \ln \frac{g_{y|x}}{p_{y|x}}.
\end{eqnarray*}
For the non-rare $x$, the $n_{x,y}$ are already within $\epsilon_1$ of
$g_{y|x}$ and thus, for $y_1^d \in J_{\vec{x}}^{\epsilon_1,
\epsilon_2}$, 
\begin{eqnarray*}
\frac{{\cal P}_P(Y_1^d = y_1^d | X_1^d = x_1^d)}{{\cal P}_G(Y_1^d = y_1^d | X_1^d
= x_1^d)} 
& = & \exp(-d \sum_{x \in {\cal X}} r_x \sum_{y \in {\cal Y}}
\frac{n_{x,y}}{r_x d}\ln \frac{g_{y|x}}{p_{y|x}}) \\
& \geq & 
\exp(-d [\sum_{x \in {\cal X}} r_x \sum_{y \in {\cal Y}}
g_{y|x} \ln \frac{g_{y|x}}{p_{y|x}} + 
(2\epsilon_2 + \epsilon_1) \sum_{x,y : r_x \neq 0, g_{y|x} \neq 0}
|\ln\frac{g_{y|x}}{p_{y|x}}|] \\
& = & 
\exp(-d [D(G || P | \vec{r}) + 
(2\epsilon_2 + \epsilon_1) \sum_{x,y : r_x \neq 0, g_{y|x} \neq 0}
|\ln\frac{g_{y|x}}{p_{y|x}}|]
\end{eqnarray*}
which establishes (\ref{eqn:typicalratio}). To get
(\ref{eqn:typicalratioworst}), just bound by the worst possible
$\vec{r}$.  \hfill $\Diamond$ \vspace{0.15in}

\subsubsection{Proof of Lemma~\ref{lem:measurechange} itself}
If $g_{y|x} \neq 0$, then it is safe to assume $p_{y|x}
\neq 0$ as well since otherwise the divergence is infinite and the
Lemma is trivially true. 

The finite sum $\sum_{x,y : r_x \neq 0, g_{y|x} \neq 0}
|\ln\frac{g_{y|x}}{p_{y|x}}|$ is thus just some finite constant $K'$
that depends only on $G$ and $P$. By choosing $\epsilon_1, \epsilon_2$
small enough, it is possible to satisfy $(2\epsilon_2 + \epsilon_1)
\sum_{x,y : r_x \neq 0, g_{y|x} \neq 0} |\ln\frac{g_{y|x}}{p_{y|x}}| <
\epsilon$.

The event $A$ has a substantial conditional probability $\delta$ when
channel $G$ is used and this probability does not diminish with
$d$. Consequently, Lemma~\ref{lem:typicalsequenceprob} implies that
for the chosen $\epsilon_1, \epsilon_2 > 0$, there exists a constant
$K$ so that ${\cal P}_G(\vec{Y} \in J_{\vec{x}}^{\epsilon_1,
  \epsilon_2} | \vec{X} = \vec{x}) \geq 1 - |{\cal X}||{\cal Y}|
\exp(-K d)$. 

Pick a $d_0(G,\delta, \epsilon_1,\epsilon_2) > 0$ large enough so that
$|{\cal X}||{\cal Y}| \exp(-K d_0) < \frac{\delta}{2}$. Thus ${\cal
  P}_G(A \cap J_{\vec{x}}^{\epsilon_1, \epsilon_2} | \vec{X} =
\vec{x}) \geq \frac{\delta}{2}$. The immediate application of the
second part of Lemma~\ref{lem:typicalsequenceprob} gives
\begin{eqnarray*}
{\cal P}_P(A | \vec{X} = \vec{x}) & \geq & {\cal P}_P(A \cap
J_{\vec{x}}^{\epsilon_1, \epsilon_2} | \vec{X} = \vec{x}) \\
       & \geq &
\frac{\delta}{2} \exp\left(-d [D(G || P | \vec{r}(\vec{x})) + 
(2\epsilon_2 + \epsilon_1) \sum_{x,y : n_x(\vec{x}) \neq 0, g_{y|x} \neq 0}
|\ln\frac{g_{y|x}}{p_{y|x}}|]\right) \\
       & \geq & 
\frac{\delta}{2} \exp(-d [D(G || P | \vec{r}(\vec{x})) + \epsilon]) 
\end{eqnarray*}
which is the desired result. \hfill $\Diamond$ \vspace{0.15in}

\subsection{Expressing the symmetric uncertainty-focusing bound in parametric form} \label{app:focussymmetric}
\begin{eqnarray*}
E_{a,s}(R)  & = & 
\inf_{0 \leq \lambda < 1} \frac{E^+(\lambda R)}{1-\lambda} \\
& = & \inf_{0 \leq \lambda < 1} \max_{\rho \geq 0} \frac{E_0(\rho) - \rho
\lambda R}{1-\lambda}.
\end{eqnarray*}
To find the minimizing $\lambda$, first observe that given
$\lambda$, the maximizing $\rho$ is the solution to 
\begin{equation} \label{eqn:rhosolve}
\frac{\partial E_0(\rho)}{\partial \rho} = \lambda R.
\end{equation}
A solution exists because $E_0$ is concave $\cap$ \cite{gallager}. If the
solution is not unique, just pick the smallest solution. Call this
solution to (\ref{eqn:rhosolve}) as $\rho(\lambda,R)$. Let
$$g(\lambda,R) = E_0(\rho(\lambda,R)) - \rho(\lambda,R)\lambda R.$$
Now, the goal is to minimize $\frac{g(\lambda,R)}{1 - \lambda}$ with
respect to $\lambda$. Take a derivative and set it to zero:
$$g(\lambda,R) + (1 - \lambda) \frac{\partial g(\lambda,R)}{\partial
\lambda} = 0.$$
But
\begin{eqnarray*}
\frac{\partial g(\lambda,R)}{\partial \lambda}
& = & 
\frac{\partial E_0(\rho(\lambda,R))}{\partial \rho} 
\frac{\partial \rho(\lambda,R)}{\partial \lambda} 
- 
\lambda R 
\frac{\partial \rho(\lambda,R)}{\partial \lambda} 
 - 
\rho(\lambda,R) R \\
& = & \frac{\partial \rho(\lambda,R)}{\partial \lambda} (
\frac{\partial E_0(\rho(\lambda,R))}{\partial \rho} - 
\lambda R) - \rho(\lambda,R) R  \\
& = & - \rho(\lambda,R) R.
\end{eqnarray*}
So, solve for $\lambda^*$ in
$$ g(\lambda^*,R) = \rho(\lambda^*,R) R (1 - \lambda^*).$$
Plugging in the definition of $g$ gives
$$\frac{E_0(\rho(\lambda^*,R))}{\rho(\lambda^*,R) R} - \lambda^* =
1 - \lambda^*.$$
Which implies
$$\frac{E_0(\rho(\lambda^*,R))}{\rho(\lambda^*,R) R} = 1$$
or $R = \frac{E_0(\rho(\lambda^*,R))}{\rho(\lambda^*,R)}$. For the
other part, just notice
\begin{eqnarray*}
E_a(R) & = & \frac{1}{1 - \lambda^*} g(\lambda^*,R) \\
& = & \frac{1}{1 - \lambda^*} \rho(\lambda^*,R) R (1 - \lambda^*) \\
& = & \rho(\lambda^*,R) R \\
& = & E_0(\rho(\lambda^*,R)).
\end{eqnarray*}
Setting $\eta = \rho(\lambda^*,R)$ gives 
(\ref{eqn:symmetricfeedbackbound}).  \hfill $\Box$ \vspace{0.20in}

\subsection{Proof of the low-rate approximation in
  Theorem~\ref{thm:erasurecase}} \label{app:erasurelowrate}

First, solve for $2^\eta$ in (\ref{eqn:becreliability}) in terms of
the reliability $E_{a}^{bec}(R') = \alpha$. This gives $2^{\eta} =
\frac{1-\beta}{2^{-\alpha} - \beta}$ and so $\eta = \alpha +
\log_2({1-\beta}{1 - 2^\alpha \beta})$. Plugging into the $R'$
expression (\ref{eqn:becreliability}) gives the desired $C'(\alpha)$
tradeoff.

It is worthwhile to investigate the behavior of this $C'(\alpha)$ for
values of reliability $\alpha$ close (within a factor of 2) to the
fundamental upper limit of $-\log_2 \beta$. Consider $0 \leq \epsilon
\leq -\frac{\log_2 \beta}{2}$.  When $\alpha = (-\log_2 \beta -
\epsilon)$,  
\begin{eqnarray*} 
C'(-\log_2 \beta - \epsilon) 
& = & \frac{-\log_2 \beta - \epsilon}{-\log_2 \beta - \epsilon +
\log_2 (1-\beta) - \log_2(1-\beta 2^{-\epsilon - \log_2 \beta})} \\
& = & \left(1 + \frac{\log_2 (1-\beta) - \log_2(1-
    2^{-\epsilon})}{-\log_2 \beta - \epsilon}\right)^{-1}.
\end{eqnarray*}
$(1-2^{-\epsilon})$ is a concave $\cap$ function of $\epsilon\in[0,1]$
and can thus be lower-bounded by $\frac{1}{2} \epsilon$. This gives
\begin{eqnarray*} 
C'(-\log_2 \beta - \epsilon) 
& \geq & \left(1 + 
\frac{\log_2 (1-\beta) + \log_2(2 \epsilon^{-1})}{-\log_2 \beta - \epsilon}\right)^{-1} \\
& > & \left(1 +
\frac{\log_2(2 \epsilon^{-1})}{\log_2(\beta^{-1}) -
  \epsilon}\right)^{-1} \\
& \geq & \left(1 +
2\frac{\log_2(2 \epsilon^{-1})}{\log_2(\beta^{-1})}\right)^{-1} .
\end{eqnarray*}
Plugging in $\epsilon = 2 \beta^r$ is valid as long as 
$r \geq \frac{2 - \log_2 \log_2 \beta^{-1}}{\log_2
  \beta^{-1}}$. This gives
\begin{equation} \label{eqn:smallRbehaviour}
C'\left((-\log_2 \beta) - 2\beta^r\right) \geq \frac{1}{1+2r}.
\end{equation}

\subsection{Proof of
  Lemma~\ref{lem:notthereyet}} \label{app:lemnotthereyet}

An error can occur only when there have not been enough successful
transmissions to get the $i$-th bit out in time. Applying the union
bound to such events gives
\begin{eqnarray*}
{\cal P}(\widehat{B}_i(\lceil \frac{i}{R'} \rceil + d) \neq B_i)
& \leq & 
\sum_{k=1}^i {\cal P}\left(\sum_{t=\lceil \frac{k}{R'} \rceil}^{\lceil \frac{i}{R'}
\rceil + d} Z_t \leq i-k\right) \\
& = & \sum_{k=1}^i {\cal P}(\frac{\sum_{t=\lceil \frac{k}{R'} \rceil}^{\lceil
\frac{i}{R'} \rceil + d} Z_t}{d + \lceil \frac{i}{R'} \rceil - \lceil
\frac{k}{R'} \rceil} \leq \frac{i-k}{d + \lceil \frac{i}{R'} \rceil - \lceil
\frac{k}{R'} \rceil}).
\end{eqnarray*}
This establishes the desired result.\hfill  $\Diamond$ \vspace{0.15in}

\subsection{The details in the proof of
  Theorem~\ref{thm:erasurecase}} \label{app:erasuredetails}

Notice that the event $\{\frac{\sum_{t=\lceil \frac{k}{R'}
\rceil}^{\lceil \frac{i}{R'} \rceil + d} Z_t}{d + \lceil \frac{i}{R'}
\rceil - \lceil \frac{k}{R'} \rceil} \leq \frac{i-k}{d + \lceil
\frac{i}{R'} \rceil - \lceil \frac{k}{R'} \rceil} \}$ is just the error
event for an ideal erasure-channel block code with block length 
$n(k) = d + \lceil \frac{i}{R'} \rceil - \lceil \frac{k}{R'} \rceil$ and
a bit rate of $R'(k) = \frac{i-k + 1}{d + \lceil \frac{i}{R'} \rceil - \lceil
\frac{k}{R'} \rceil}$. This is because it represents the event that
the channel erases too many symbols. Let $\lambda(k) = 1 - \frac{d}{d + \lceil
\frac{i}{R'} \rceil - \lceil \frac{k}{R'} \rceil}$. Then $n(k) = \lambda(k)
n(k) + d$ and $R'(k) \in (\lambda(k) R', \lambda(k) R' +
\frac{2}{n(k)})$. Thus, for every $\epsilon_1 > 0$, there
exists a $d_1(\epsilon_1)$ so that for all $d > d_1(\epsilon_1)$,
\begin{eqnarray*}
{\cal P}(\frac{\sum_{t=\lceil \frac{k}{R'}
\rceil}^{\lceil \frac{i}{R'} \rceil + d} Z_t}{d + \lceil \frac{i}{R'}
\rceil - \lceil \frac{k}{R'} \rceil} \leq \frac{i-k}{d + \lceil
\frac{i}{R'} \rceil - \lceil \frac{k}{R'} \rceil}) 
& \leq &
\exp(-n(k) \left(D(R'(k) || 1 - \beta) - \epsilon_1\right)) \\
& < &
\exp(-n(k) \left(D(\lambda(k)R' + \frac{2}{n(k)} || 1 - \beta) - \epsilon_1\right)) \\
& = &
\exp(-d \left[\frac{D(\lambda(k)R' + \frac{2}{n(k)} || 1 - \beta) - \epsilon_1}{1-\lambda(k)}\right]).
\end{eqnarray*}

Now, divide the events in (\ref{eqn:sumbound}) into two categories
(illustrated in Figure~\ref{fig:erasurebounding}) based on a critical
value for $\lambda(k)$ and $k$. Let $\lambda^*$ from
(\ref{eqn:optimumlambda}) be the $\lambda$ that minimizes the exponent
$\frac{D(\lambda R' || 1 - \beta)}{1 - \lambda}$. Set $\bar{n} = d
\frac{D(\lambda^*R' || 1 - \beta)}{(1-\lambda^*)D(R' || 1 -
\beta)}$. Let $\bar{k}$ be the largest $k$ for which $n(k) >
\bar{n}$. For all $k < \bar{k}$,
\begin{eqnarray}
& & {\cal P}(\frac{\sum_{t=\lceil \frac{k}{R'}
\rceil}^{\lceil \frac{i}{R'} \rceil + d} Z_t}{d + \lceil \frac{i}{R'}
\rceil - \lceil \frac{k}{R'} \rceil} \leq \frac{i-k}{d + \lceil
\frac{i}{R'} \rceil - \lceil \frac{k}{R'} \rceil}) \nonumber\\
& < &
\exp(-n(k) 
     (D\bigg(\lambda(k)R' + \frac{2}{n(k)} || 1 - \beta \bigg) - \epsilon_1)) \nonumber\\
& < &
\exp(-n(k) 
     (D\bigg(R' + \frac{2}{\bar{n}} || 1 - \beta \bigg) - \epsilon_1)) \nonumber\\
& = &
\exp(-(\bar{n} + (n(k) - \bar{n})) (D(R' + \frac{2}{\bar{n}} || 1 - \beta) -
\epsilon_1)) \nonumber\\
& = &
\exp\left(-d \frac{D(\lambda^*R' || 1 - \beta)}{1-\lambda^*}(\frac{D(R' +
\frac{2}{\bar{n}} || 1 - \beta) - \epsilon_1}{D(R' || 1 - \beta)})\right)
\exp\left(-(n(k) - \bar{n})(D(R' + \frac{2}{\bar{n}} || 1 - \beta) -
  \epsilon_1)\right)  \label{eqn:caseksmall}.
\end{eqnarray}
Meanwhile, for $k \geq \bar{k}$, 
\begin{eqnarray}
{\cal P}(\frac{\sum_{t=\lceil \frac{k}{R'}
\rceil}^{\lceil \frac{i}{R'} \rceil + d} Z_t}{d + \lceil \frac{i}{R'}
\rceil - \lceil \frac{k}{R'} \rceil} \leq \frac{i-k}{d + \lceil
\frac{i}{R'} \rceil - \lceil \frac{k}{R'} \rceil}) 
& < &
\exp(-d \frac{D(\lambda(k)R' + \frac{2}{n(k)} || 1 - \beta) -
\epsilon_1}{1-\lambda(k)}) \nonumber\\
& \leq &
\exp(-d [\frac{D(\lambda(k)R' + \frac{2}{d} || 1 - \beta)}{1 - \lambda(k)} -
\frac{\epsilon_1}{1-\lambda^*}]). \label{eqn:caseklarge}
\end{eqnarray}
If there were no $\frac{2}{d}$ term above, then the terms
(\ref{eqn:caseklarge}) could be bounded by using $\lambda^*$ in place
of $\lambda(k)$ since $\lambda^*$ is the worst possible $\lambda$. But
since the divergence is continuous in its first argument and
$\frac{2}{d}$ is small, we bound them all by allowing for a small slop
$\epsilon_2$. Explicitly, for every $\epsilon_2 > 0$, it is clear
there exists a 
$d_2(\epsilon_2) > 0$ so that for all $d > d_2(\epsilon_2)$ and $k$
such that $n(k) \leq \bar{n}$, we have
\begin{eqnarray}
{\cal P}(\frac{\sum_{t=\lceil \frac{k}{R'}
\rceil}^{\lceil \frac{i}{R'} \rceil + d} Z_t}{d + \lceil \frac{i}{R'}
\rceil - \lceil \frac{k}{R'} \rceil} \leq \frac{i-k}{d + \lceil
\frac{i}{R'} \rceil - \lceil \frac{k}{R'} \rceil}) 
& < &
\exp(-d [(1-\epsilon_2) \frac{D(\lambda^*R' || 1 - \beta)}{1 - \lambda^*} -
\frac{\epsilon_1}{1-\lambda^*}]). \label{eqn:caseklarge2}
\end{eqnarray}
Putting the two bounds (\ref{eqn:caseksmall}) and
(\ref{eqn:caseklarge2}) together for $d > \max(d_1(\epsilon_1),
d_2(\epsilon_2))$ gives
\begin{eqnarray*}
& & {\cal P}(\widehat{B}_i(\lceil \frac{i}{R'} \rceil + d) \neq B_i) \\
& \leq &
\sum_{k=1}^i {\cal P}\left(\frac{1}{d + \lceil \frac{i}{R'} \rceil - 
\lceil \frac{k}{R'} \rceil} 
\sum_{t=\lceil \frac{k}{R'} \rceil}^{\lceil \frac{i}{R'} \rceil + d} 
Z_t \leq \frac{i-k}
{d + \lceil \frac{i}{R'} \rceil - \lceil \frac{k}{R'} \rceil}
\right) \\
& < &
[\sum_{k=1}^{\bar{k}} \exp(-d \frac{D(\lambda^*R' || 1 -
\beta)}{1-\lambda^*}(\frac{D(R' + \frac{2}{\bar{n}} || 1 - \beta) -
\epsilon_1}{D(R' || 1 - \beta)}))
  \exp(-(n(k) - \bar{n})(D(R' + \frac{2}{\bar{n}} || 1 -
 \beta) - \epsilon_1))] \\
& & 
+ (i - \bar{k} + 1) \exp(-d [(1-\epsilon_2) \frac{D(\lambda^*R' || 1 -
\beta)}{1 - \lambda^*} - \frac{\epsilon_1}{1-\lambda^*}]) \\
& < &
\exp(-d \frac{D(\lambda^* R' || 1 - \beta)}{1-\lambda^*} 
(\frac{D(R' + \frac{2}{\bar{n}} || 1 - \beta) - \epsilon_1}
      {D(R' || 1 - \beta)}
))
 \left[ \lceil \frac{1}{R'} \rceil 
 \sum_{l=0}^{\infty} 
 \exp(-l (D(R' + 2\bar{n}^{-1} || 1 - \beta) - \epsilon_1)) \right] \\ 
& & 
+ d(\frac{D(\lambda^*R' || 1 - \beta)}
         {(1-\lambda^*)D(R' || 1 - \beta)})
  \exp(-d \bigg((1-\epsilon_2) \frac{D(\lambda^*R' || 1 - \beta)}
                               {1 - \lambda^*} 
       - \frac{\epsilon_1}{1-\lambda^*} \bigg)).
\end{eqnarray*}

\subsection{Proof of Lemma
  \ref{lem:transmissiontimes}} \label{app:lemtransmissiondelay} 
That the transmission times $T_j$ are iid is obvious since each
depends on disjoint channel uses and the channel is memoryless and
stationary. Before proving (\ref{eqn:bettertransmissionbound}), it is
useful to first establish 
\begin{equation} \label{eqn:transmissionbound}
{\cal P}(T_j > t ck) \leq \exp(\rho R nck)
[\exp(-ck E_0(\rho,\vec{q}))]^t.
\end{equation}
The only way that the transmission time can be longer than $tck$ for
some integer $t \geq 1$ is if the  block-length-$tck$ code cannot be
correctly decoded to within a list of size $l$. The effective rate of
the block code in nats is thus
\begin{eqnarray*}
\frac{\frac{nck}{R}}{tck} 
& = & R\frac{n}{t}.
\end{eqnarray*}
Applying the list-decoding upper-bound (\ref{eqn:listrhorandom}) on
the probability of error for random block coding gives 
\begin{eqnarray*}
{\cal P}(T_j > t ck) 
& \leq & 
\exp(-tck[E_0(\rho,\vec{q}) - \rho R\frac{n}{t}]) \\
& = &
\exp(\rho R nck)
\exp(-tck E_0(\rho,\vec{q})) \\
& = &
\exp(\rho R nck)
[\exp(-ck E_0(\rho,\vec{q}))]^t
\end{eqnarray*}
where this holds for all $0 \leq \rho \leq 2^l$. Pulling the constant
into the exponent gives
\begin{eqnarray*}
{\cal P}(T_j > t ck) 
& \leq & 
[\exp(-ck E_0(\rho,\vec{q}))]^{t - \frac{\rho R nck}{ck
    E_0(\rho,\vec{q})}} \\
& = & 
[\exp(-ck E_0(\rho,\vec{q}))]^{t - \frac{\rho R n}{E_0(\rho,\vec{q})}}
  \\
& \leq & 
[\exp(-ck E_0(\rho,\vec{q}))]^{t - \lceil \frac{\rho R
    n}{E_0(\rho,\vec{q})} \rceil} \\
& = & 
[\exp(-ck E_0(\rho,\vec{q}))]^{t - \lceil \widetilde{t}(\rho,R,n,\vec{q}) \rceil}
\end{eqnarray*}
and this proves the desired result. \hfill $\Diamond$ \vspace{0.15in}

\bibliographystyle{./IEEEtran}
\bibliography{./IEEEabrv,./MyMainBibliography} \end{document}